\documentclass[aps,preprint,noshowpacs,superscriptaddress,groupedaddress]{revtex4}
\usepackage{newtxtext,newtxmath}

\usepackage{graphicx}
\usepackage[a4paper,margin=25.4mm]{geometry}

\usepackage[dvipsnames]{xcolor}
\usepackage[normalem]{ulem}
\definecolor{goodgreen}{rgb}{0.1,0.5,0}
\definecolor{goodred}{rgb}{0.7,0,0}
\usepackage{float}
\usepackage{multirow}
\usepackage{bm}
\usepackage{siunitx}
\usepackage{xcolor}

\usepackage{bibunits}

\usepackage{braket}
\usepackage{booktabs} 
\usepackage{enumitem}

\usepackage{setspace}

\usepackage{ragged2e}
\usepackage{caption}
\newcommand{\customcaption}[1]{%
  \caption{\justifying #1}%
}
\newcommand{\customtablecaption}[1]{%
  \captionof{table}{\justifying #1}%
}

\makeatletter
\renewcommand{\p@subsection}{}
\renewcommand{\p@subsubsection}{\thesubsection.}
\makeatother

\makeatletter
\renewcommand{\l@subsubsection}[2]{}
\makeatother

\makeatletter
\newcommand{\SectionNoPage}[1]{%
  \refstepcounter{section}% increment counter for references
  \addtocontents{toc}{%
    \protect\contentsline{section}{\textbf{#1}}{}{}%
  }%
  {\large\bfseries #1\par}%
  \vspace{1ex}
}
\makeatother

\makeatletter
\renewcommand{\l@subsection}[2]{%
  \setlength\@tempdima{2.5em}
  \begingroup
    \leavevmode
    \advance\leftskip\@tempdima
    \hskip -\leftskip
    #1\leaders\hbox{.\kern0.5em}\hfill
    \nobreak
    \hb@xt@\@pnumwidth{\hss #2}\par
  \endgroup
}
\makeatother

\makeatletter
\newcommand{\SectionNoTOC}[1]{%
  {\large\bfseries #1\par}%
  \vspace{1ex}
}
\makeatother

\makeatletter
\newcommand{\putbibNoTOC}[1]{%
  \begingroup
  \let\addcontentsline\@gobblethree
  \putbib[#1]%
  \endgroup
}
\makeatother



\DeclareUnicodeCharacter{202F}{\,}

\newcommand{\sx}{\sigma_x}

\newcommand{\sz}{\sigma_z}

\newcommand{\im}{\mathrm{Im}} 
\newcommand{\re}{\mathrm{Re}}

\newcommand{\C}{\mathrm{C}}

\newcommand{\D}{\mathrm{D}} 

\newcommand{\M}{\mathrm{M}} 
\newcommand{\EM}{\mathrm{EM}} 

\newcommand{\EC}{\mathrm{EC}}  
\newcommand{\E}{\mathrm{E}} 
\newcommand{\dif}{\mathrm{d}}

\newcommand{\rchi}{\raisebox{0.4ex}{$\chi$}}

\makeatletter
\renewcommand{\fnum@figure}{\textbf{Figure \thefigure}}
\renewcommand{\fnum@table}{\textbf{Table \thetable}}
\makeatother

\usepackage{url}
\usepackage[colorlinks,urlcolor=goodgreen,citecolor=blue,linkcolor=goodred]{hyperref}

\makeatletter
\def\hyper@natlinkstart#1{%
  \Hy@backout{#1}%
  \hyper@linkstart{cite}{cite.\the\@bibunitauxcnt.#1}%
  \def\hyper@nat@current{#1}%
}

\def\hyper@natanchorstart#1{%
  \Hy@raisedlink{%
    \hyper@anchorstart{cite.\the\@bibunitauxcnt.#1}%
  }%
}
\makeatother

\newcommand{\ICFO}{\affiliation{\small ICFO -- Institut de Ci\`encies Fot\`oniques, The Barcelona Institute of Science and Technology, 08860 Castelldefels (Barcelona), Spain}}

\newcommand{\NBI}{\affiliation{\small Center for Hybrid Quantum Networks (Hy-Q), Niels Bohr Institute,\\
University of Copenhagen, DK-2200 Copenhagen, Denmark}}

\newcommand{\CHIRAL}{\affiliation{\small Chiral Nano AG, Ueberlandstrasse 129, 8600 Dübendorf, Switzerland}}

\newcommand{\UCHICAGO}{\affiliation{\small Pritzker School of Molecular Engineering, University of Chicago, IL 60439, USA}}

\newcommand{\ANL}{\affiliation{\small Center for Nanoscale Materials, Argonne National Laboratory, Argonne, IL 60439, USA\\ 
\footnotesize
\normalfont
$^{*}$ These authors contributed equally\\
$^{\dagger}$ Correspondence: christoffer.moller@nbi.ku.dk; roger.tormo@icfo.eu; adrian.bachtold@icfo.eu\\
\phantom{space}}
}

\newcommand{\UBORDEAUX}{\affiliation{\small Universit\'{e} de Bordeaux, CNRS, LOMA, UMR 5798, F-33400 Talence, France}}

\newcommand{\TITLE}{\Large 
Tunable nonlinear electromechanics \\at the zero-point motion scale
}

\begin{document}

\title{\TITLE}

\author{C. B. M{\o}ller*$\dagger$}
\ICFO
\NBI
\author{R. Tormo-Queralt*$\dagger$}
\ICFO
\author{E. Vázquez-Rodríguez}
\ICFO
\author{V. Román-Rodríguez}
\ICFO
\author{M. Cagetti}
\ICFO
\author{E. Mateos-Madinabeitia}
\ICFO
\author{J. C. Franz}
\UBORDEAUX
\author{S. Forstner}
\ICFO
\author{S. L. De Bonis}
\ICFO
\author{L. Ornago}
\CHIRAL
\author{M. El Abbassi}
\CHIRAL
\author{S. Jung}
\CHIRAL
\author{A. N. Cleland}
\UCHICAGO
\author{D. A. Czaplewski}
\ANL
\author{F. Pistolesi}
\UBORDEAUX
\author{A. Bachtold$\dagger$}
\ICFO

\begin{abstract} \bfseries \boldmath
\vspace{0cm}
Nonlinearity at the scale of zero-point motion opens new possibilities for the control and readout of nanomechanical systems, but achieving this remains a formidable challenge. Here we demonstrate that ultrastrong coupling (USC) between a nanotube mechanical oscillator and a double-quantum-dot electronic two-level system enables a mechanical Kerr (Duffing) nonlinearity at the zero-point motion scale. In the dispersive regime, this large coupling yields a mechanical anharmonicity of $\alpha = 1.4\%$—three orders of magnitude larger than in previous work—while preserving the predominantly mechanical nature of the lowest energy states. We further demonstrate a purely quadratic cavity-based continuous readout of the mechanical motion. This continuous nonlinear optomechanical readout is enforced by a double-quantum dot symmetry, which can be broken by gate tuning to introduce a large linear transduction. These results establish a tunable USC platform that enables strong mechanical anharmonicity and nonlinear continuous readout at the zero-point motion scale.
\end{abstract}

\maketitle

\newpage
\begin{bibunit}[naturemag_modified]
% \begin{bibunit}[apsrev4-2]

In a nonlinear mechanical resonator, the restoring force is no longer simply proportional to displacement~\cite{bachtold2022mesoscopic,eichler2023classical}. As a result, the resonance frequency can depend on displacement amplitude, the driven response can become bistable, and nonlinear coupling can transfer energy between different modes. This physics has been used to access fluctuation-induced switching and bifurcation dynamics~\cite{AldridgeClelandPRL2005,KozinskyPRL2007,HanPRR2024}, phase-sensitive amplification and squeezing~\cite{Pirkkalainen2015a,HuberPRX2020}, phonon lasing~\cite{MahboobYamaguchiPRL2013,WendtNature2026} and frequency-comb generation~\cite{CzaplewskiPRL2018,KeskeklerAlijaniNanoLett2022,OchsWeigPRX2022}. 
However, in most micro- and nanomechanical resonators the intrinsic nonlinearities are weak, so these effects become visible only at large motional amplitudes, as discussed in Ref.~\cite{samanta2023nonlinear}.
Accessing nonlinear mechanics near the zero-point motion scale requires sizeable anharmonicity of the mechanical spectrum, creating new opportunities for controlling mechanical motion.

Coupling mechanical motion to a two-level system~\cite{chu2017quantum,Satzinger2018,Kettler2021,AresNatureCom2025,yuksel2026intrinsic} offers a way to engineer such nonlinearities, since the mechanical oscillator inherits the anharmonicity of the two-level system. This approach has recently been demonstrated by coupling a large-mass, bulk-acoustic-wave mechanical resonator to a superconducting qubit~\cite{Yang2024Science_MechanicalQubit}, but the achieved mechanical anharmonicity remained small and 
the readout of the nonlinear mechanical spectrum was indirect, relying on resonant-interaction pulsed qubit spectroscopy measurements.
Carbon nanotube mechanical oscillators are especially attractive in this context: their low mass gives rise to a very large zero-point motion, which enhances the coupling between vibrations and the two-level system. In suspended nanotubes, the capacitive interaction between mechanical displacement and single-electron dynamics can be particularly strong, reflecting the large Coulomb forces generated between the electron and nearby gates~\cite{Pistolesi2015}. Previous nanotube electromechanical experiments have shown electron-mediated frequency softening~\cite{Lassagne2009,Steele2009,Benyamini2014}, self-oscillation~\cite{Wen2020}, and ultrastrong coupling between mechanical motion and tunneling electrons~\cite{samanta2023nonlinear,Vigneau2022_PRR}.

Here we realize a nanotube electromechanical platform operating in the dispersive ultrastrong-coupling regime. The system consists of a carbon nanotube mechanical oscillator dispersively coupled to an electronic two-level system (ETLS), characterized by a bare mechanical frequency $\omega_\mathrm{M}^0/2\pi = \SI{0.80}{GHz}$, an electromechanical coupling strength $g_\mathrm{EM}/2\pi = \SI{0.46}{GHz}$, and an ETLS frequency tuned electrostatically around $\omega_\mathrm{E}^0/2\pi \sim \SI{10}{GHz}$. This parameter set satisfies both the far-detuned dispersive condition $| \omega_\mathrm{M}^0 - \omega_\mathrm{E}^0 | \gg g_\mathrm{EM}$ and the ultrastrong coupling criterion $g_\mathrm{EM} > 0.1\cdot \omega_\mathrm{M}^0$~\cite{Forn-Diaz2019_RMP,Kockum2019_NatRevPhys}. 
In agreement with the system parameters measured independently, we show strong nonlinearity in both the mechanical dynamics and its continuous readout at the zero-point-motion scale.

\vspace{1.0\baselineskip}
\textbf{System description}

\noindent We consider the hybrid system where a mechanical oscillator is coupled to an ETLS, which is also coupled to a readout cavity (Figs.~1a,b).
The full system Hamiltonian is described by
\begin{equation}
H/\hbar =
\omega^0_\mathrm{M} a^\dagger a + \omega^0_\mathrm{C} b^\dagger b + 2t_\mathrm{E} \frac{\sigma_\mathrm{z} }{2} +
\left[\varepsilon + 2g_\mathrm{EM} (a + a^\dagger) + 2g_\mathrm{EC}\left(b + b^\dagger\right)
\right]\frac{\sigma_\mathrm{x} }{2}, 
    \label{eq:hamiltonian}
\end{equation}
where $a$, $b$ are annihilation operators of the mechanical and cavity modes, $\sigma_\mathrm{x,z}$ are the Pauli operators for the ETLS, $\omega^0_\mathrm{C}$ is the bare cavity frequency, $\omega_\mathrm{E}^0=\sqrt{(2t_\mathrm{E})^2+\varepsilon^2}$ is the ETLS transition frequency with $2t_\mathrm{E}$ the tunnel coupling rate as discussed below, and $g_\mathrm{EC}$ 
denotes the coupling strength for the ETLS–cavity interaction. 
In the limit $\omega^0_\mathrm{M},\omega^0_\mathrm{C}\ll 2t_\mathrm{E}$, one can diagonalize the Hamiltonian in a Born-Oppenheimer picture, by considering the mechanical and cavity degrees of freedom classically and quantizing them afterwards (see SI~2B and 2G). 
The mechanical oscillator acquires an effective Kerr nonlinearity arising from the coupling to the ETLS, which remains sizeable at the zero-point motion scale~\cite{Pistolesi2021}.
The origin of this nonlinearity can be understood intuitively from symmetry. While the mechanical motion $x=x_\mathrm{zpf}(a+a^\dag)$ linearly modulates the ETLS detuning as $\varepsilon \rightarrow \varepsilon + 2g_\mathrm{EM} x/x_\mathrm{zpf}$, the ETLS energy depends on the square of this effective detuning. Thus, at zero detuning ($\varepsilon=0$), opposite mechanical displacements produce the same change in energy and the mechanical frequency shift scales quadratically with small displacements. 
For driven motion, we get
\begin{equation}
    \Delta\omega_\mathrm{M} =  3\frac{g_\mathrm{EM}^4}{(2t_\mathrm{E})^3} \left(\frac{x_\mathrm{0}}{x_\mathrm{zpf}}\right)^2,
    \label{eq:mechanicalfrequencyshift}
\end{equation}
with the linear contribution being suppressed to zero. Here $x_0$ denotes the amplitude of the driven mechanical motion, and it can be smaller than the zero-point motion $x_\mathrm{zpf}$. 

Similarly, the mechanical displacement readout via the cavity exhibits a frequency shift of the form 
\begin{equation}
    \Delta\omega_\mathrm{C} = 6\frac{g_\mathrm{EC}^2g_\mathrm{EM}^2}{(2t_\mathrm{E})^3} \left(\frac{x_\mathrm{0}}{x_\mathrm{zpf}}\right)^2,
    \label{eq:cavityfrequencyshift}
\end{equation}
while the usual linear displacement readout is suppressed to zero by symmetry for $\varepsilon=0$. 
The purely quadratic dependence of the readout provides direct access to the oscillator’s energy and thus to its averaged phonon number.
An additional appealing feature is that the cavity frequency shift serves as a direct, absolute measure of the displacement oscillation amplitude in units of the zero-point motion.

Central to our work is the parameter regime where $\omega_\mathrm{E}^0=2t_\mathrm{E}$ is about one order of magnitude larger than $g_\mathrm{EM}$ and $\omega_\mathrm{M}$. This configuration enables both a large Kerr nonlinearity and a purely quadratic readout, which scale as $g_\mathrm{EM}^4/(2t_\mathrm{E})^3$ and $g_\mathrm{EM}^2/(2t_\mathrm{E})^3$, respectively (Eqs.~2 and 3), while preserving the validity of the dispersive approximation.  
Keeping the ETLS frequency about an order of magnitude above the mechanical frequency ensures that the low-lying eigenstates remain in the ETLS ground-state branch and are well separated from the excited branch (see SI~2B). Thus, the lowest states are predominantly mechanical with only weak electrical admixture.

Our device consists of a suspended carbon nanotube hosting a double-quantum dot, which also acts as a mechanical oscillator (Fig.~1a). The nanotube~\cite{Wen2020,urgell2020cooling,neukelmance2025microsecond,riechert2025carbon} is a small band gap semiconductor whose chemical potential is tuned using five underlying gate electrodes. The gates exploit the nanotube’s band gap to form tunnel barriers, confining electrons into two adjacent quantum dots coupled through a tunable central barrier~\cite{Tormo2022}. When the potential difference between the two dots is set to $\varepsilon=0$, a single electron becomes delocalized across the two sites through the tunnel coupling $2t_\mathrm{E}$, forming the bonding and anti-bonding eigenstates of the ETLS. By placing the quantum dots symmetrically at the center of the suspended section, the coupling to the second (antisymmetric) flexural mechanical mode is maximized (Fig.~1a). 
The second flexural mode is experimentally identified by its prominent frequency softening at zero detuning, its frequency being about twice that of the first mode (the first five modes scaling as $\omega_i\simeq i\cdot \omega_1$ for integers $i>0$), and its distinctive optomechanical red-sideband response (see SI 1E).
The electromechanical coupling strength $g_\mathrm{EM}$ increases with the number of electrons confined in the quantum dots \cite{Pistolesi2021}.
The electromechanical system is read out dispersively via a superconducting microwave resonator fabricated on a nearby chip. The compact device design ensures that the mechanical mode frequency is high enough for the oscillator to approach the quantum regime upon cooling in a dilution cryostat. 

The double-quantum dot system allows for precise control of both charge occupation and tunnel coupling with the gates. Electrons can be added one by one to either of the two dots, starting from the band gap, where the system is in the  $(N_\mathrm{L},N_\mathrm{R})=(0,0)$ charge configuration (Fig.~1c). An interdot charge transition (ICT) at $(N_\mathrm{L}+1,N_\mathrm{R})\leftrightarrow(N_\mathrm{L},N_\mathrm{R}+1)$ indicates that an electron is delocalized between the two dots (Fig.~1d). 
These ICTs exhibit a finite quantum capacitance, which shifts the cavity resonance frequency and is detected through the change in phase and amplitude of a weak microwave probe tone (Fig.~1e). 
The largest shift occurs at 
$\varepsilon =0$, allowing this point to be clearly identified. The tunnel coupling $2t_\mathrm{E}$, entering in the ETLS frequency, is widely tunable via the central gate $\text{G3}$, which controls the potential barrier between the two dots. We extract $2t_\mathrm{E}$ from the cavity response by sweeping the detuning $\varepsilon$ or the temperature, predominantly altering the ETLS frequency or its thermal population, respectively (Figs.~2a and 2b, see SI~1D). Both methods yield consistent results within measurement uncertainty (Fig.~2c).

The ultrastrong coupling $g_\mathrm{EM}$ between the ETLS and the mechanical oscillator leads to a significant suppression of the mechanical frequency, nearly half of which arises from the Bloch–Siegert shift in the far off-resonant coupling regime. The measured mechanical frequency suppression is maximal at $\varepsilon = 0$ and vanishes at large detuning, in agreement with theoretical predictions (inset of Fig.~3a; see SI~2B). The zero-amplitude mechanical frequency ${\omega}_\mathrm{M}/2\pi = \SI{0.74}{GHz}$ at $\varepsilon = 0$ is quantified by extrapolating the measured response to zero drive amplitude.
We extract the coupling strength $g_\mathrm{EM}/2\pi = \SI{0.46\pm0.05}{GHz}$, from ${\omega}_\mathrm{M}$ measured at different $2t_\mathrm{E}$ values using
\begin{equation}
{\omega}_\mathrm{M} = \omega_\mathrm{M}^0 \left[1 - \frac{g_\mathrm{EM}^2}{ \omega_\mathrm{M}^0}\left(\frac{1}{2t_\mathrm{E}-\omega_\mathrm{M}^0}+\frac{1}{2t_\mathrm{E}+\omega_\mathrm{M}^0}\right)\right].
\label{eq:mechanicalfrequencyrenormalization}
\end{equation}
The terms $\frac{1}{2t_\mathrm{E}-\omega_\mathrm{M}^0}$ and $\frac{1}{2t_\mathrm{E}+\omega_\mathrm{M}^0}$ correspond to the resonant (rotating-wave) and anti-resonant (Bloch–Siegert) contributions, respectively. In the adiabatic limit $2t_\mathrm{E} \gg \omega_\mathrm{M}^0$, these contributions become comparable, implying that the observed mechanical frequency suppression in our measurements with large $2t_\mathrm{E}$ arises equally from both processes.

\vspace{1.0\baselineskip}
\textbf{Nonlinearity in both mechanical dynamics and readout at the zero-point motion scale}

\noindent Dispersive USC enables a continuous displacement readout with purely quadratic character for driven mechanical amplitudes near the zero-point motion. 
The scaling $\Delta\omega_\mathrm{C} \propto x_\mathrm{0}^2$ is experimentally verified by measuring the cavity frequency shift $\Delta\omega_\mathrm{C}$ as a function of a coherent ETLS detuning drive amplitude $\varepsilon_\mathrm{D}$ at the mechanical resonance frequency (Fig.~3b). 
The observed quadratic scaling $\Delta\omega_\mathrm{C} \propto \varepsilon_\mathrm{D}^2$ over more than one order of magnitude at low drive confirms the $x^2$-sensitivity of the readout, as the  mechanical displacement amplitude is expected to scale linearly with the drive amplitude.
In this regime, using Eq.~3 to infer $x_\mathrm{0}$ from the measured $\Delta\omega_\mathrm{C}$, we find that the smallest resolved driven displacements fall below the zero-point motion. 
This represents a significant advance, as previous demonstrations of $x^2$-based readout required mechanical displacements far exceeding the zero-point motion~\cite{thompson2008strong,lee2015multimode,brawley2016nonlinear,leijssen2017nonlinear}.

The driven mechanical oscillator exhibits nonlinear dynamics at displacements approaching the zero-point motion (Fig.~3c). At small amplitudes, the observed mechanical frequency shift scales quadratically with displacement, in agreement with Eq.~2. A fit to the data (dashed black line) yields a mechanical Kerr nonlinearity of $K = 2\pi\cdot\SI{2.4\pm0.3}{MHz}$ (see SI~2.H). This value agrees with the theoretical prediction $K = 12g_\mathrm{EM}^4/(2t_\mathrm{E})^3 = 2\pi\cdot\SI{1.3\pm0.8}{MHz}$ obtained from independently extracted cavity, electronic, and mechanical response parameters (see SI~1A), and corresponds to an anharmonicity of $\alpha=\SI{0.3}{\percent}$. So far, only the system in Ref.~\cite{Yang2024Science_MechanicalQubit} reported a Kerr nonlinearity at the zero-point motion scale, while in other systems nonlinear effects are observed at amplitudes at least one order of magnitude larger than the zero-point motion (see Ref.~\cite{samanta2023nonlinear}). In the present system, the anharmonicity can be tuned up to $\alpha=\SI{1.4}{\percent}$ by varying $2t_\mathrm{E}$ (Fig.~4b), corresponding to an increase of three orders of magnitude compared to Ref.~\cite{Yang2024Science_MechanicalQubit}. We evaluate $\alpha$ as $\alpha =(\omega_\mathrm{21}-\omega_\mathrm{10})/\omega_\mathrm{10}=K/\omega_\mathrm{10}$, with $\omega_\mathrm{ij}$ the transition frequency between eigenstates of Eq.~1 for phonon numbers $i$ and $j$. 

At larger displacement amplitudes, the measured shifts in both the mechanical and cavity frequencies deviate from the expected quadratic $x^2$ dependence (Figs.~3b,d). This deviation becomes significant when $2g_\mathrm{EM}x_0/x_\mathrm{zpf}$ approaches $2t_\mathrm{E}$
as the ETLS becomes strongly modified by the mechanical motion. 
The impact of a finite mechanical amplitude $x_0$ on the 
cavity shift is thus evaluated numerically by simulating both the oscillator dynamics and the resulting cavity response, while for small displacements it reduces to the analytical form given in Eq.~3 (see SI~1H).
Remarkably, despite this non-quadratic behavior, the two frequency shifts $\Delta{\omega}_\mathrm{M}$ and $\Delta\omega_\mathrm{C}$ measured with the same $x_0$ maintain a linear relationship over a broad range of amplitudes (Fig.~3e). This reveals a common nonlinear mechanism in which mechanical oscillations shift the ETLS frequency, thereby inducing nonlinear frequency shifts in both the mechanical and cavity resonances (Eqs.~2 and~3). The observed slope of $0.015\pm0.001$ is in reasonable agreement with the theoretical slope expected in the small-displacement regime, $2\left(g_\mathrm{EC}/g_\mathrm{EM}\right)^2 = 0.023\pm0.006$, using $g_\mathrm{EC}/2\pi= \SI{49.8\pm2.5}{MHz}$. 
All these measured nonlinearities were observed in a second device (see SI~1I).

\vspace{1.0\baselineskip}
\textbf{Highly tunable hybrid system}

\noindent Gate voltages provide broad tunability, allowing control over the mechanical frequency $\omega_\mathrm{M}$, the electromechanical coupling $g_\mathrm{EM}$, the ETLS frequency $\omega_\mathrm{E}^0 = 2t_\mathrm{E}$, and the mechanical nonlinearity $K$—with the latter two spanning more than one and three orders of magnitude, respectively (Figs.~2c and 4a,b). 
Remarkably, all of these variations are quantitatively captured by the system Hamiltonian, without requiring additional fitting parameters (see SI~1A). The readout interaction is equally tunable.
At zero detuning ($\varepsilon = 0$), the system supports a purely quadratic optomechanical response, $\Delta\omega_\mathrm{C}^{(2)} = g_\mathrm{OM}^{(2)} \cdot (x/x_\mathrm{zpf})^2$, since the linear term $\Delta\omega_\mathrm{C}^{(1)} = g_\mathrm{OM}^{(1)} \cdot (x/x_\mathrm{zpf})$ is suppressed by the symmetric frequency dispersion of the ETLS around $\varepsilon = 0$ (see SI~2G). Introducing a finite detuning $\varepsilon$ reestablishes the conventional linear optomechanical coupling. 
The linear term gives rise to optomechanical blue and red sidebands at $\omega_\mathrm{C} \pm {\omega}_\mathrm{M}(\varepsilon)$. 
Since the mechanical frequency is maximally renormalized by the ETLS at $\varepsilon=0$, as quantified by Eq.~4, the sideband frequencies exhibit the corresponding peak and dip observed in the two-tone spectroscopy data (Fig.~4d).
Both linear and quadratic couplings can be continuously tuned via $\varepsilon$ (Fig.~4e): the quadratic term maximizes at $\varepsilon = 0$, while increasing detuning enhances the linear term. This provides smooth control over the optomechanical interaction.

Despite its high tunability and strong nonlinearities, the system remains robust against noise. We use the mechanical linewidth $\delta {\omega}_\mathrm{M}$ as a proxy for dephasing (Fig.~4c). It is expected that electrical detuning noise and thermally induced mechanical frequency fluctuations both lead to a linewidth scaling as $1/(2t_\mathrm{E})^3$, but with different dependence on the coupling strength $g_\mathrm{EM}$: the linewidth contribution from electrical detuning noise scales as $g_\mathrm{EM}^2$, while that of thermal fluctuations combined with the Kerr nonlinearity scales as $g_\mathrm{EM}^4$ (see SI~2I). We distinguish these mechanisms by comparing two ICTs with different electromechanical couplings, $g_{\mathrm{EM}}^\mathrm{ICT45}/2\pi = \SI{0.46\pm0.05}{GHz}$ and $g_{\mathrm{EM}}^\mathrm{ICT01}/2\pi = \SI{0.18\pm0.02}{GHz}$. The measured linewidth ratio, extracted from the fitted $1/(2t_\mathrm{E})^3$ prefactors, is $46\pm1$, in agreement with the expected quartic scaling $\left(g_{\mathrm{EM}}^\mathrm{ICT45} / g_{\mathrm{EM}}^\mathrm{ICT01}\right)^4 = 43\pm27$. This identifies thermal fluctuations of mechanical modes as the dominant source of dephasing. Using independently measured self- and cross-Kerr contributions, we estimate a mechanical thermalization temperature of $47\pm\SI{5}{mK}$, consistent with both ICT configurations (see SI~1E). The finite thermal occupation broadens the mechanical linewidth but does not affect the Kerr nonlinearity inferred from the driven mechanical response.

Several technical improvements provide a clear route to reducing dephasing. Operating the system at lower temperature and increasing mechanical frequencies using shorter and mechanically tensioned nanotubes will lower thermal occupations. Additional fine tuning of the gate array further decouples the cross‑Kerr noise. Together these improvements should enable resolved single phonon transitions by boosting $K/\delta \omega_M$.
Our ultraclean suspended nanotubes, positioned far away from dielectrics to reduce charge noise~\cite{Tormo2022}, have the potential to significantly enhance the ETLS coherence time, as recently demonstrated in state-of-the-art charge qubits~\cite{zhou2022single,zhou2024electron}.

\vspace{0.75\baselineskip}
\textbf{Outlook}

\noindent 
Our nonlinear USC platform opens several new directions for quantum nanomechanics. The strong tunability of the mechanical frequency and nonlinearity provides a route toward a quantum bifurcation of the mechanical potential into a double-well configuration as $\omega_\mathrm{M}$ approaches zero~\cite{roda2024macroscopic}. Related protocols are being pursued with levitated particles~\cite{gonzalez2021levitodynamics}, where engineering double-well potentials with nanometer-scale separation between minima remains a major challenge. In contrast, our platform allows continuous tuning of the separation between the potential minima, providing a route toward spatially delocalized mechanical states.

The dispersive USC regime also provides a route toward a long-lived mechanical qubit~\cite{Pistolesi2021}. While the first mechanical qubit was demonstrated using near-resonant coupling~\cite{Yang2024Science_MechanicalQubit}, the dispersive regime in our platform is predicted to significantly enhance mechanical-qubit coherence by increasing $2t_\mathrm{E}/\omega_\mathrm{M}$~\cite{Pistolesi2021}. At the same time, the purely quadratic readout combined with the far-detuned regime could enable the observation of quantum jumps between different phonon states and engineered two-phonon processes, providing new routes toward the preparation and stabilization of nonclassical mechanical states~\cite{dykman1973quantum,tan2013generation,brunelli2018unconditional}.

\newpage

\begin{figure}
	\begin{center}
		\includegraphics[width=1\textwidth]{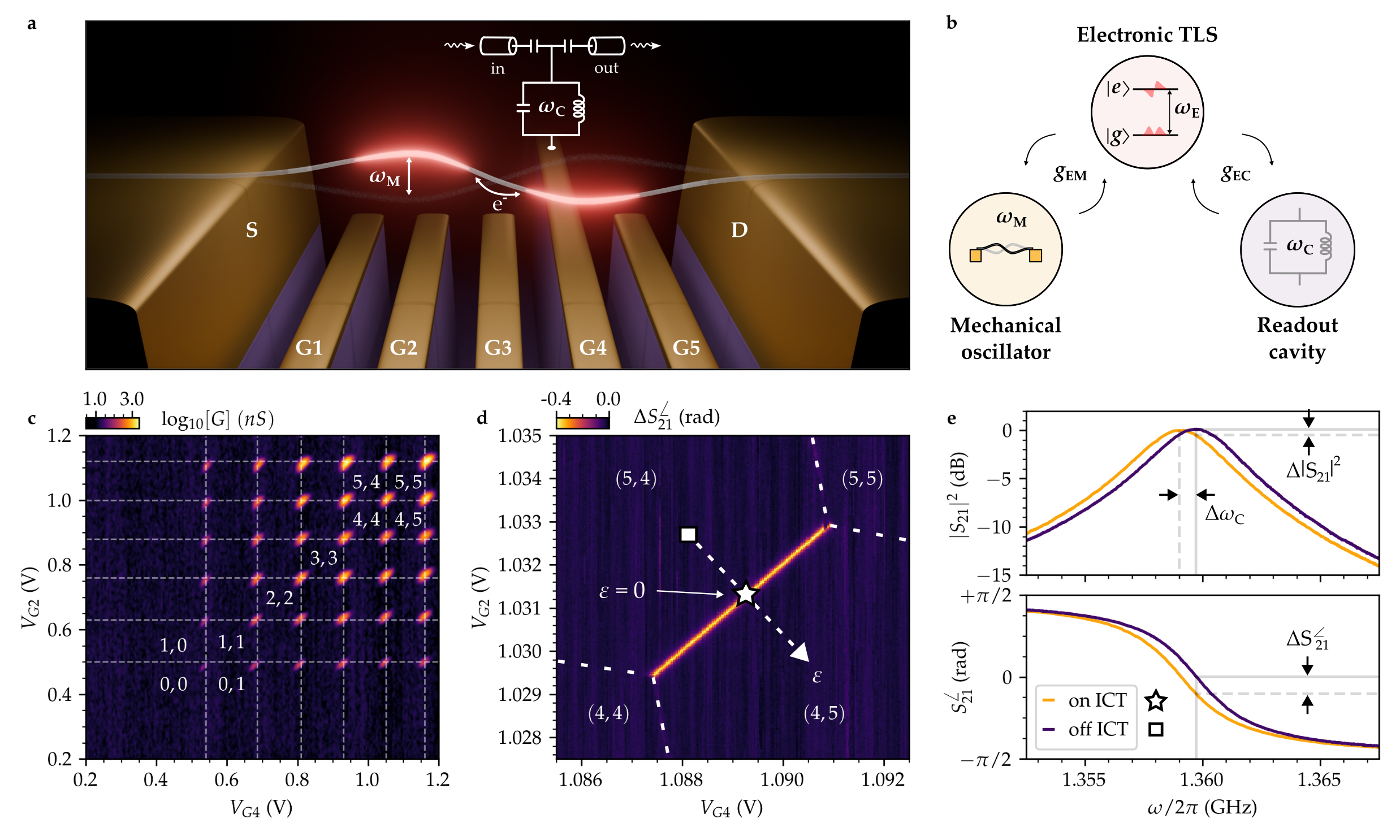}
        \customcaption{\textbf{Hybrid system overview.} \textbf{a.} Simplified representation of a double-quantum dot (red) embedded in a vibrating carbon nanotube suspended above five gate electrodes G1-G5, one of which is galvanically connected to a superconducting microwave resonator used for cavity readout. \textbf{b.} Schematic of the tri-partite hybrid system comprising a mechanical oscillator, an electronic two-level system (ETLS), and superconducting microwave cavity. \textbf{c.} 
        Charge stability diagram showing the double-quantum dot electron configuration controlled by the gate voltages $V_\text{G2}$ and $V_\text{G4}$, and measured via the electrical conductance $G$ at \SI{10}{K} with an applied source-drain bias $V_\text{SD}=\SI{3}{mV}$. 
        We continuously compensate gate cross capacitances and reduce the $G$ variation at the triple points by tuning $V_\text{G3} = 0.41-0.41(V_\text{G2} + V_\text{G4})/2$.
        \textbf{d}. Zoom-in of an interdot charge transition (ICT) in the charge stability diagram measured via the dispersive shift of the microwave cavity at a fixed $V_\text{G3} = \SI{-120}{mV}$. No $V_\text{SD}$ bias is applied and the cryostat is operated at 
        the \SI{20}{mK} cryostat base temperature.
        \textbf{e.} The cavity transmission (magnitude--top, phase--bottom) probed on and off the ICT.}
\label{fig1}
\end{center}
\end{figure}

\newpage
\phantom{bla}

\begin{figure}[tb!]
	\begin{center}
		\includegraphics[width=0.6\textwidth]{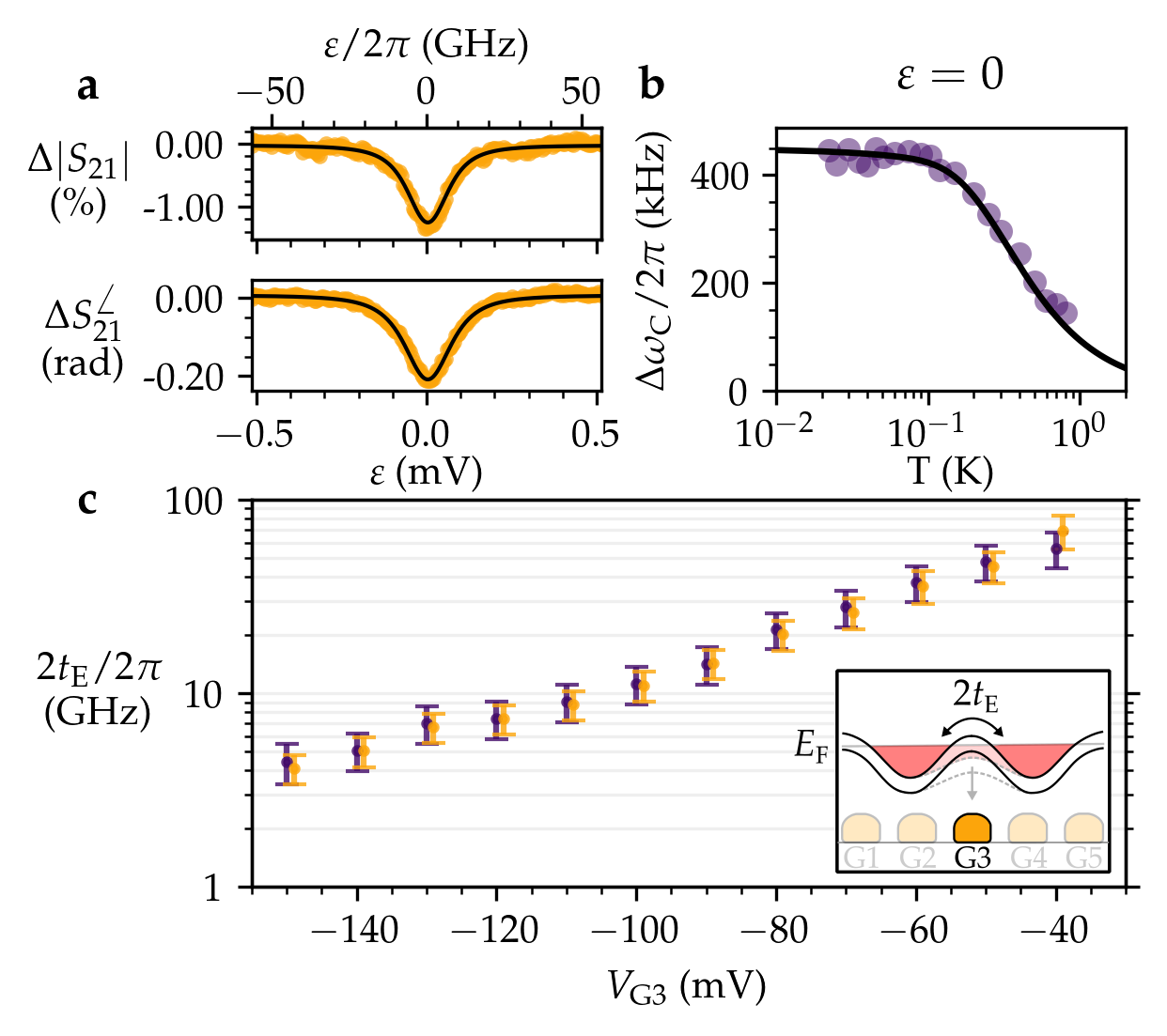}
        \customcaption{\textbf{Highly tunable electronic two-level system.} \textbf{a.} Magnitude and phase shift of the cavity transmission as a function of the ETLS detuning $\varepsilon$ measured at base temperature with $V_\text{G3}=\SI{-100}{mV}$. \textbf{b.} ETLS induced cavity frequency shift as a function of temperature measured at $\varepsilon=0$ and $V_\text{G3}=\SI{-100}{mV}$.
        \textbf{c.} ETLS frequency at $\varepsilon=0$ as a function of the control gate voltage $V_\text{G3}$. The ETLS frequency is quantified from the two independent measurement methods shown in (\textbf{a}) and (\textbf{b}). 
        The inset is a schematic of the gated nanotube electrostatic potential, relative to the Fermi energy $E_\mathrm{F}$, with the two quantum dots (red) formed by three tunnel barriers enabled by the nanotube bandgap. The tunnel coupling $2t_\text{E}$ is tuned with gate G3.  
        }
\label{fig2}
\end{center}
\end{figure}

\newpage
\phantom{bla}

\begin{figure}[tb!]
	\begin{center}
		\includegraphics[width=1\textwidth]{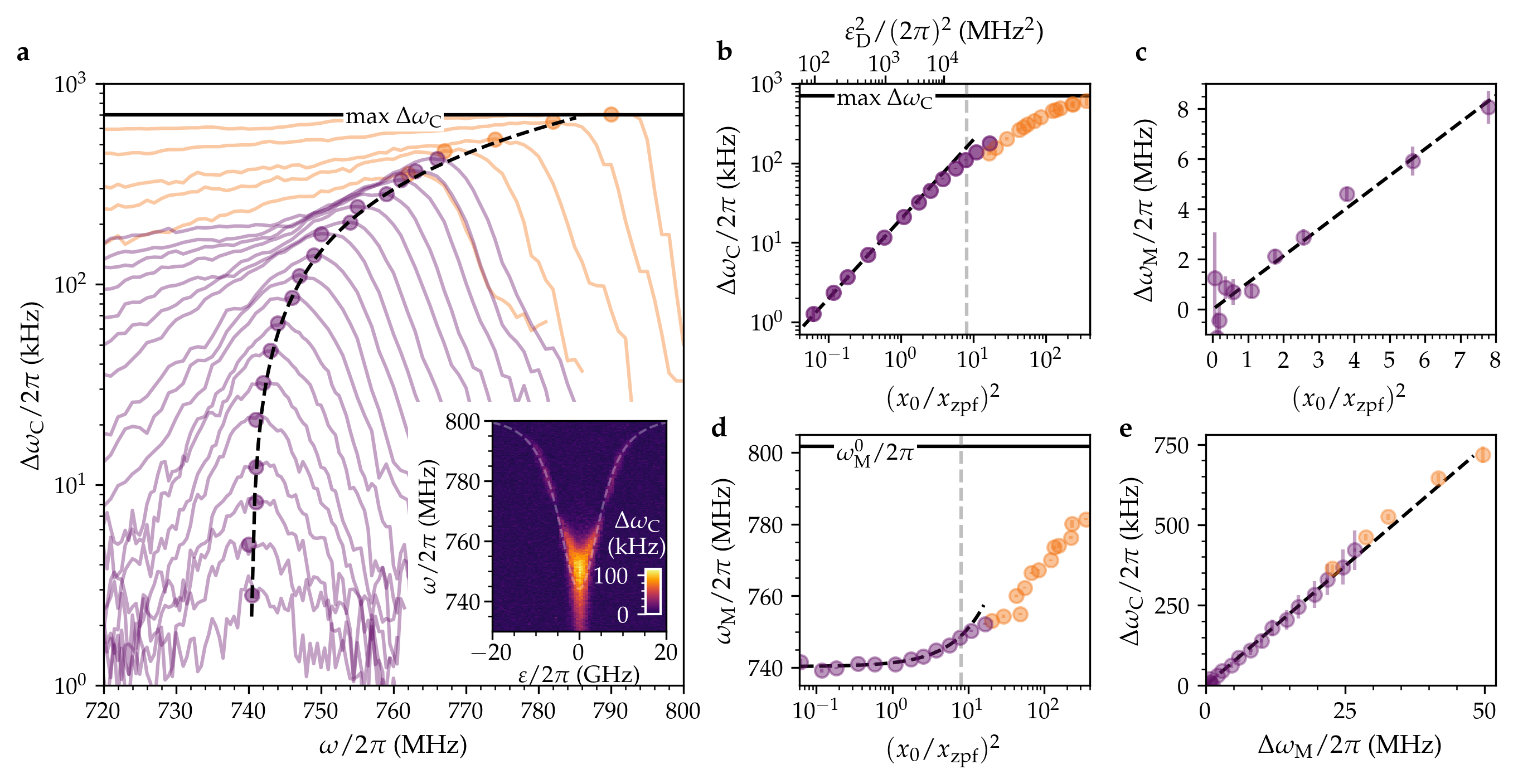}
		\customcaption{\textbf{Nonlinearity in both the mechanical motion and its readout.} 
        \textbf{a.} Mechanical spectra for different drives at $\varepsilon=0$ measured by two-tone spectroscopy with an AM (purple) and CW (orange) homodyne detection scheme. 
        Solid horizontal line is given by the maximum cavity frequency shift $\Delta\omega_\mathrm{C}$.
        Inset shows the mechanical frequency $\omega_\mathrm{M}$ suppression as a function of detuning. Dashed white line shows theoretical prediction of the suppression without any Kerr contribution.      
        \textbf{b.} Dependence of $\Delta\omega_\mathrm{C}$ on calibrated displacement in units of the zero-point motion. The measured $\Delta\omega_\mathrm{C}$ depends quadratically on the detuning drive (dashed line) consistent with a $x^2$-readout. The scaling persists until $2g_\mathrm{EM}x/x_\mathrm{zpf}\simeq 2t_\mathrm{E}$ (gray dashed line) where the mechanical oscillation washes out the effect of the ETLS.       
        \textbf{c, d.} $x^2$-dependence of $\Delta\omega_\mathrm{M}$ at low displacements (dashed line, Eq.~2) indicating Kerr nonlinearity at the zero-point motion scale. $\Delta\omega_\mathrm{M}$ is no longer described by Eq.~2 at large drive when $2g_\mathrm{EM}x/x_\mathrm{zpf}> 2t_\mathrm{E}$.
        \textbf{e.} Dependence of $\Delta\omega_\mathrm{C}$ on $\Delta\omega_\mathrm{M}$ quantified at the peaks of the mechanical spectra in (\textbf{a}) at $\varepsilon=0$. 
        This linear dependence is also represented in (\textbf{a}) (dashed black line) with $\omega_\mathrm{M}$ reaching \SI{0.740}{GHz} at small drive where the Kerr contribution becomes zero.   
        Data in all panels are recorded at base temperature \SI{20}{mK}, $g_\mathrm{EM}/2\pi=\SI{0.46}{GHz}$, and $2t_\mathrm{E}/2\pi=\SI{7.4}{GHz}$.
        }
\label{fig3}
\end{center}
\end{figure}
 
\newpage
\phantom{bla}

\begin{figure}[tb!]
	\begin{center}
		\includegraphics[width=1\textwidth]{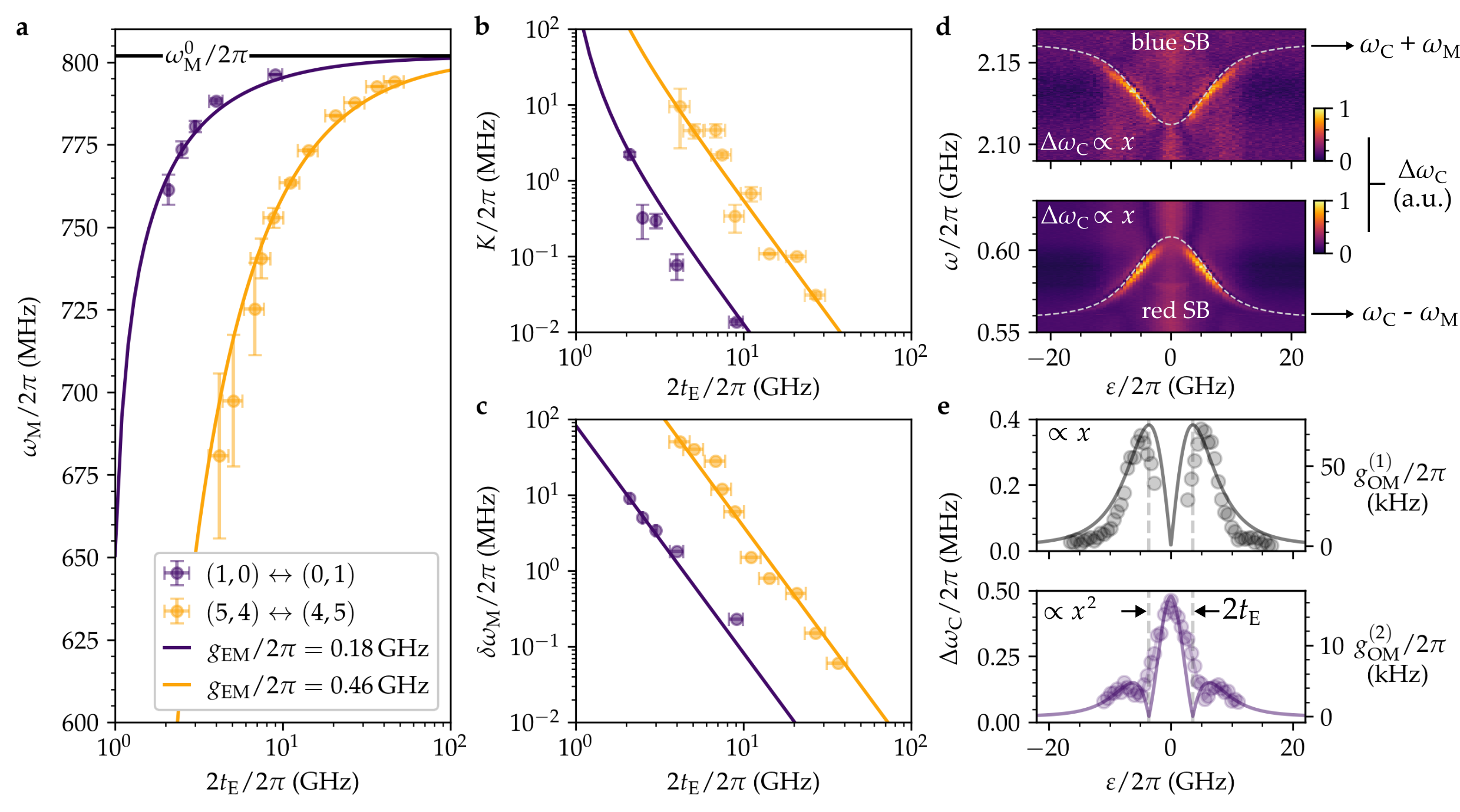}
		\customcaption{\textbf{Highly tunable hybrid system.} \textbf{a-c.} Mechanical frequency, Kerr nonlinearity, and linewidth of the mechanical resonance measured at $\varepsilon=0$ as a function of the ETLS tunnel coupling $2t_\mathrm{E}$ for two different electromechanical couplings reached at two ICT electron configurations. Solid lines are theory predictions in (\textbf{a}) \& (\textbf{b}), and fit to a $1/(2t_\mathrm{E})^3$ dependence in (\textbf{c}). \textbf{d.} Two-tone spectroscopy of the upper and lower optomechanical sidebands as a function of ETLS detuning at a fixed detuning drive amplitude. The shift of the measured sideband frequencies $\omega_\mathrm{C}\pm\omega_\mathrm{M}$ are consistent with the measured $\varepsilon$ dependence of $\omega_\mathrm{M}$.
        \textbf{e.} Observed cavity frequency shift as a function of the ETLS detuning as well as the theoretical linear and quadratic effective optomechanical coupling rates. Solid lines are theory and points are extracted from peak values of (\textbf{d}). Deviations between data and theory are attributed to the $\varepsilon$ dependent effective driving force and mechanical susceptibility.  
        Data in all panels are recorded at base temperature \SI{20}{mK}.
        }
\label{fig4}
\end{center}
\end{figure}

\clearpage

\putbibNoTOC{bibliography}

\end{bibunit}

%%%%%%%%%%%%%%%% ACKNOWLEDGEMENTS %%%%%%%%%%%%%%%
\newpage

\vspace{0.5cm}
\noindent \textbf{Acknowledgements}\\
We thank discussions with Victor Champain. 
AB and CBM acknowledge support from ERC Advanced Grant 101198268-QTube and Marie Sklodowska-Curie grant agreement No. 101023289. Work performed by DAC at the Center for Nanoscale Materials, a U.S. Department of Energy Office of Science User Facility, was supported by the U.S. DOE, Office of Basic Energy Sciences, under Contract No. DE-AC02-06CH11357. AB acknowledges support from MICINN Grant No. RTI2018-097953-B-I00 and PID2021-122813OB-I00, the Quantera grant (PCI2022-132951), the Fondo Europeo de Desarrollo, the Spanish Ministry of Economy and Competitiveness through Quantum CCAA, TED2021-129654B-I00, EUR2022-134050, and CEX2019-000910-S [MCIN/AEI/10.13039/501100011033], MCIN with funding from European Union NextGenerationEU(PRTR-C17.I1), Fundacio Cellex, Fundacio Mir-Puig, Generalitat de Catalunya through CERCA, 2021 SGR 01441, the predoctoral program AGAUR-FI ajuts (2025 FI-2 00011) Joan Oró and the predoctoral program FI-STEP (2025 STEP 00040), which are backed by the Secretariat of Universities and Research of the Department of Research and Universities of the Generalitat of Catalonia, as well as the European Social Plus Fund. FP acknowledges the French ANR through contract MORETOME (ANR-22-CE24-0020), the GPR and from the French government in the framework of the University of Bordeaux's France 2030 program/GPR LIGHT. ANC acknowledges support from a Vannevar Bush Faculty Fellowship (ONR N000142512032) and the U.S. Department of Energy Office of Science National Quantum Information Science Research Center Q-NEXT.
%
% \vspace{-0.5cm}

\vspace{0.5cm}
\noindent \textbf{Author contributions:}\\
CBM and RTQ carried out the measurements with support of ANC, AB, EVR, VRR, EMM and SLDB. 
CBM, RTQ, VRR, EVR, EMM and MC designed and characterized the devices with support from SF. 
DAC, LO, MEA and SJ fabricated the devices. 
FP and JCF developed the theory and wrote the theory sections of the SI. RTQ, CBM and AB analyzed the data. CBM, RTQ and AB wrote the manuscript with inputs from the other authors. CBM, RTQ and AB supervised the work.

\vspace{0.5cm}
\noindent \textbf{Competing interests:}
There is potential competing interest. Luca Ornago, Maria El Abbassi, and Seoho Jung are shareholders of Chiral Nano, a startup company developing carbon nanotube devices. Chiral Nano supplied one of the devices used in this work. The remaining authors declare no competing interests.

\clearpage
\newgeometry{textwidth=468pt, top=30mm, bottom=22mm}
\renewcommand{\baselinestretch}{1.31}  
\normalsize
\renewcommand{\rmdefault}{cmr}
% \normalfont

%%%%%%%%%%%%%%%% START OF SUPPLEMENT %%%%%%%%%%%%%%%

% Figures, tables, equations and pages in the supplement are numbered S1, S2 etc.
\renewcommand{\thefigure}{S\arabic{figure}}
\renewcommand{\figurename}{Fig.}

\renewcommand{\thetable}{S\arabic{table}}
\renewcommand{\theequation}{S\arabic{equation}}
\renewcommand{\thepage}{S\arabic{page}}
\setcounter{figure}{0}
\setcounter{table}{0}
\setcounter{equation}{0}
\setcounter{page}{1} % not 0 as \newpage already started a supplementary page

% Section lettering
\renewcommand{\thesection}{\arabic{section}}
\renewcommand{\thesubsection}{\thesection\Alph{subsection}}
\renewcommand{\thesubsubsection}{\roman{subsubsection}}

% Reset citations referencing
\makeatletter
\setcounter{NAT@ctr}{0}
\makeatother

%%%%%%%%%%%%%%%% SUPPLEMENT TITLE PAGE %%%%%%%%%%%%%%%
\begin{bibunit}[naturemag_modified]

\begin{center}
\SectionNoTOC{Supplementary Information for:\\ Tunable nonlinear electromechanics at the zero-point motion scale
}

\normalsize C.~B.~M{\o}ller$^{\ast\dagger}$,
R.~Tormo-Queralt$^{\ast\dagger}$,
E.~Vázquez-Rodríguez,
V.~Román-Rodríguez,
M.~Cagetti,
E.~Mateos-Madinabeitia,
J.~Franz,
S.~Forstner,
S.~L.~De Bonis,
L.~Ornago,
M.~El~Abbassi,
S.~Jung,
A.~N.~Cleland,
D.~A.~Czaplewski,
F.~Pistolesi,
A.~Bachtold$^{\dagger}$\\
\phantom{space} \\
\small$^{\ast}$These authors contributed equally to this work.
\\

\small$^{\dagger}$Correspondence to: christoffer.moller@nbi.ku.dk; roger.tormo@icfo.eu; adrian.bachtold@icfo.eu
\end{center}

\begingroup
\parindent=0pt % remove indentation just for TOC title
\tableofcontents
\endgroup

\newpage

% ====================================================================================================
% ====================================================================================================
\clearpage
\newpage

\SectionNoPage{Experimental Section}

\subsection{Summary of device parameters and their estimation}

\begin{table}[!ht]
    \vspace{-5mm}
    \centering
    \footnotesize
    \renewcommand{\arraystretch}{1.5}
    \setlength{\tabcolsep}{4pt}
    \begin{tabular}{%
    >{\raggedright\arraybackslash}p{3cm} |
    >{\raggedright\arraybackslash}p{4cm} |
    >{\centering\arraybackslash}p{3.4cm} |
    >{\centering\arraybackslash}p{0.9cm} |
    >{\centering\arraybackslash}p{3.8cm}%
    }

    % Mechanical oscillator header framed
    \cline{1-5}
    \multicolumn{5}{|c|}{\textbf{Mechanical oscillator}} \\ \cline{1-5}
    ~ & Resonance frequency & ~ & $\omega_\mathrm{M}^0$ & $2\pi \times \SI{0.801}{\giga\hertz}$  \\ 
    @ $\varepsilon = 0$ & Measured linewidth$^\mathrm{(i)}$ & ~ & $\delta \omega_\mathrm{M}$ & $2\pi \times \left(\SI{60}{\kilo\hertz}-\SI{50}{\mega\hertz}\right)$  \\ 
    @ $\varepsilon \gg 10\times 2t_\mathrm{E}$ & Intrinsic quality factor & $\omega_\M^0 / \delta \omega_\M$ & $Q_\M$ & $>10^5$\\  
    ~ & Thermal occ. (\SI{47}{\milli\kelvin})& $\left[\exp\!\left(\hbar\omega_\mathrm{M}/k_\mathrm{B} T\right)-1\right]^{-1}$ & $\bar{n}_\mathrm{M}^\mathrm{th}$ & $\sim 0.9$  \\ 
  ~ & Mechanical nonlinearity & ~ & $K$ & $2\pi \times \left( \SI{30}{\kilo\hertz} - \SI{10}{\mega\hertz}\right)$\\
    ~ & Length$^\mathrm{(ii)}$ & ~ & $L_\mathrm{M}$ & \SI{1.1}{\micro\meter}  \\ 
    ~ & Diameter$^\mathrm{(ii)}$ & ~ & $d_\mathrm{M}$ & $\sim\SI{1}{\nano\meter}$  \\ 

    % Double-quantum dot ETLS header framed
    \cline{1-5}
    \multicolumn{5}{|c|}{\textbf{Double-quantum dot ETLS}} \\ \cline{1-5}
    @ $\varepsilon=0$ & Transition frequency & $\sqrt{(2t_\mathrm{E})^2+\varepsilon^2}$ & $\omega_\mathrm{E}^\mathrm{0}$ & $2\pi \times \left(4 - 70 \right) \SI{}{\giga\hertz}$ \\ 
    @ $2t_\mathrm{E} = \SI{7.4}{\giga\hertz}$ & GS prob.$^\mathrm{(iii)}~(\SI{47}{\milli\kelvin})$ & $1- \bar{n}_\mathrm{E}^\mathrm{th}$ & $p_\mathrm{g}$ & $> \SI{99.9}{\percent}$  \\ 
    ~ & Detuning lever arm & ~ & $\alpha_\mathrm{\varepsilon}$ & $0.45\pm\SI{0.05}{\electronvolt/\volt}$ \\ 
    
    % Microwave cavity header framed
    \cline{1-5}
    \multicolumn{5}{|c|}{\textbf{Microwave cavity}} \\ \cline{1-5}
    ~ & Resonance frequency  & ~ & $\omega_\mathrm{C}^0$ & $2\pi \times \SI{1.359}{\giga\hertz}$  \\ 
    ~ & Total loss rate & ~ & $\kappa$ & $2\pi \times \SI{3.7}{\mega\hertz}$  \\ 
    ~ & Loaded quality factor & $\omega_\mathrm{C}^0 / \kappa$ & $Q_\mathrm{C}$ & $\sim 370$  \\ 
    ~ & Char. impedance & $\sqrt{L/C}$ & $Z$ & \SI{715}{\ohm}  \\ 
    ~ & Thermal occ. $(\SI{20}{\milli\kelvin})$  & $\left[\exp\!\left(\hbar\omega_\mathrm{C}/k_\mathrm{B} T\right)-1\right]^{-1}$ & $\bar{n}_\mathrm{C}^\mathrm{th}$ & $\sim 0.04$  \\ 

    % Coupling rates header framed
    \cline{1-5}
    \multicolumn{5}{|c|}{\textbf{Coupling rates}} \\ \cline{1-5}
    ~ & Electron–photon & ~ & $g_\mathrm{EC}$ & $2\pi \times 49.8\pm\SI{2.5}{\mega\hertz}$  \\ 
    ~ & Electron–phonon & ~ & $g_\mathrm{EM}$ & $2\pi \times 0.46\pm\SI{0.05}{\giga\hertz}$  \\ 
    \multirow{2}{*}{@ $2t_\mathrm{E} = \SI{7.4}{\giga\hertz}$} & \multirow{2}{*}{Photon–phonon} & Linear (Eq.~\ref{gom_1}) & $g_\mathrm{OM}^{(1)}$ & $2\pi \times \SI{70}{\kilo\hertz}$  \\ 
     ~ & ~ & Quadratic (Eq.~\ref{gom_2}) & $g_\mathrm{OM}^{(2)}$ & $2\pi \times \SI{15}{\kilo\hertz}$ \\ \cline{1-5} 
    \end{tabular}
    \customtablecaption{\textbf{Summary of relevant experimental parameters.} (i) Evaluated at $\varepsilon = 0$. (ii) The nanotube length is taken as the nominal source–drain separation, assuming a typical nanotube diameter. 
    (iii) Ground state (GS) probability calculated using the thermal ETLS excited state population of $\bar{n}_\mathrm{E}^\mathrm{th} = \left[\exp\!\left(\hbar \omega_\mathrm{E}/k_\mathrm{B} T\right)-1\right]^{-1}$ at $\varepsilon = 0$.
    }
    \label{table_parameters}
\end{table}

\newpage
Key experimental parameters are estimated using a variety of distinct and independent measurements. The bare cavity and bare mechanical mode frequencies are determined from driven response spectroscopy, such as Fig.~\ref{fig:SI_mechanical_mode_spectrum}, performed at very large ETLS detuning where the three subsystems effectively decouple.

The ETLS parameters are obtained from a global fit to the ETLS induced cavity response as a function of double-quantum dot detuning and temperature (Fig.~\ref{fig:SI_collage_fitting_results},~\ref{fig:SI_results_gec_lever_arm_fitting}), without involving any mechanical measurements. Specifically, this fit yields the ETLS tunneling rate $2t_\mathrm{E}$, ETLS-cavity coupling $g_\mathrm{EC}$, and the gate lever arm $\alpha_\varepsilon$, which converts applied gate voltages to double-quantum dot detuning. The extracted lever arm is furthermore in excellent agreement with independent measurements from electrostatic charge stability diagrams (Fig.~\ref{fig:SI_bias_triangles}).

The electromechanical coupling $g_\mathrm{EM}$ is determined independently from the suppression of the mechanical frequency as a function of ETLS tunneling rate (main text Fig. 4a), which does not rely on displacement calibration. The mechanical Kerr nonlinearity is estimated from fits to the shift of the mechanical resonance frequency with increasing drive-induced displacement amplitudes (main text Fig. 3a), where the amplitude is inferred from the measured displacement induced cavity frequency shift. This Kerr nonlinearity  (main text Fig. 4b) is in good agreement with the theoretical prediction combining the separately measured $2t_\mathrm{E}$ and $g_\mathrm{EM}$ over a very wide parameter range.

% ====================================================================================================
% ====================================================================================================
\clearpage
\newpage
\vspace{-1mm}
\subsection{Device fabrication}\label{Device geometry}
\vspace{-3mm}

The devices studied in this work are assembled in a side-by-side two-chip configuration, combining a multielectrode chip hosting a suspended carbon nanotube with a superconducting spiral microwave resonator chip. 
Results from two nanotube devices are presented in this work, each fabricated using a different approach. 
In one approach, nanotubes are grown off-chip on a separate SOI wafer and subsequently stamped onto pre-fabricated electrodes on a high-resistivity silicon substrate with a SiN$_\mathrm{x}$ layer. All results in the main text and SI are from this device, unless stated otherwise.
In a second approach, carbon nanotubes are grown directly on electrodes pre-fabricated on a lightly p-doped Si wafer covered with a low-stress SiN$_\mathrm{x}$ layer, as shown in our earlier work employing SiO$_2$ on Si substrates~\cite{Tormo2022}. The measurements for this device are presented in Section~\ref{sec:second_device}, and despite the different fabrication processes, the two devices exhibit very similar behavior.

The two-chip architecture enables galvanic coupling between the nanotube and the superconducting resonator, allowing integration of devices fabricated on separate substrates while maintaining a high resonator quality factor. The superconducting resonator used in the experiment, inspired by Ref.~\cite{HorstigPRA2024}, is capacitively coupled to a readout line at a rate of $\kappa_\mathrm{R}/2\pi = \SI{1.8}{MHz}$ and weakly coupled to a drive line at $\kappa_\D/2\pi = \SI{0.01}{MHz}$. With an internal loss rate of $\kappa_\mathrm{I}/2\pi = \SI{1.0}{MHz}$, the superconducting resonator is overcoupled through the readout port.
\vspace{-5mm}
\subsubsection{Transferred nanotube devices}\label{Off-chip grown}
\vspace{-2mm}
The fabrication of wafer CN05 starts with a high-resistivity silicon wafer 
($> \SI{10}{ \kilo\ohm \cdot \centi\metre}$), covered with a 
$\SI{100}{\nano\metre}$ stoichiometric SiN$_\mathrm{x}$ layer. 
Alignment marks are patterned using a JEOL 8100FS electron-beam lithography tool, 
followed by a lift-off of $\SI{5}{\nano\metre}$ titanium and 
$\SI{60}{\nano\metre}$ platinum. Due to the small dimensions of the gates, two separate patterning and lift-off steps are used to create the source and drain electrodes along with five gates. Half of the non-adjacent structures are patterned first, followed by the remaining structures. All structures are defined by electron-beam lithography followed by a $\SI{5}{\nano\metre}/\SI{60}{\nano\metre}$ Ti/Pt lift-off process. Direct-write optical lithography, followed by a $\SI{5}{\nano\metre}/\SI{100}{\nano\metre}$ Ti/Pt lift-off process, is used to pattern bond pads for electrical connections. To create a separation between the suspended nanotube and the underlying control gates, the source and drain electrodes are patterned again using electron-beam lithography, followed by a $\SI{5}{\nano\metre}/\SI{90}{\nano\metre}$ Ti/Pd lift-off process. Direct-write optical lithography is then used to define trenches around the devices, enabling the carbon nanotube stamping process. The silicon nitride is etched by reactive ion etching, and the silicon substrate is etched using a Bosch process to a depth of approximately $\SI{55}{\micro\metre}$. Finally, after removing the patterned resist, a fresh resist layer is applied to protect the devices during transportation to ICFO and subsequent singulation.
\vspace{0mm}
\begin{figure}[t!]
	\centering
	\includegraphics[width=1\linewidth]{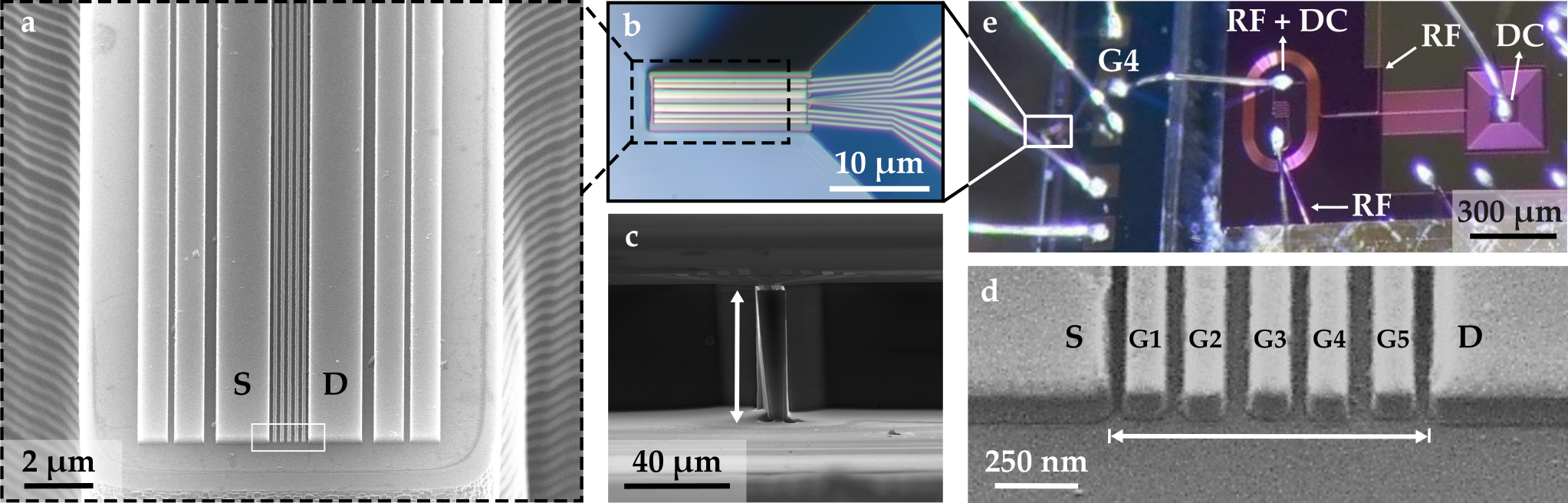}
    \vspace{-7mm}
    \customcaption{\textbf{Transferred nanotube device.} \textbf{a.}~Tilted scanning electron micrograph (SEM) of a multielectrode device prior to carbon nanotube transfer (tilt angle: \SI{39}{\degree} relative to the chip surface). The source S and drain D electrodes are labeled in the image. 
    \textbf{b.}~Optical micrograph of a device similar to the one shown in (\textbf{a}). 
    \textbf{c.}~Tilted SEM of a different device (tilt angle: \SI{85}{\degree} relative to the chip surface). The white double-headed arrow denotes the~\SI{50}{\micro\meter} substrate etching required for the nanotube fork-transfer.
    \textbf{d.}~Zoomed-in SEM of the area marked by the white rectangle in (\textbf{a}). The white double-headed arrow indicates the separation of source and drain of ca.~\SI{1}{\micro\meter} for this specific device. The gate electrodes G1–G5 are labeled in the image. 
    \textbf{e.}~Optical image of the bonded double-chip assembly discussed in the main text. On the left-hand side the multielectrode chip with the carbon nanotube; on the right-hand side is the chip with the superconducting microwave spiral resonator. Gate G4 of the multigate chip is connected to the RF+DC port of the spiral via a bonding wire. The ports used to apply independently a DC or RF voltage on the spiral are labeled in the image.}
	\label{fig:SI_geometry_device_chiral}
\end{figure}
\newpage
\vspace{20mm}
After wafer dicing, the carbon nanotubes (CNTs) are grown, characterized, and transferred by Chiral Nano AG. First, CNTs are grown using chemical vapour deposition onto a set of forks fabricated on an SOI wafer with multiple cantilevers each. The nanotubes are characterized using Raman spectroscopy for both localization and quality assessment prior to transfer, ensuring that only mechanically and structurally pristine tubes are selected. Using a high-speed measurement protocol together with a deep-learning identification algorithm allows to automatically scan and analyse the data of a full growth substrate in only a few hours \cite{Zhang2022}, enabling efficient large-scale screening. Only defect-free, single-walled and individual nanotubes are retained for transfer.

The target devices are cleaned immediately before transfer according to an optimized protocol \cite{Jung2021}, which ensures good adhesion and minimizes contamination at the contact interface. Finally, the nanotubes are mechanically transferred from the growth substrate onto the wafer CN05 described above using an automatic, robotic assembly machine \cite{Butzerin2024}, providing reproducible placement with micrometre accuracy.

\subsubsection{On-chip grown nanotube devices}\label{On-chip grown}

The fabrication of wafer B81 starts with a regular resistivity silicon wafer $\SIrange{10}{20}{\ohm \cdot \centi\metre}$, covered with a 
$\SI{300}{\nano\metre}$ low-stress SiN$_\mathrm{x}$ layer. 
Alignment marks are patterned using a JEOL 8100FS electron-beam lithography tool, 
followed by a lift-off of $\SI{5}{\nano\metre}$ titanium and 
$\SI{60}{\nano\metre}$ platinum. Due to the small dimensions of the gates, two separate patterning and metal lift-off steps are used to define the source and drain electrodes along with five gates. Half of the non-adjacent structures are patterned first, followed by the remaining structures. All structures are patterned using electron-beam lithography and a $\SI{5}{\nano\metre}/\SI{60}{\nano\metre}$ Ti/Pt lift-off process. Direct-write optical lithography, followed by a $\SI{5}{\nano\metre}/\SI{100}{\nano\metre}$ Ti/Pt lift-off process, is used to pattern bond pads for electrical connections. To create a separation between the suspended nanotube and the underlying control gates, the source and drain electrodes are patterned again using electron-beam lithography, followed by a $\SI{5}{\nano\metre}/\SI{45}{\nano\metre}$ Ti/Pt lift-off process. The wafer surface is etched by approximately $\SI{200}{\nano\metre}$ to remove possible defects in the bulk film, simultaneously creating the pillar structures visible in cross-sectional SEM images. Electron-beam lithography is then used to define areas near the gates where a protective silicon dioxide layer is deposited, preventing CNTs from shorting the electrodes outside the suspended regions. Finally, electron-beam lithography is used to open holes for the deposition of catalyst for CNT growth. CNTs are grown using chemical vapor deposition as described in Ref.~\cite{Tormo2022}. 

\begin{figure}[h!]
	\vspace{10mm}
    \centering
	\includegraphics[width=1\linewidth]{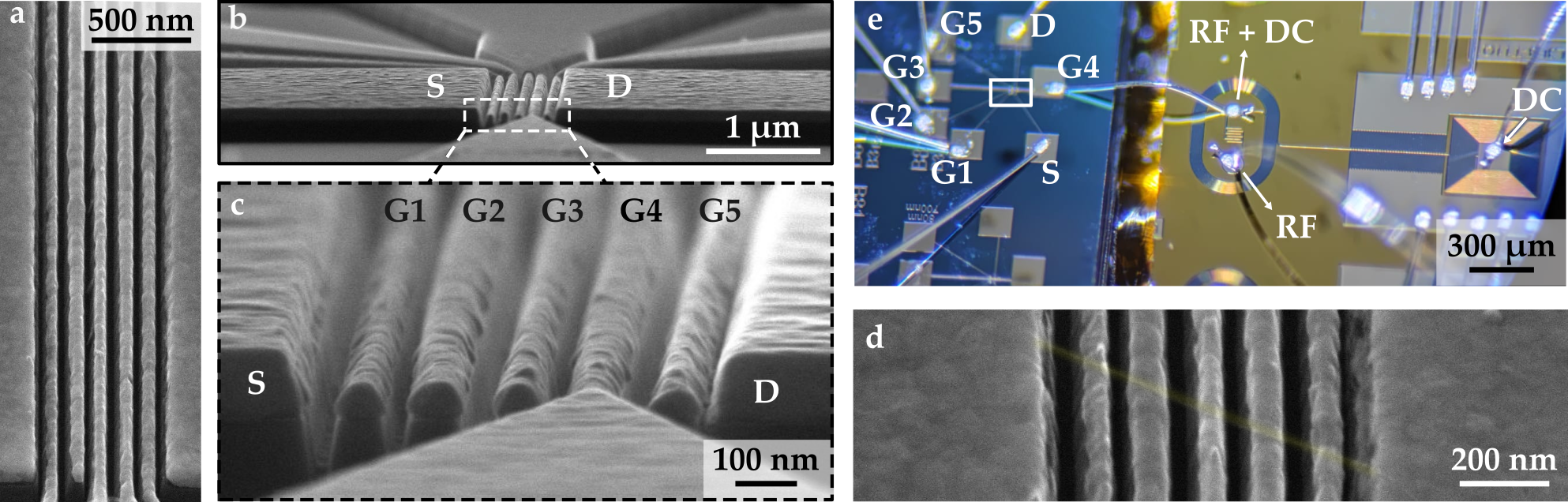}
	\customcaption{\textbf{On-chip CVD grown nanotube device.}
    \textbf{a.}~Tilted scanning electron micrograph (SEM) of a multielectrode device after a CVD nanotube growth (tilt angle: \SI{45}{\degree} relative to the chip surface).
    \textbf{b.}~SEM of a similar device (tilt angle: \SI{87}{\degree} relative to the chip surface). The source S and drain D electrodes are labeled in the image. 
    \textbf{c.}~Zoomed-in SEM of the area marked by the white dashed rectangle in (\textbf{b}). The gate electrodes G1–G5 are labeled in the image. 
    \textbf{d.}~SEM of a device with a suspended carbon nanotube indicated by the yellow shaded line. 
    \textbf{e.}~Optical image of the bonded double-chip assembly of the device presented in Section~\ref{sec:second_device}. On the left-hand side the multielectrode chip with the carbon nanotube; on the right-hand side is the chip containing the superconducting microwave spiral resonator. Gate G4 of the multielectrode chip is connected to the RF+DC port of the spiral via a bonding wire. The ports used to apply independently a DC or RF voltage on the spiral are labeled in the image.}
	\label{fig:SI_geometry_device_on-chip_grown}
\end{figure}

\subsubsection{Superconducting spiral resonator}\label{Superconducting spiral resonator}
The fabrication of wafer BSF01 starts with a high-resistivity silicon wafer covered with a $\SI{316}{\nano\metre}$ SiO$_2$ layer. A $\SI{100}{\nano\metre}$ niobium film is deposited by DC sputtering and patterned by optical lithography (MLA150) using S1805 resist. After development, the Nb layer is etched in an Oxford reactive ion etcher, leaving a remaining SiO$_2$ thickness of $\sim\SI{210}{\nano\metre}$. The resist is stripped in acetone with sonication, followed by IPA rinse and N$_2$ drying. Finally, a protective S1813 resist layer is spin-coated to preserve the wafer for subsequent processing steps. The full wafer is diced in chips of $8\times \SI{2}{mm}$ and each chip is partially diced in two sections of $4\times \SI{2}{mm}$ for subsequent singularization. The protective resist layer is lifted-off in acetone at $\SI{55}{\celsius}$, followed by sonication.

% \vspace{-10mm}
\subsubsection{Two-chip assembly}\label{Double-chip assembly}
\vspace{-3mm}
Both the chips containing the superconducting resonator and the multielectrode carbon nanotube device are glued onto a gold-plated OFC base using a very thin layer of CMR GE varnish to maximize the thermalization to the mixing chamber. The two chips are typically placed $300$–$\SI{500}{\micro\meter}$ apart and electrically connected with aluminum bonding wires (Figs.~\ref{fig:SI_geometry_device_chiral}e and \ref{fig:SI_geometry_device_on-chip_grown}e).
% ====================================================================================================
% ====================================================================================================
\subsection{Mechanical oscillator readout scheme}\label{mechanical oscillator measurements}

The mechanical oscillator is characterized using a two-tone spectroscopy scheme, in which the cavity is continuously probed while the mechanical oscillator is driven through gate~G2 with an oscillating voltage $V\!(t)$. In this scheme, the displacement of the mechanical oscillator modulates the double-quantum dot ETLS wavefunction. This modulation is detected using a cavity-based readout, where the cavity is galvanically connected to gate G4 (see Section~\ref{Device geometry}), located directly beneath one of the quantum dots. The displacement-dependent cavity response is amplified through a chain of cryogenic and room-temperature amplifiers, and subsequently demodulated. We distinguish between two measurement modalities, determined by the specific driving scheme.

\begin{figure}[h!]
	\centering	\includegraphics[width=0.95\linewidth]{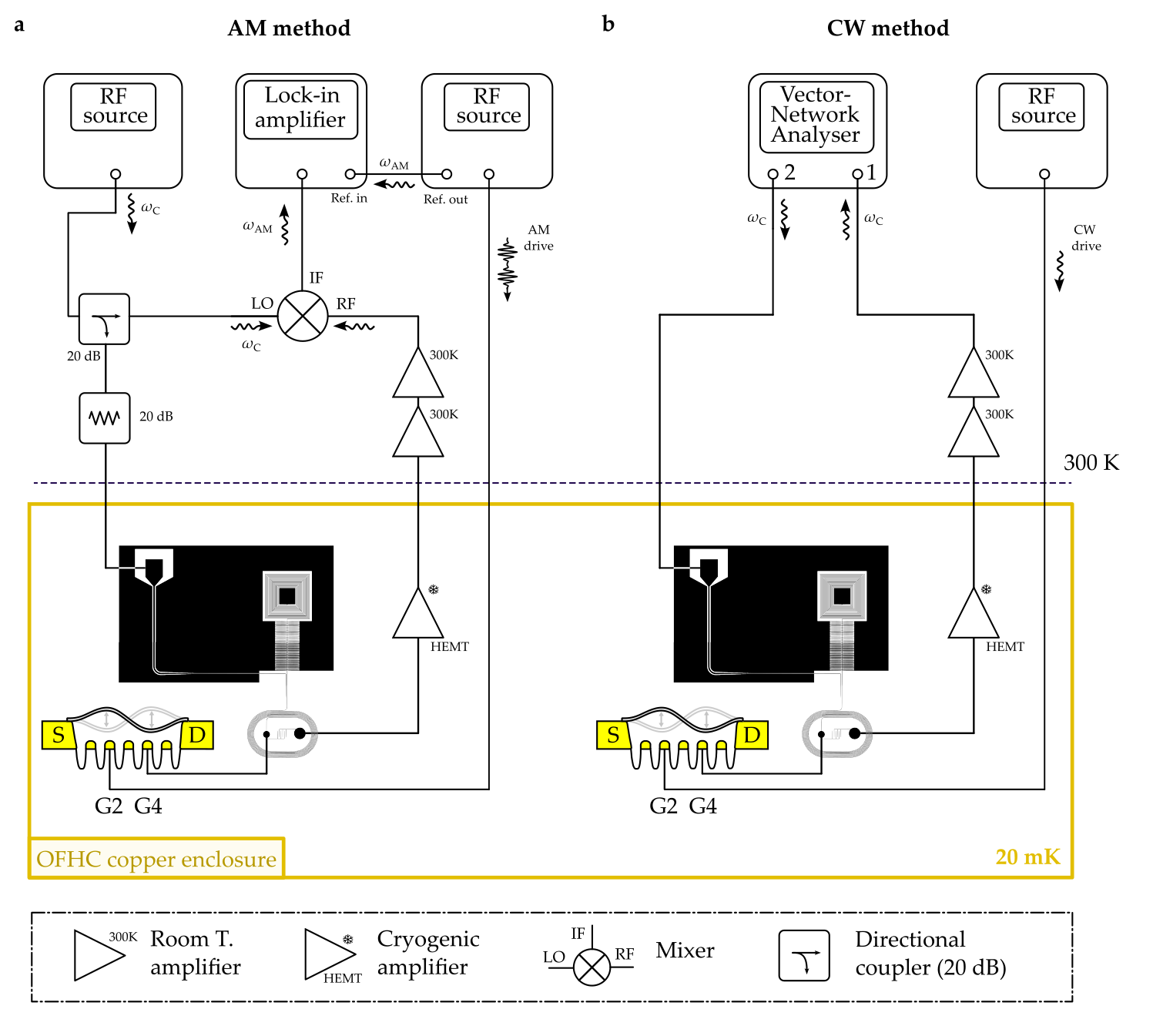}
	\customcaption{
		\textbf{Cryogenic two-tone spectroscopy measurement setup.}
		Schematic of the measurement setups for the (AM) amplitude-modulated scheme in (\textbf{a})  and the continuous-wave (CW) driving scheme in (\textbf{b}). The diagrams show simplified electrical layouts; the complete attenuation and filtering stages are provided in Fig.~\ref{fig:SI_full_electrical_circuit}.
        Electrical components are identified in the legend.
	}
	\label{fig:SI_AM_and_cond_electronic_setup}
\end{figure}

In the first, an amplitude-modulated (AM) drive is applied to gate~G2. This generates a detuning drive since $\varepsilon (t) = \alpha_\mathrm{\varepsilon} V (t)$, with $\alpha_\mathrm{\varepsilon}$ the detuning lever arm. The resulting AM drive is given by $\varepsilon (t) = \varepsilon_\D[1 + \cos(\omega_\mathrm{AM} t)]\cos(\omega_\D t)$, where $\varepsilon_\D$ is the amplitude of the signal, $\omega_\mathrm{AM}$ denotes the modulation frequency and $\omega_\D$ the drive carrier frequency. After mix-down, the cavity output is demodulated at the reference modulation frequency $\omega_\mathrm{AM}/2\pi = \SI{2}{\kilo\hertz}$ using a lock-in amplifier, allowing access to both quadratures of the signal. We refer to this scheme as the AM method (Fig.~\ref{fig:SI_AM_and_cond_electronic_setup}a).

\newpage
In the second modality, a continuous (unmodulated) drive
$\varepsilon(t) = \varepsilon_\mathrm{D}\cos(\omega_\mathrm{D} t)$
is applied to the mechanical oscillator. In this case, the cavity response is monitored directly at the probe tone near the cavity frequency, $\omega_\mathrm{P} \simeq \omega_\mathrm{C}/2\pi=\SI{1.359}{\giga\hertz}$, using a vector network analyzer (VNA). The VNA records the complex transmission coefficient $S_{21}$, enabling direct extraction of amplitude and phase variations induced by the mechanical motion. This approach, which avoids low-frequency modulation but offers high spectral resolution around the cavity resonance, is referred to as the CW method (Fig.~\ref{fig:SI_AM_and_cond_electronic_setup}b).

\vspace{3mm}

\begin{figure}[h!]
	\centering
	\includegraphics[width=1\linewidth]{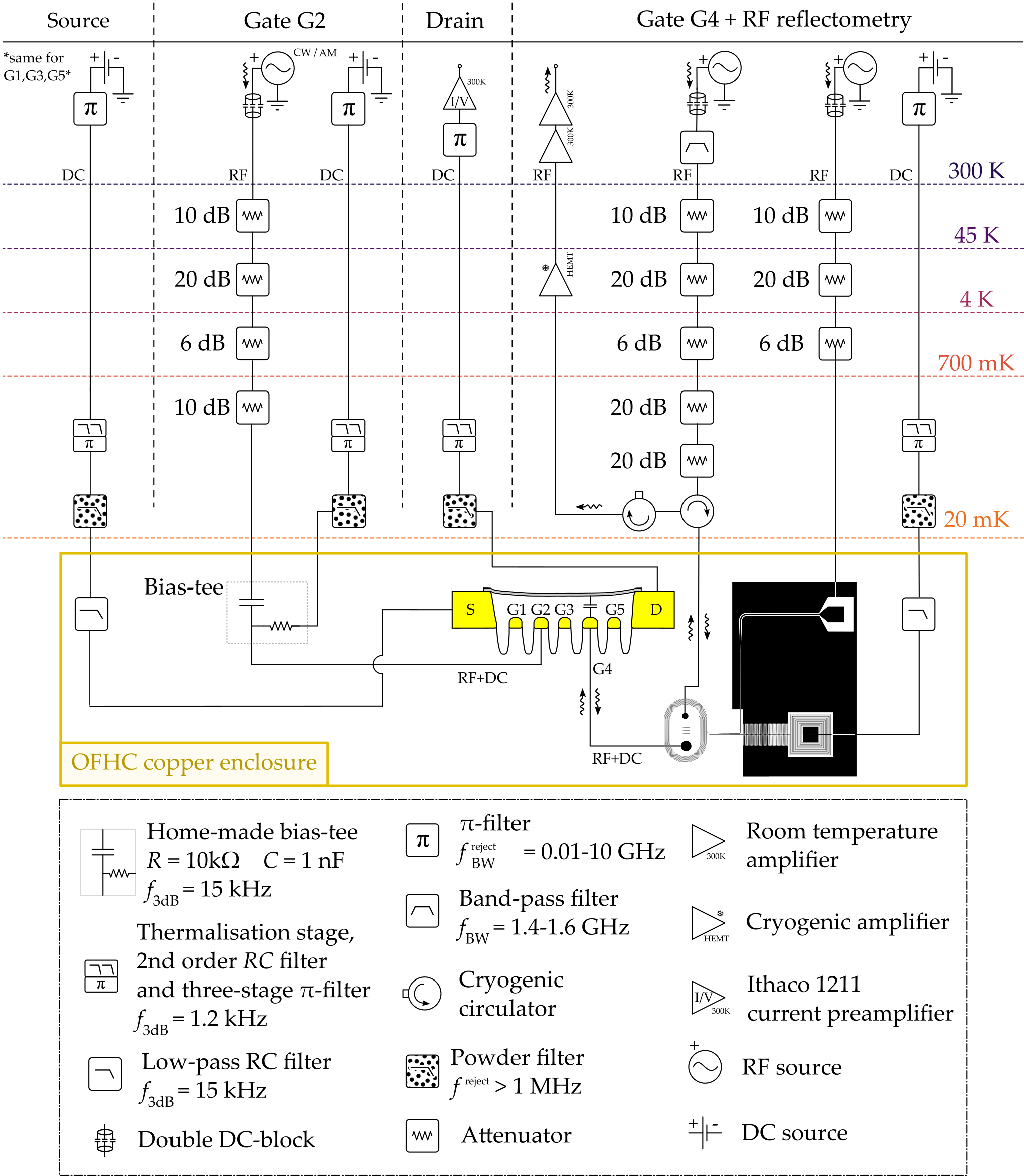}
	\customcaption{
		\textbf{Electrical scheme of the cryogenic lines.}
		Schematic of the electrical components in the RF and DC lines connected to gates G1-5, source and drain as well as the microwave resonator at each plate of the dilution refrigerator. The microwave resonator can be driven via two RF ports, enabling both reflection and transmission measurements. DC lines are shown for gates G2, G4, source and drain; the remaining gates share the same DC filtering as the source. A simplified schematic of the multigate carbon nanotube device and the microwave resonator is also included. The different electrical components and the cut-off frequencies of the various filters are detailed in the legend.
        }
	\label{fig:SI_full_electrical_circuit}
\end{figure}

\clearpage
\newpage

% ====================================================================================================
% ====================================================================================================
\clearpage
\newpage
\subsection{Characterization and control: double quantum dot ETLS}
\label{sec:Characterization_and_control}
\vspace{-0.3cm}
\subsubsection{Charge stability diagram}
\begin{figure}[h!]
    \centering
    \includegraphics[width=0.92\linewidth]{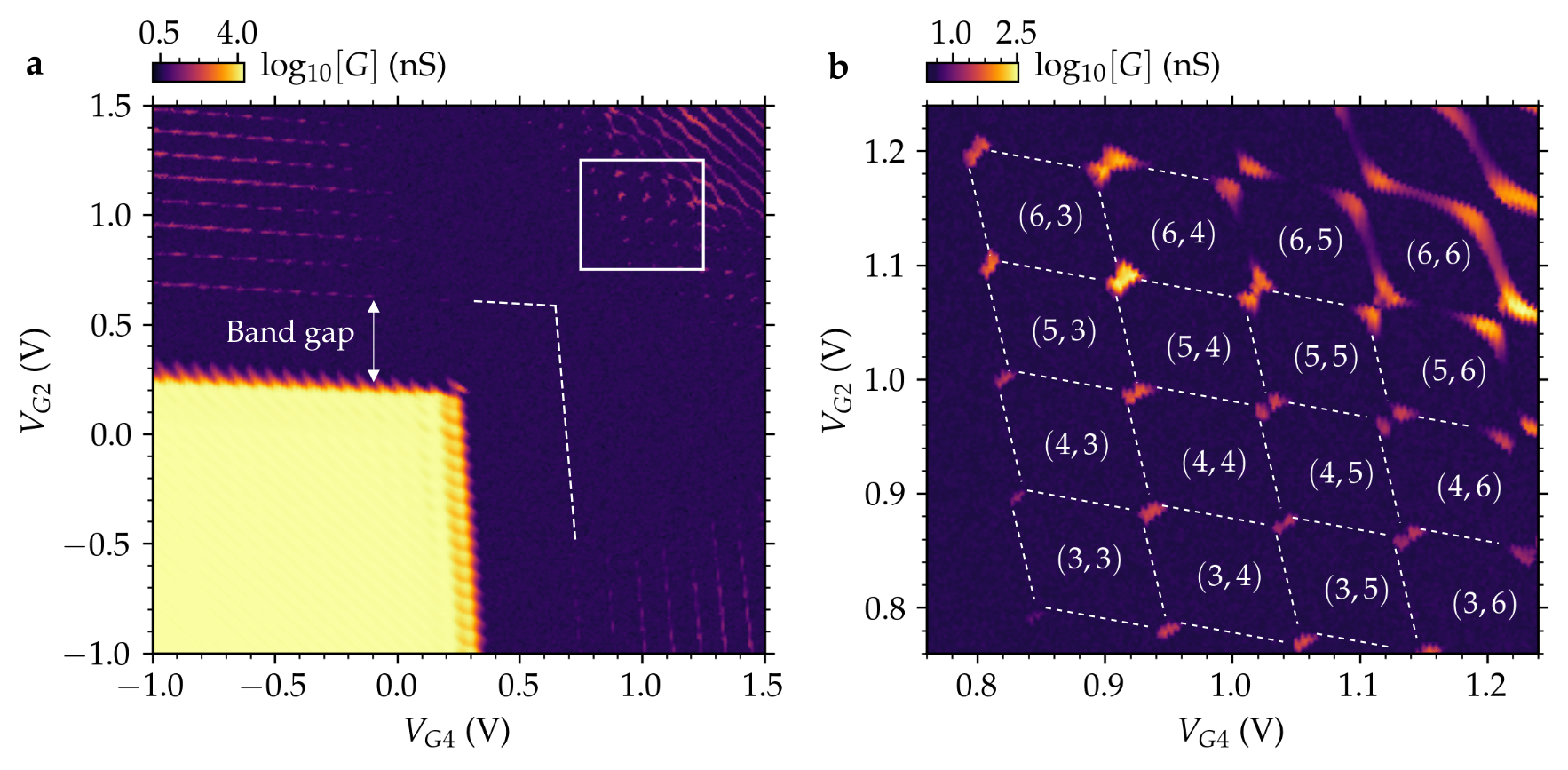}
    \vspace{-0.5cm}
    \customcaption{
    \textbf{Charge stability diagram.}
    \textbf{a.}~Charge stability diagram showing electron transport through the DQD, measured via the conductance $G$, as a function of the plunger gate voltages $V_\mathrm{G2}$ and $V_\mathrm{G4}$. The white dashed lines indicate the transition of the first electron into each dot.
    \textbf{b.}~Zoom-in of the region highlighted by the white rectangle in \textbf{(a)}. The electron occupation numbers in the left and right quantum dots are indicated in parentheses $(N_\mathrm{L},N_\mathrm{R})$.
    The voltage barrier configuration for both panels is $[V_\text{G1},\, V_\text{G3},\, V_\text{G5}] = [-1,\ 0,\ -1]\,\text{V}$.
    Data in all panels are recorded at \SI{20}{\milli\kelvin} with an applied DC bias of $V_\mathrm{SD} = \SI{3}{\milli\volt}$.}
    \label{fig:SI_charge_stability_diagram}
    \vspace{-8mm}
\end{figure}

\vspace{-0.3cm}
We record the conductance $G$ of the device discussed in the main text as a function of the plunger gate voltages $V_\mathrm{G2}$ and $V_\mathrm{G4}$ at $\SI{20}{\milli\kelvin}$ (Figs.~\ref{fig:SI_charge_stability_diagram}a,b). The charge stability diagram in Fig.~\ref{fig:SI_charge_stability_diagram}a exhibits a high degree of mirror symmetry along the $V_\mathrm{G2} = V_\mathrm{G4}$ diagonal axis, suggesting that the nanotube has similar electron transport properties along its entire length. This behavior was also reported in our previous work Ref.~\cite{Tormo2022} and in Ref.~\cite{Waissman2013}, highlighting the robustness of the pristine carbon nanotube device fabrication, whether using a transferred nanotube (Section~\ref{Off-chip grown}) or an in-situ grown nanotube (Section~\ref{On-chip grown}).

The clean electrostatic landscape of the double dot allows identification of the nanotube band gap and the electron transition for the left ($N_\mathrm{L}$) and right ($N_\mathrm{R}$) quantum dots, as highlighted by the white dashed lines in Fig.~\ref{fig:SI_charge_stability_diagram}a. The determination of the double-quantum-dot configuration $(N_\mathrm{L},N_\mathrm{R})$ (Fig.~\ref{fig:SI_charge_stability_diagram}b) demonstrates carbon nanotubes as a spin-qubit platform.

The regular charge stability diagram in Fig.~\ref{fig:SI_charge_stability_diagram}b allows us to compensate for cross-capacitive effects associated to the $V_\mathrm{G2}$ and $V_\mathrm{G4}$ sweeps by adjusting $V_\mathrm{G3}$, producing vertical and horizontal charging lines. An example is shown in Fig.~1c of the main text, obtained using
$V_\mathrm{G3} = 0.41 - 0.41(V_\mathrm{G2} + V_\mathrm{G4})/2$ while keeping $V_\mathrm{G1}=V_\mathrm{G5}=\SI{-1}{\volt}$.
This compensation also keeps the tunnel coupling $2t_\mathrm{E}$ comparatively constant across a wide range of $V_\mathrm{G2}$ and $V_\mathrm{G4}$, which would otherwise vary by tens of gigahertz (see Section~\ref{Double quantum dot ETLS characterization}).
\vspace{-3mm}

\clearpage 
\newpage
\subsubsection{Interdot charge transition tunability}\label{Interdot charge transition tunability}

\begin{figure}[h!]
    \centering
    \vspace{-5mm}
    \includegraphics[width=0.95\linewidth]{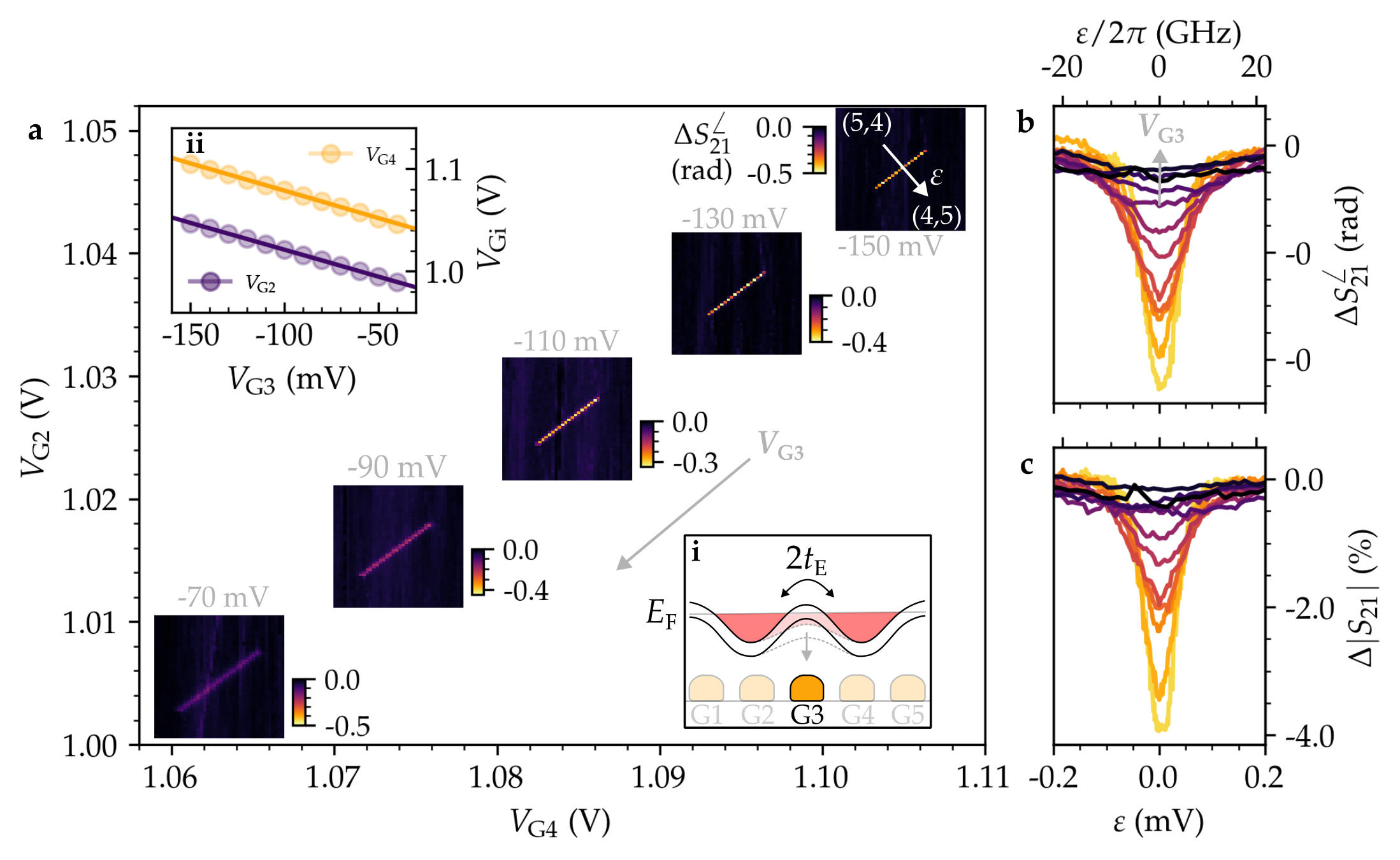}
    \vspace{-5mm}
    \customcaption{
    \textbf{Voltage tuning of the $\mathbf{(5,4) \leftrightarrow (4,5)}$ interdot charge transition.}
    \textbf{a.}~Charge stability diagrams acquired for various interdot barrier gate voltages $V_\mathrm{G3}$, measured via the cavity phase response $\Delta S_\mathrm{21}^{\angle}$ as a function of plunger gate voltages $V_\mathrm{G2}$ and $V_\mathrm{G4}$. The other gates are held fixed at $[V_\mathrm{G1}, V_\mathrm{G5}] = [-1,\ -1]\text{V}$. (Inset i)~Schematic illustration of the nanotube electrostatic potential profile. As $V_\mathrm{G3}$ is increased the potential barrier between the two dots decreases, leading to an increasing interdot tunnel coupling $2t_\mathrm{E}$. (Inset ii) Adjusted plunger gate voltages $V_\mathrm{Gi}$ ($i=2,4$ for gates G2 and G4) as a function of $V_\mathrm{G3}$. 
    \textbf{b,c.}~Linecuts along the detuning axis $\varepsilon$ for each $V_\mathrm{G3}$ configuration, showing the cavity phase $\Delta S_\mathrm{21}^{\angle}$ and amplitude $\Delta|S_\mathrm{21}|$ responses, respectively. The interdot barrier voltage increases from yellow to black.
    Data in all panels are recorded at \SI{20}{\milli\kelvin}.
    }
    \label{fig:SI_ICT_tunability_Vg3}
    \vspace{-10mm}
\end{figure}

In the double-quantum dot configuration, we study the evolution of the $(5,4)\leftrightarrow(4,5)$ interdot charge transition (ICT) as a function of the interdot barrier voltage $V_\mathrm{G3}$ (Figs.~\ref{fig:SI_ICT_tunability_Vg3}a,b). Increasing $V_\mathrm{G3}$ from $-150$ to \SI{-40}{\milli\volt} induces two main effects.

First, the cavity phase response $\Delta S_{21}^{\angle}$ decreases while the ICT linewidth broadens, both suggesting an increase of the interdot tunnel coupling $2t_\mathrm{E}$. This trend is expected, as raising $V_\mathrm{G3}$ lowers the interdot barrier potential and thus increases the tunnel coupling (inset i). Figures~\ref{fig:SI_ICT_tunability_Vg3}b,c corroborate this behavior, showing a pronounced suppression of both phase $\Delta S_{21}^{\angle}$ and magnitude $\Delta |S_{21}|$ responses along the detuning axis $\varepsilon$ as $V_\mathrm{G3}$ is increased. Second, inscreasing $V_\mathrm{G3}$ shifts the ICT center $(V_\mathrm{G2},V_\mathrm{G4})$ to lower values, reflecting the cross-capacitance of gate G3 to both dots. To first order, the ICT follows a linear trajectory described by $V_\mathrm{G2}=-0.54V_\mathrm{G3}+0.97$ and $V_\mathrm{G4}=-0.55V_\mathrm{G3}+1.02$ (inset ii), enabling us to track its position while continuously tuning $2t_\mathrm{E}$. A similar behavior is observed for the $(1,0)\leftrightarrow(0,1)$ ICT, confirming that the system behaves as a highly tunable, textbook double quantum dot \cite{Wiel2003}.

\vspace{-7mm}
\subsubsection{Double quantum dot ETLS characterization}\label{Double quantum dot ETLS characterization}
\begin{figure}[H]
	\centering
	\includegraphics[width=.99\linewidth]{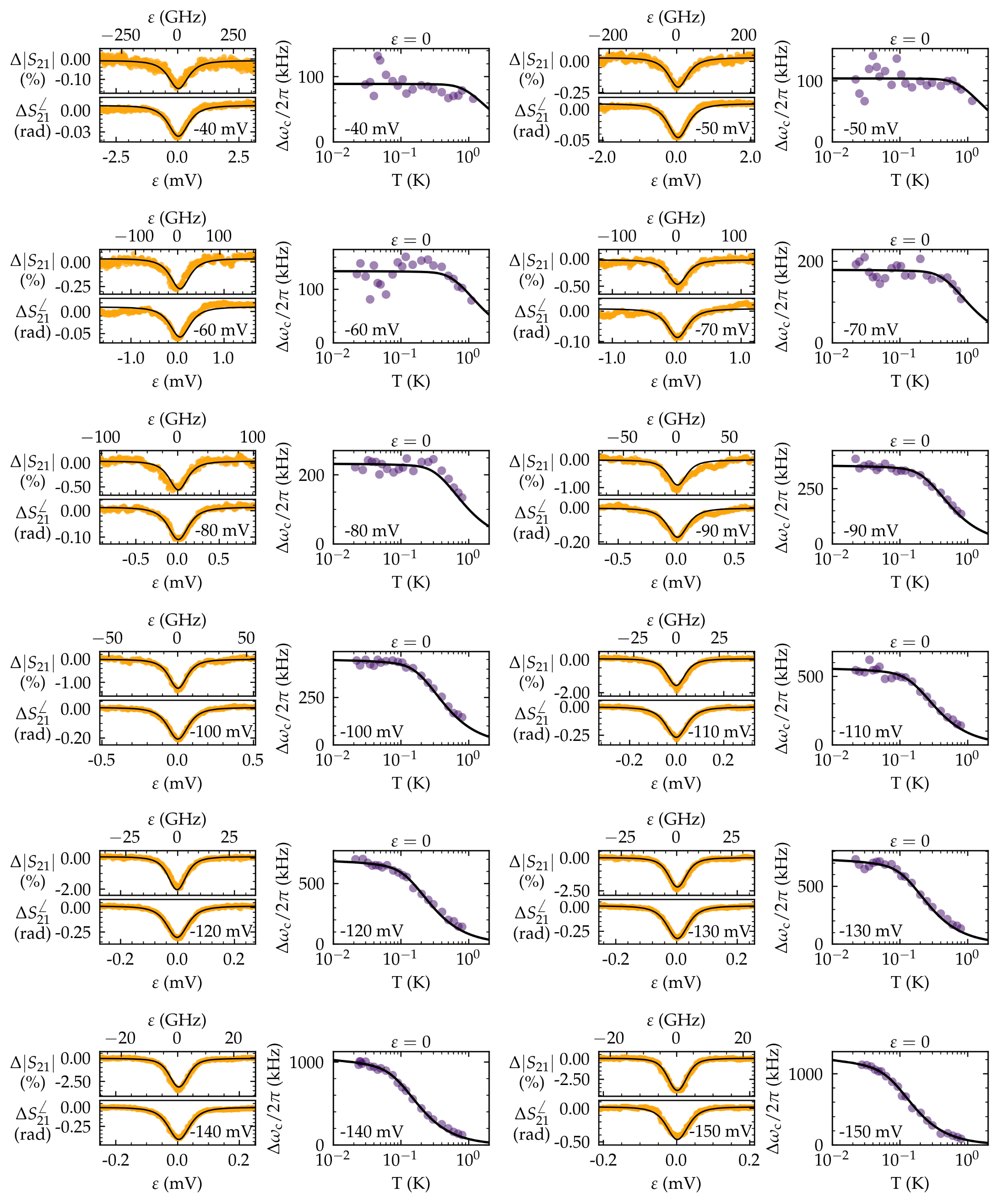}
	\vspace{-.3cm}
	\customcaption{
        \textbf{Characterization the $\mathbf{(5,4) \leftrightarrow (4,5)}$ double quantum dot ETLS.} 
        Each pair of plots displays (1) the change in cavity phase $\Delta S_\mathrm{21}^{\angle}$ and magnitude $\Delta |S_\mathrm{21}|$ response as a function of detuning $\varepsilon$, together with (2) the cavity frequency shift $\Delta \omega_\mathrm{C}$ as a function of temperature $T$, for different interdot barrier gate voltages $V_\mathrm{G3}$ indicated in the right panels.}
	\label{fig:SI_collage_fitting_results}
\end{figure}

In our experiments, the cavity is dispersively coupled to the  double-quantum dot electronic two-level system (ETLS). This interaction is mediated through the electron-photon coupling rate, which is small compared to the rest of the frequencies of the problem $g_\mathrm{EC} \ll 2t_\mathrm{E}, \omega_\mathrm{C}^0$. This allows us to treat the interaction as a small perturbation to the total energy of the system and incorporate it into the equations of motion of the cavity (Section~\ref{Equations of motions}). From the resulting ETLS-dependent cavity response (Section \ref{Cavity response to the drive}) we extract key parameters of the hybrid system, including $2t_\mathrm{E}$, $g_\mathrm{EC}$, and $\alpha_\mathrm{\varepsilon}$. The ETLS dissipation rate $\Gamma$ can also be bounded by this method, however, we find $\Gamma \ll \omega_\mathrm{E}$, and so it is neglected in the following analysis.

To determine the relevant parameters, we fit the measured cavity response using Eq.~\ref{eq:S21_thermal}, as derived in Section~\ref{sec:S21_thermal}. For physical insight, we restate below the complex cavity response probed at a frequency $\omega_\mathrm{P} \sim \omega_\mathrm{C}^0$ introduced in Section~\ref{Cavity response to the drive}, neglecting the ETLS dissipation
\vspace{-2mm}
\begin{align}
    S_{21}
    &\propto
    \frac{\kappa/2 }
         {  \kappa/2 + i( \omega_\mathrm{C}^0 - \omega_\mathrm{P}^{ })
            -g_\EC^2 \langle \sigma_\mathrm{z} \rangle_0 
            \left(\frac{2t_\E}{\omega_\E}\right)^2\left[
                {1\over  i \left( \omega_\mathrm{E}-\omega_\mathrm{P}\right)}  - 
                {1\over - i \left( \omega_\mathrm{E}-\omega_\mathrm{P}\right)} 
            \right]
         }.
    \label{eq:cavity_response_fit}
\end{align}

We vary two independent control parameters in Fig.~\ref{fig:SI_collage_fitting_results}: the ETLS detuning $\varepsilon$, which defines the ETLS transition frequency $\omega_\mathrm{E}=\sqrt{(2t_\mathrm{E})^2+\varepsilon^2}$, and the temperature \textit{T}, which determines the thermal expectation value $\langle \sigma_\mathrm{z} \rangle_0  \approx -\left(2n_\mathrm{B}+1\right)^{-1}$ via the ETLS thermal excitation $ n_\mathrm{B} = \left[\exp\!\left(\hbar \omega_\mathrm{E}/k_\mathrm{B} T\right)-1\right]^{-1}$.

\begin{figure}[h!]
	\centering
	\includegraphics[width=1\linewidth]{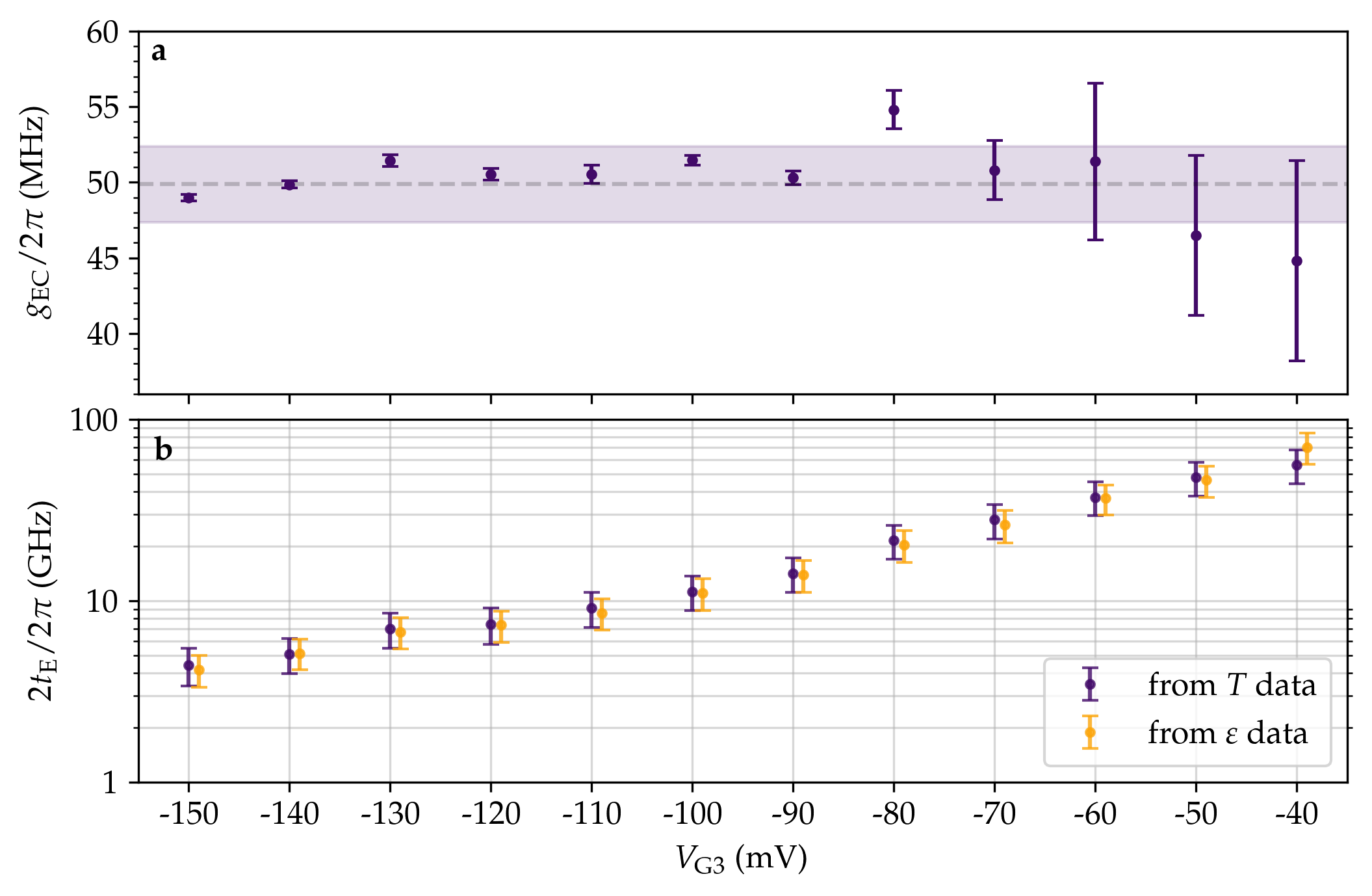}
	\customcaption{
        \textbf{\boldmath Evolution of $g_\mathrm{EC}$ and $2t_\mathrm{E}$ when tunning the interdot barrier voltage.}
		\textbf{a.}~Electron-photon coupling rates $g_\mathrm{EC}$ for different $V_\mathrm{G3}$ configurations. The values are obtained from fitting the cavity frequency shift $\Delta \omega_\mathrm{C}$ as a function of temperature $T$, shown in Fig.~\ref{fig:SI_collage_fitting_results}. The gray dashed line indicates the weighted mean value. The purple shaded area indicates the weighted uncertainty.
		\textbf{b.}~Double quantum dot ETLS frequency $\omega_\mathrm{E}=2t_\mathrm{E}$ at zero detuning estimated for different $V_\mathrm{G3}$ configurations. The values are estimated from the fits of the cavity response as a function of detuning $\varepsilon$ (orange) and temperature $T$ (purple) in Fig.~\ref{fig:SI_collage_fitting_results}. The orange points are offset in $V_\mathrm{G3}$ for visualization.
	}
	\label{fig:SI_results_gec_lever_arm_fitting}
\end{figure}

The first step consists in fitting the cavity frequency shift as a function of temperature at $\varepsilon = 0$ while treating $g_\mathrm{EC}$ as a free parameter. From the fitted $g_\mathrm{EC}$ values (Fig.~\ref{fig:SI_results_gec_lever_arm_fitting}a) we compute the weighted mean and weighted standard deviation from the individual fitting errors, yielding $g_\mathrm{EC}/2\pi = 49.8 \pm \SI{2.5}{\mega\hertz}$. This value is subsequently fixed for the remainder of the fitting procedure. With $g_\mathrm{EC}$ constrained, the temperature-dependent data are refitted to extract $2t_\mathrm{E}$ as described in Ref.~\cite{Yu2023} (purple dataset in Fig.~\ref{fig:SI_collage_fitting_results}).

For the detuning-dependent traces, the temperature is fixed at the cryostat base temperature. With $g_\mathrm{EC}$ fixed, we first allow $\alpha_\mathrm{\varepsilon}$ and $2t_\mathrm{E}$ to vary, then compute the weighted mean and weighted standard deviation, obtaining $\alpha_\mathrm{\varepsilon} = 0.45 \pm \SI{0.05}{\electronvolt/\volt}$. This value is consistent with the independent estimate obtained from conductance measurements of DC bias triangles in Section~\ref{DC-bias triangles}. Once $\alpha_\mathrm{\varepsilon}$ is fixed, we refit the detuning traces to extract $2t_\mathrm{E}$ (orange dataset in Fig.~\ref{fig:SI_collage_fitting_results}).

Figure~\ref{fig:SI_results_gec_lever_arm_fitting}b summarizes the extracted values of $2t_\mathrm{E}$ obtained from both detuning- and temperature-dependent datasets (orange and purple in Fig.~\ref{fig:SI_collage_fitting_results}), which are in excellent agreement. The data further reveal an exponential dependence of $2t_\mathrm{E}$ on the interdot gate voltage $V_\mathrm{G3}$, spanning one order of magnitude over roughly one hundred millivolts. We neglect variations in $g_\mathrm{EC}$ and $\alpha_\mathrm{\varepsilon}$ due to changes in dot size or position induced by $V_\mathrm{G3}$, as these effects are expected to be small over the voltage range studied.
\vspace{5mm}

\clearpage
\newpage
\subsubsection{Triple point bias triangles}\label{DC-bias triangles}
\begin{figure}[h!]
    \centering    \includegraphics[width=0.82\linewidth]{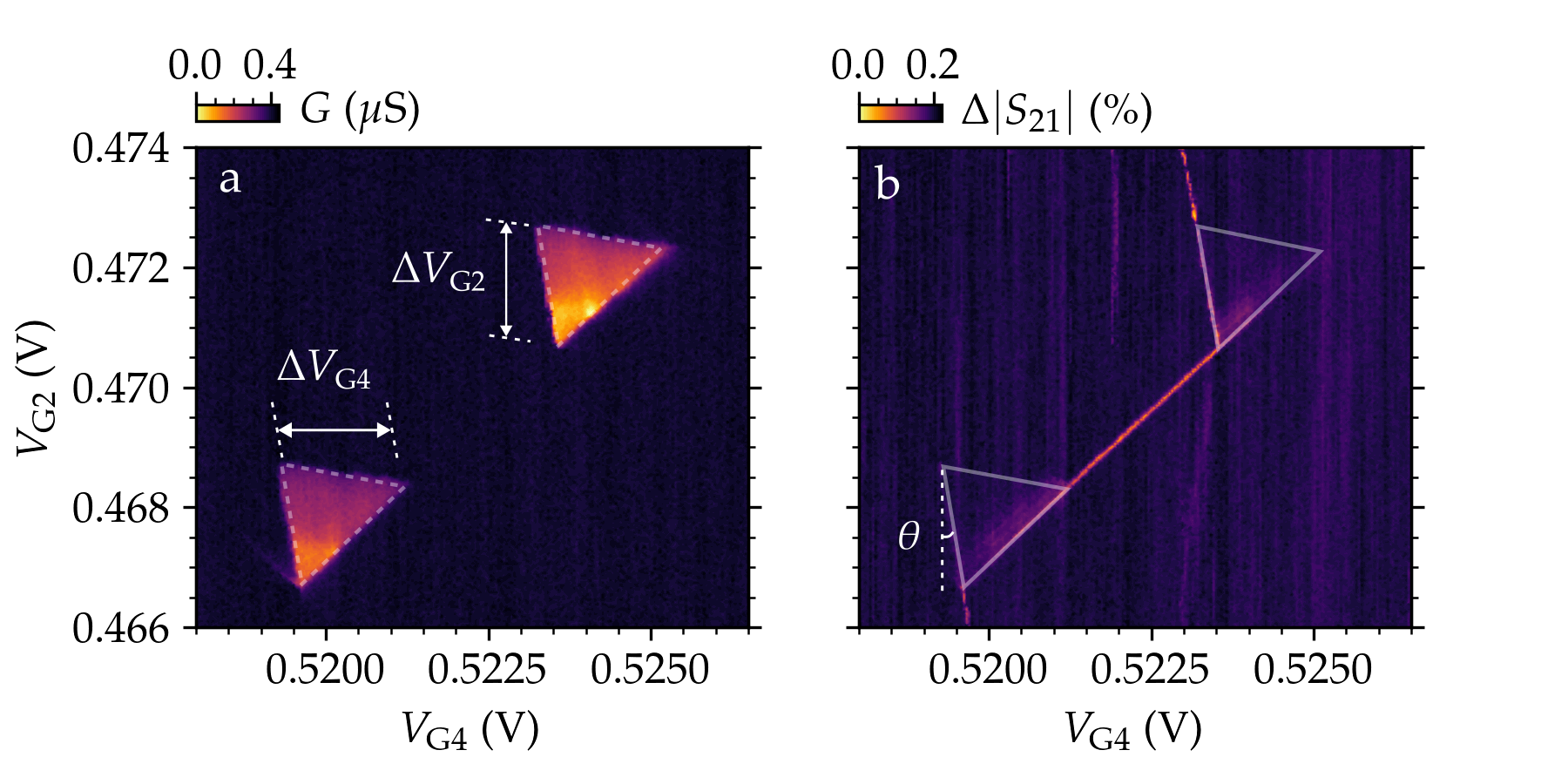}
    \vspace{-0.5cm}
	\customcaption{
		\textbf{DC-bias triangles of the $\mathbf{(1,0) \leftrightarrow (0,1)}$ interdot charge transition.} 
		\textbf{a,b.} Charge stability diagrams showing electron transport through the double quantum dot, measured via \textbf{(a)}, conductance $G$ and \textbf{(b)} the cavity magnitude response $\Delta |S_{21}|$, as a function of the plunger gate voltages $V_\text{G2}$ and $V_\text{G4}$. The voltage barrier configuration $[V_\text{G1}, V_\text{G3}, V_\text{G5}] = [-1,\, 0.185,\, -1]$V resulted in a ETLS frequency at zero detuning of $2t_\mathrm{E}/2\pi = \SI{3}{\giga\hertz}$. A DC bias voltage of $V_\mathrm{SD} = \SI{0.84}{\milli\volt}$ is applied to the source electrode. White lines are added as guides to the triangle dimensions $\Delta V_\mathrm{G2}$, $\Delta V_\mathrm{G4}$ and angles $\theta$ following the methodology in Ref.~\cite{Penfold2017}. Data recorded at $\SI{20}{\milli\kelvin}$.
	}
    \label{fig:SI_bias_triangles}
\end{figure}

A crucial parameter for determining the interdot tunnel coupling $2t_\mathrm{E}$ is the detuning lever arm $\alpha_\mathrm{\varepsilon}$.
From the fitting procedure described in Section~\ref{Double quantum dot ETLS characterization} we obtained $\alpha_\mathrm{\varepsilon} = 0.45 \pm \SI{0.05}{\electronvolt / \volt}$. To independently validate this estimate, we analyse the DC bias triangles \cite{Wiel2003} of the $(1,0) \leftrightarrow (0,1)$ ICT, recorded with a DC bias of $V_\mathrm{SD} = \SI{0.84}{\milli\volt}$. The nanotube conductance $G$ reveals well-defined bias triangles (Fig.~\ref{fig:SI_bias_triangles}a). We use the simultaneously recorded cavity magnitude response $\Delta |S_{21}|$ to determine the charging line slopes and define the extent of the triangles (Fig.~\ref{fig:SI_bias_triangles}b).

Adapting the methodology outlined in Ref.~\cite{Penfold2017}, we determine (1) the lever arm of gate G2 to the left quantum dot, $\alpha_\mathrm{G2\text{-}L}$, from $\Delta V_\mathrm{G2}$; (2) the lever arm of gate G4 to the right quantum dot, $\alpha_\mathrm{G4\text{-}R}$, from $\Delta V_\mathrm{G4}$; and (3) the cross-lever arm of gate G4 to the left quantum dot, $\alpha_\mathrm{G4\text{-}L}$, from the angle $\theta$ corresponding to the $(N_\mathrm{R} \leftrightarrow N_\mathrm{R}+1)$ transition. From these values and assuming a symmetric DQD and gates, the detuning lever arm $\alpha_\mathrm{\varepsilon}$ can be calculated using a modified version of the expression provided in Ref.~\cite{Gonzalez-Zalba2016}
\vspace{-3mm}
\begingroup
\setlength{\belowdisplayskip}{1.8pt}%
\setlength{\belowdisplayshortskip}{4pt}%
\begin{equation}
   \alpha_\varepsilon
   \approx  \sqrt{2}\left( \alpha_\mathrm{G4\text{-}R} - \alpha_\mathrm{G2\text{-}R} \right)
   = \sqrt{2}V_\mathrm{SD} / \left( \Delta V_\mathrm{G4} - \Delta V_\mathrm{G2} \tan{\theta} \right)
   \label{eq:lever_arm}
\end{equation}
\endgroup
From this analysis, we obtain $\alpha_\mathrm{\varepsilon} = 0.48 \pm \SI{0.19}{\electronvolt\per\volt}$, where the uncertainty is obtained via error propagation in Eq.~\ref{eq:lever_arm} using the FWHM of the triangle boundaries extracted from the derivative of the measurement in Fig.~\ref{fig:SI_bias_triangles}a (not shown). This value is consistent with that obtained from the fits to the $(5,4)\leftrightarrow(4,5)$ ICT, with minor deviations likely arising from changes in the lever arm caused by voltage-induced shifts in the quantum dot position.
\vspace{-0.45cm}

% ====================================================================================================
% ====================================================================================================
\clearpage
\newpage
\subsection{Mechanical mode spectroscopy and thermalization}\label{Two-tone spectroscopy map}

\begin{figure}[h!]
    \centering
    \includegraphics[width=.9\linewidth]{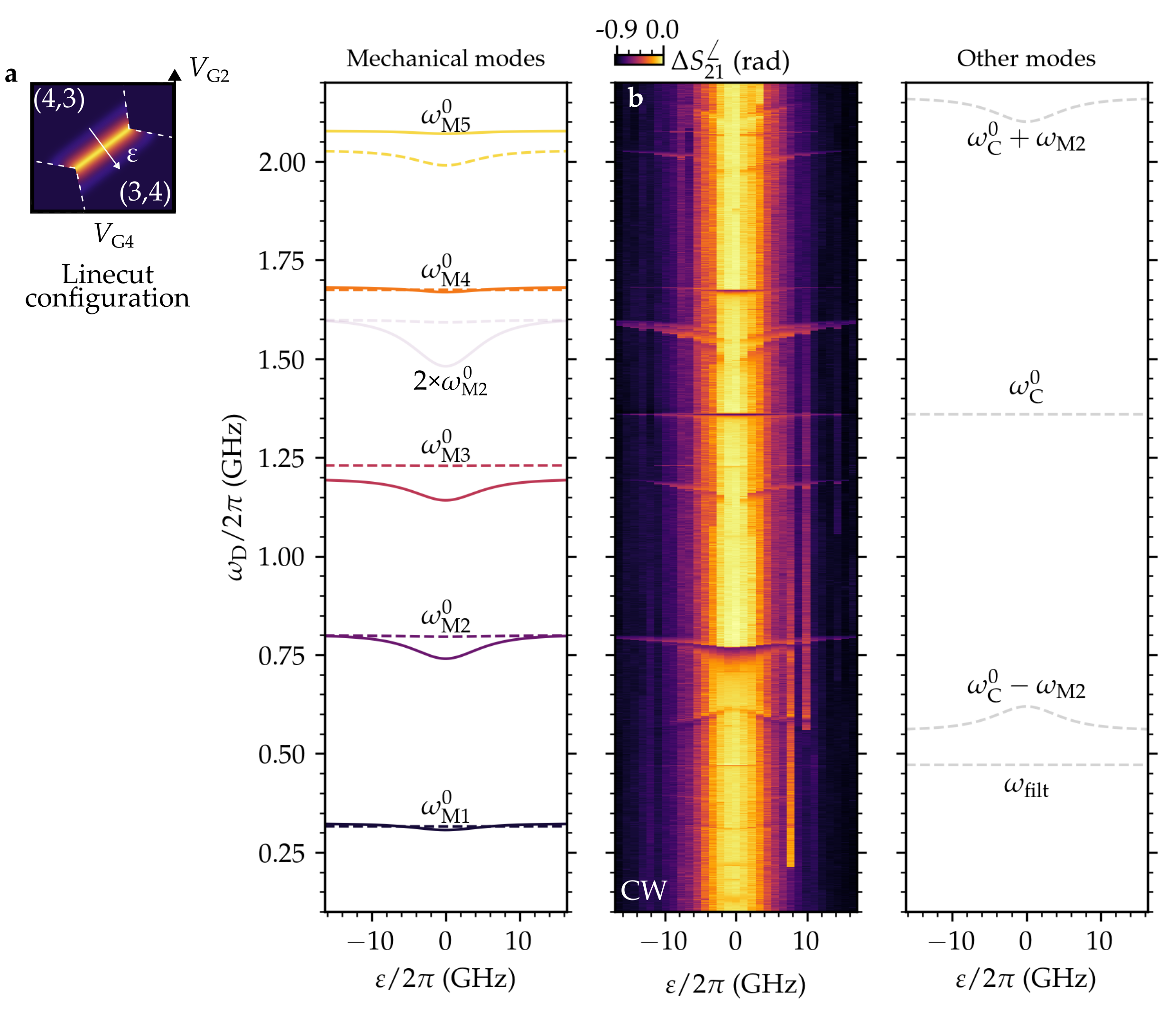}
    \customcaption{
        \textbf{Mechanical mode spectrum of device in the main text.}
        \textbf{a.}~Schematic of the measurement configuration.
        \textbf{b.}~Two-tone CW spectroscopy at the $(4,3) \leftrightarrow (3,4)$ ICT as a function of ETLS detuning $\varepsilon$.
        The barrier gate configuration $[V_\mathrm{G1},\, V_\mathrm{G3},\, V_\mathrm{G5}] = [-1,\ 0,\ -1]\,\text{V}$ results in a ETLS frequency at zero detuning of $2t_\mathrm{E}/2\pi = \SI{4.4}{\giga\hertz}$. 
        Multiple resonances are observed in the cavity phase response, $\Delta S_{21}^{\angle}$, arising from the two polarizations of the mechanical resonances $\omega_{\mathrm{M}i}^0$ ($i = 1,2,3,4,5$) corresponding to the first five flexural modes. The left panel shows the calculated mechanical modes using the parameters listed in Tab.~\ref{tab:all_modes_overview}. The right panel highlights the modes associated with the cavity resonance $\omega_\mathrm{C}^0$, the cavity filter $\omega_\mathrm{filt}$, and the red $(\omega_\mathrm{C}^0-\omega_{\mathrm{M2}})$ and blue $(\omega_\mathrm{C}^0+\omega_{\mathrm{M2}})$ optomechanical sidebands.}
    \label{fig:SI_mechanical_mode_spectrum}
    \vspace{-7mm}
\end{figure}

We identify the nanotube’s mechanical modes by performing two-tone CW spectroscopy along the detuning axis $\varepsilon$ at the $(4,3)\leftrightarrow(3,4)$ ICT (Figs.~\ref{fig:SI_mechanical_mode_spectrum}a,b). In this scheme, a continuous drive is applied to gate G2 while probing the cavity response, see section~$\ref{mechanical oscillator measurements}$. Several flexural modes are visible at frequencies $\omega^0_\mathrm{Mi}$ ($i=1,2,3,4, 5$), corresponding to the first five flexural modes. The characteristic red and blue optomechanical sidebands are also revealed symmetrically placed around the cavity resonance $\omega_\mathrm{C}^0$. The mechanical response of the second flexural mode at $\omega_\mathrm{M}^0/2\pi = \SI{0.801}{\giga\hertz}$ displays the largest cavity shift, indicating its strongest coupling to the ETLS.

\vspace{-4mm}
\newpage

\begin{figure}[b]
    \vspace{-5mm}
    \centering
    % Left: Figure
    \begin{minipage}[t]{0.56\linewidth}
        \vspace{3mm}
        \centering
        \includegraphics[width=\linewidth]{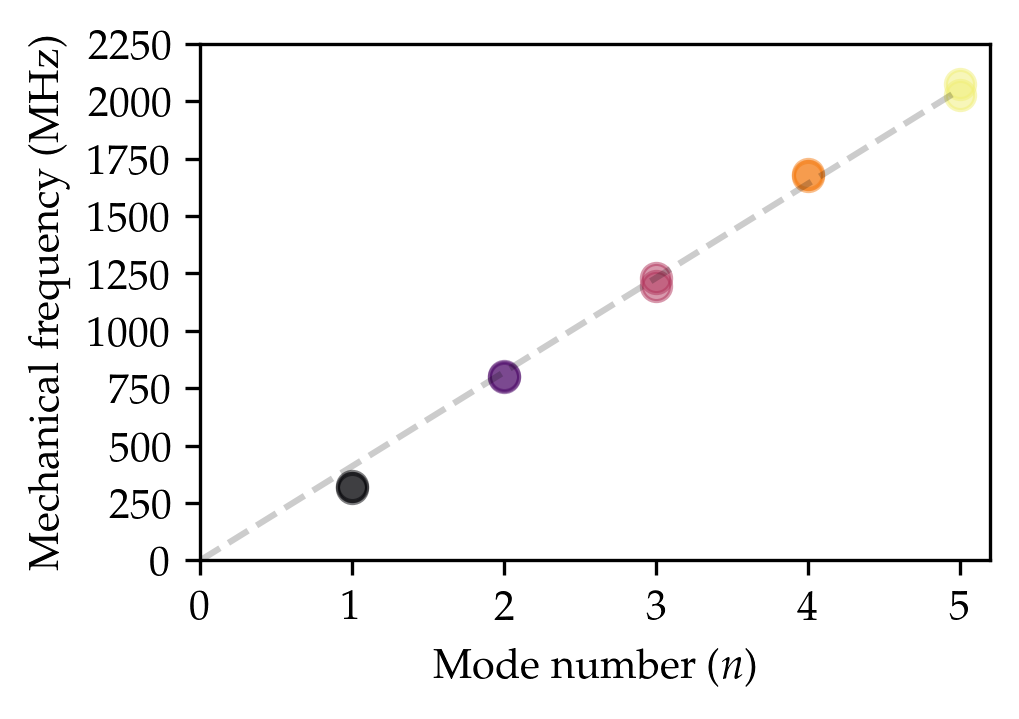}
        \vspace{-10mm}
        \customcaption{
        \textbf{Mechanical frequency dispersion.}
        Bare mechanical frequencies as a function of mode number $n$ extracted from spectroscopy at large ETLS detuning. Each mechanical mode number has two near-degenerate spatial polarizations. Straight line is a zero-offset linear fit with a slope of $\SI{410}{MHz}/n$.}
        \label{fig:SI_mechanical_freq_dispersion}
    \end{minipage}
    \hfill
    % Right: Table
    \begin{minipage}[t]{0.4\linewidth}
        \vspace{4mm}
        \centering
        \setlength{\tabcolsep}{7pt}
        \begin{tabular}{ccc}
        Mode & Frequency (MHz) & $g_{\mathrm{EM}}$ \\
        \hline
        M1 P1 & 316  & 0.1 \\
        M1 P2 & 323  & 0.5 \\
        M2 P1 & 799  & 0.2 \\
        M2 P2 & 802  & 1.0 \\
        M3 P1 & 1230 & 0.1 \\
        M3 P2 & 1196 & 0.9 \\
        M4 P1 & 1675 & 0.1 \\
        M4 P2 & 1681 & 0.4 \\
        M5 P1 & 2028 & 0.8 \\
        M5 P2 & 2077 & 0.3 \\
        \hline
        \end{tabular}
        \vspace{4mm}
        \customtablecaption{
        \textbf{Mechanical resonance frequencies and electromechanical couplings.}
        Bare mechanical mode frequencies and electromechanical couplings $g_{\mathrm{EM}}$ relative to the second flexural mode M2 P2.
        }
        \label{tab:all_modes_overview}
    \end{minipage}
\end{figure}

The bare mechanical mode frequencies display a linear dispersion as expected for a straight tensioned string (Fig.~\ref{fig:SI_mechanical_freq_dispersion}). The second flexural mode, with bare frequency $\omega_\mathrm{M}^0/2\pi = \SI{0.801}{\giga\hertz}$, is thus easily identified and furthermore displays the most prominent Duffing-like response across a zero ETLS detuning linecut (Fig.~\ref{fig:SI_mechanical_mode_spectrum}c). Each mechanical mode number features two spatial polarizations, P1 and P2. For the device discussed in the main text, the two polarizations of the second flexural mode correspond to vibrations that are nearly in‑plane (P1, low coupling) and out‑of‑plane (P2, high coupling). The vibration axis of the in-plane polarization is offset by a only small angle of \SI{12}{\degree} relative to the chip plane as estimated by fitting the mechanical frequency softening of both modes as a function of ETLS detuning. A similarly low polarization angle was found for the first few flexural modes allowing the weakly coupled polarization to be effectively ignored in the quantitative analysis and general discussion. 
The mechanical resonance frequencies of the first five flexural modes, including both polarizations, along with their estimated coupling rates relative to the second flexural mode, are given in Table~\ref{tab:all_modes_overview}.

\begin{figure}[h!]
    \centering
    \includegraphics[width=.65\linewidth]{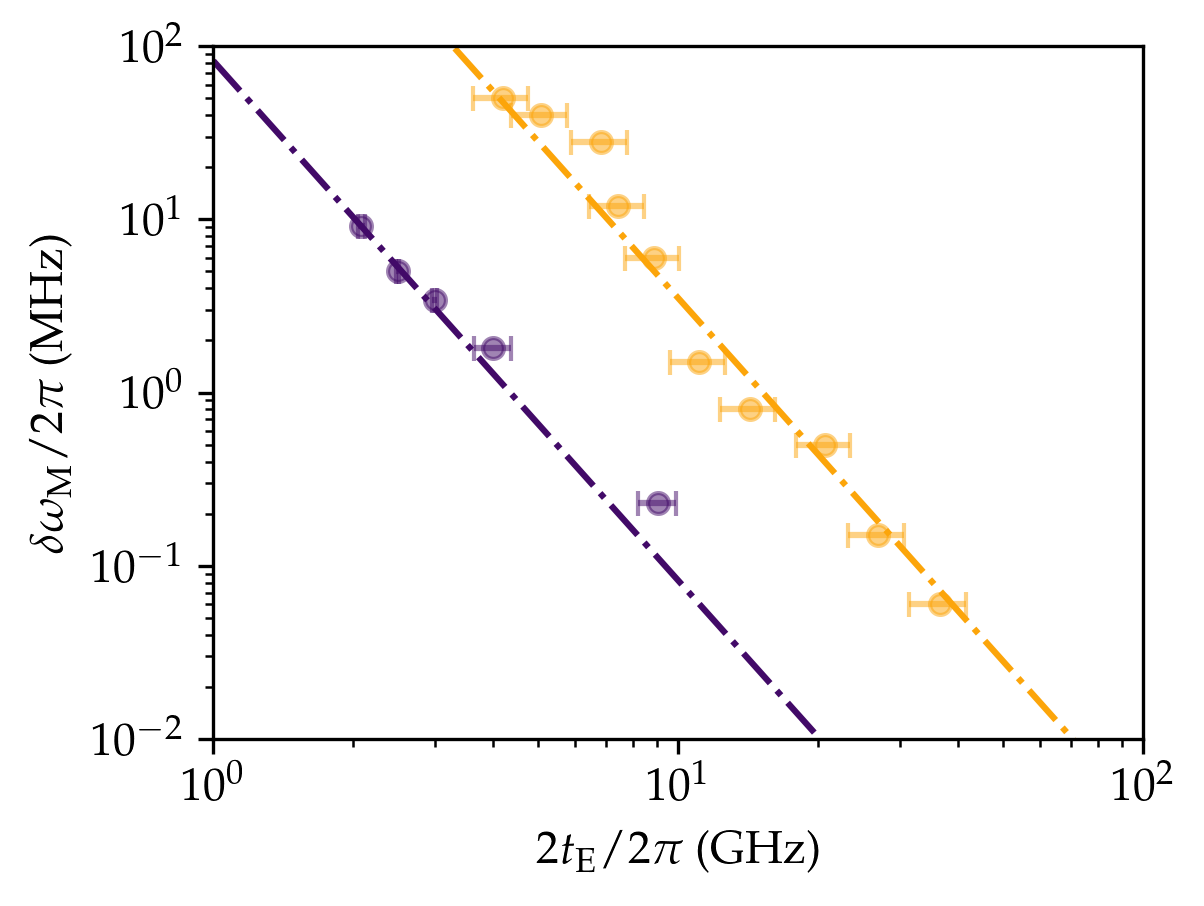}
    \vspace{-0.5cm}
    \customcaption{
        \textbf{Thermomechanical linewidth broadening.} The thermal broadening of the mechanical linewidth from Fig. 4c in the main text fit solely with a common temperature of \SI{47}{mK} for both ICT configurations according to Eq.~\ref{eq:SI_version_mechanical_thermal_broadening}.
        }
    \label{fig:SI_linewidth_thermal_noise_fit}
    \vspace{-20mm}
\end{figure}

The temperature of the mechanical resonator can be estimated from the measured mechanical linewidths of Fig.~4c in the main text. 
Due to the self- and cross-Kerr interactions described in Section~\ref{SectionBroadening}, all mechanical modes contribute to the thermal broadening of the main mechanical mode (M2 P2). 
By including all mechanical modes described in Table~\ref{tab:all_modes_overview}, the thermal broadening is fit with just a single free parameter, namely a common temperature for both ICT configurations, according to Eq.~\ref{eq:SI_version_mechanical_thermal_broadening} (see Fig.~\ref{fig:SI_linewidth_thermal_noise_fit}). This yields a mechanical temperature of $47\pm\SI{5}{mK}$, where the uncertainty predominantly arises from the roughly \SI{10}{\percent} uncertainty in the extracted electromechanical coupling rates. This temperature corresponds to a mean thermal phonon number of $0.9\pm0.1$ 
for the main mechanical mode. While not independently verified, we stress that a small residual thermal occupation does not affect the Kerr-nonlinearity inferred from the coherently driven mechanical response.

Using the experimentally determined frequencies and electromechanical coupling rates from Table~\ref{tab:all_modes_overview}, we can furthermore contrast the thermal noise dephasing contributions. The self-Kerr noise of the main mechanical mode M2 P2 contributes about \SI{50}{\percent} while the cross-Kerr contribution of all other modes contribute the remaining \SI{50}{\percent}. For the cross-Kerr contributions, the fundamental mode M1 P2 accounts for roughly \SI{90}{\percent}. Higher frequency modes have an increasingly lower thermal occupation and vanishing impact.

Currently, resolved phonon transitions are not observed spectroscopically as $K/\delta \omega_M < 1$, predominantly limited by thermal noise. This thermal noise can realistically be reduced through a combination of improved mechanical thermalization and increased mechanical frequencies. We expect to reach device temperatures of \SI{20}{mK} by operating in a better thermalized new \SI{10}{mK} cryostat and using shorter nanotubes, which additionally feature higher mechanical frequencies and thereby lower thermal phonon occupancy.
Nanotubes half the current length are feasible, as previously demonstrated \cite{Tormo2022}, and are expected to be achievable with substantially higher yield using our employed nanotube stamping technique. Importantly, stamping also enables significant nanotube tensioning, which combined with room‑temperature pre‑selection, can increase mechanical frequencies by at least a factor of two.
Additionally, electrostatic fine‑tuning of the gate array can localize the DQD such that the fundamental mechanical mode symmetrically modulates the energy detuning $\varepsilon$. This will decouple the fundamental mode, the dominant source of cross‑Kerr noise, and similarly reduces the contributions of higher modes. This mode‑symmetry‑based decoupling is expected to reduce dephasing by an additional factor of two. Taken together, these improvements should enable $K/\delta\omega_M > 10$.

\vspace{-3mm}

% ====================================================================================================
% ====================================================================================================
\clearpage
\subsection{Mechanical hysteresis}\label{Mechanical Hysteresis}

The realization of a nonlinear mechanical oscillator in the dispersive USC regime enables measuring its driven response in a direct way. This differs from previous approaches Ref.~\cite{Yang2024Science_MechanicalQubit,Arrangoiz-Arriola2019}, where the spectral properties were reconstructed indirectly via ancillary qubit manipulation, underscoring the advantage of our method in accessing the mechanical dynamics without intermediate protocols.

\begin{figure}[h!]
    \centering
    \includegraphics[width=1\linewidth]{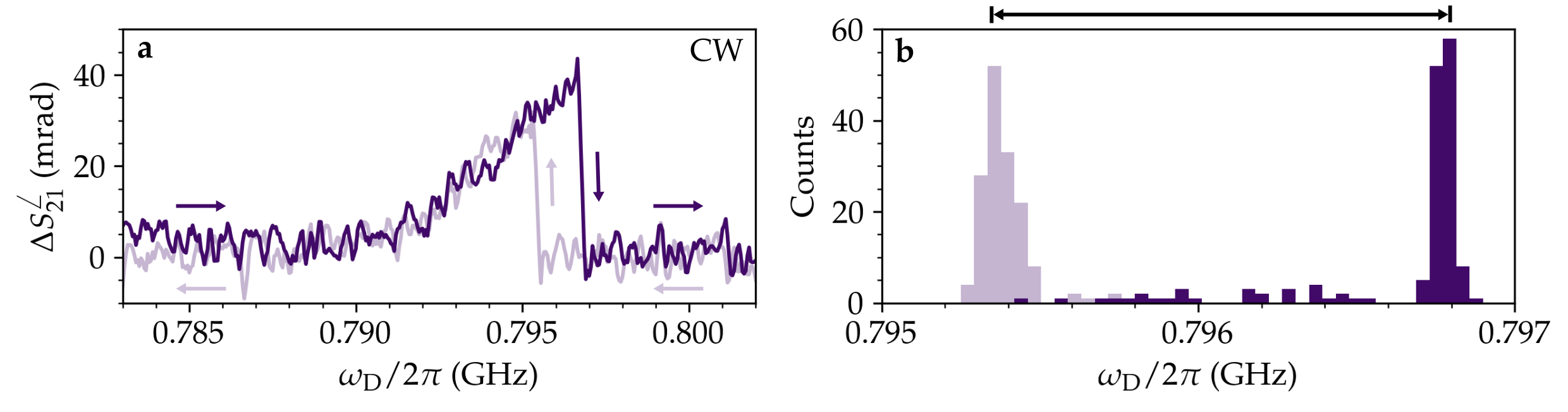}
    \customcaption{
        \textbf{Hystheretic behaviour of the Kerr oscillator.}
        \textbf{a.}~Mechanical CW response as a function of drive frequency $\omega$ at the $(4,3) \leftrightarrow (3,4)$ ICT for $\varepsilon = 0$. The barrier gate configuration $[V_\mathrm{G1},\, V_\mathrm{G3},\, V_\mathrm{G5}] = [-1,\ 0.05,\ -1]\,\text{V}$ results in a ETLS frequency at zero detuning of $2t_\mathrm{E}/2\pi = \SI{28}{\giga\hertz}$. The dark (light) purple curve indicates the up (down) frequency scan.
        \textbf{b.}~Histogram of the frequency bistability of 308 traces like the ones shown in \textbf{(a)}. The double-headed black arrow indicates the frequency difference between the switches of the up and down sweeps, which is found to be approximately $\SI{1.4}{\mega\hertz}$. Data recorded at \SI{20}{mK}.
        }
    \label{fig:SI_Duffing_hystheresis}
\end{figure}

The direct access to the driven spectrum allows us to observe hysteresis in the dynamics, a hallmark of Kerr-type nonlinear oscillators~\cite{bachtold2022mesoscopic}. The effect arises from the anharmonicity of the potential, which produces bistable solutions when the mechanical oscillator is driven near resonance. An example of the nonlinear response of our mechanical oscillator is shown in Fig. \ref{fig:SI_Duffing_hystheresis}a, where we observe a clear positive Kerr nonlinearity at $\varepsilon = 0$. The nonlinearity manifested as a hysteresis loop, where the system follows a different trajectory depending on whether the drive frequency is swept upwards or downwards, consistent with the bistable solutions predicted by Kerr (Duffing) theory. At finite detunings $\varepsilon > 2t_\mathrm{E}/2$, we also observe a negative Kerr nonlinearity as observed in our previous work with single quantum dot-based devices Ref.~\cite{samanta2023nonlinear}.

Figure \ref{fig:SI_Duffing_hystheresis}b shows a histogram compiled from 308 individual traces like the one in Fig. \ref{fig:SI_Duffing_hystheresis}a. The histogram reveals two pronounced peaks at frequencies separated by $\SI{1.4}{\mega\hertz}$. This splitting directly reflects the reproducibility of the Kerr nonlinearity, demonstrating the consistent bistable behavior of the mechanical oscillator across repeated measurements.

% ====================================================================================================
% ====================================================================================================
\clearpage
\newpage
\subsection{Optomechanical sidebands}

The optomechanical interaction is mediated by the ETLS, and is described in detail in Section~\ref{sec:NEW__Effective_cavity-oscillator_coupling}. The optomechanical sidebands (SBs) shown in main text Fig.~4d result from a linear canonical optomechanical interaction described in Eq.~\ref{eq:effective_classical_cavity_hamiltonian}. The linear optomechanical interaction emerges at finite ETLS detuning, resulting in a cavity frequency shift contribution that is linear in the mechanical displacement, $\Delta\omega_\mathrm{C}^{(1)} = g_\mathrm{OM}^{(1)} \cdot (x/x_\mathrm{zpf})$. 
Experimentally, we apply two tones to the cavity: a strong, off-resonant drive at $\omega_\C \pm \omega_\M$, far from both the cavity and the mechanical resonance frequencies, and a weak probe at the cavity resonance frequency. The drive is AM modulated allowing us to use the same two-tone AM homodyne technique described in Section~\ref{mechanical oscillator measurements}. The strong drive results in the usual linearized optomechanical interaction~\cite{AspelmeyerRMP2014}, which up/down-converts drive photons via a scattering process. The scattered drive photons are detected through interference with the weak resonant probe. This occurs when the drive is at the red (blue) sideband frequency of $\omega_\D = \omega_\mathrm{C} - \omega_\M$ ($\omega_\D = \omega_\mathrm{C} + \omega_\M$) which results in up (down) converted photons being resonant with the cavity. The interference created in such two-tone experiments can lead to optomechanically induced transparency/amplification\cite{WeisScience2010}, an effect recently demonstrated in a nanotube single quantum dot electromechanical system\cite{Blien2020}. 

\begin{figure}[h!]
    \centering
    \includegraphics[width=0.55\linewidth]{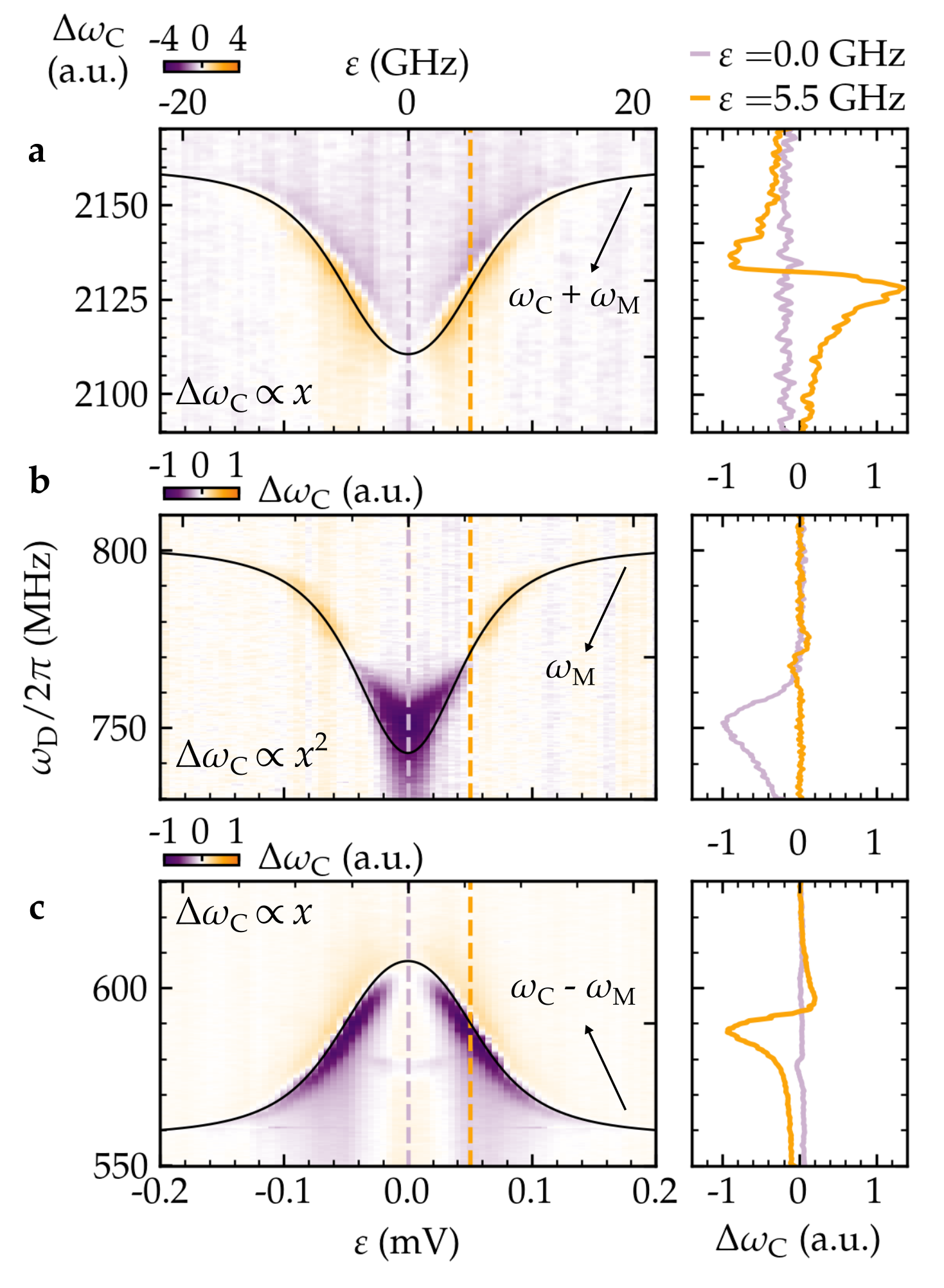}
    \customcaption{
            \textbf{Optomechanical sidebands.}
            (Left) Two-tone AM spectroscopy maps of the \textbf{(a)} blue sideband $(\omega_\D = \omega_\C + \omega_\M)$, \textbf{(c)} red sideband $(\omega_\D = \omega_\C - \omega_\M)$, and \textbf{(b)} the second flexural mechanical mode $(\omega_\M)$. The DQD ETLS configuration is the same as in the main text, corresponding to the $(5,4) \leftrightarrow (4,5)$ ICT, with $2t_\mathrm{E}/2\pi = \SI{7.4}{\giga\hertz}$ and $g_\mathrm{EM}/2\pi = \SI{0.46}{\giga\hertz}$. The color scale represents the real part of the AM demodulated cavity response, following the measurement protocol described in Section~\ref{Details on calibration of mechanical displacement}. The black solid lines are the theoretical frequencies expected at zero mechanical displacement using $\omega_\M$ from Eq.~\ref{eq:softening} and $\omega_\D = \omega_\C \pm \omega_\M$. All subplots span the same frequency range. (Right) Different line cuts of the two-tone spectroscopy maps are shown for detunings $\varepsilon_0 = 0$ (light purple) and $\SI{5.5}{\giga\hertz}$ (yellow). Data in all panels are recorded at \SI{20}{mK}.
            }        
    \label{fig:SI_OMIT_sidebands}
\end{figure}

The measured ETLS detuning ($\varepsilon$) dependence of these two optomechanical sidebands at $\omega_\C \pm \omega_\M$ is compared to the dependence of the mechanical frequency renormalization at $\omega_\M$ in Fig.~\ref{fig:SI_OMIT_sidebands}.
Both the optomechanical sideband at $\omega_\mathrm{C} + \omega_\M$ and the mechanical resonance at $\omega_\M$ feature a frequency reduction at zero detuning ($\varepsilon=0$), while the optomechanical sideband at $\omega_\mathrm{C} - \omega_\M$ displays a frequency increase because of the negative sign in $\omega_\mathrm{C} - \omega_\M$ (Fig.~\ref{fig:SI_OMIT_sidebands}a-c).
Moreover, the signal of the two sidebands disappears near zero detuning but becomes largest at finite detuning near $\varepsilon \sim t_\E$ (Fig.~\ref{fig:SI_OMIT_sidebands}~a,c), in agreement with the expected cavity frequency shift associated with the linear coupling strength $g_\mathrm{OM}^{(1)} = \left. \frac{d\omega_\mathrm{C}}{dx}\right|_{x=0} \propto \frac{\varepsilon}{(4t_\E^2+\varepsilon^2)^{5/2}}$ (Eq.~\ref{gom_1}).
Conversely, the signal of the mechanical resonance at $\omega_\M$ is largest at zero detuning and reverses sign at finite detuning, in agreement with the predicted quadratic coupling strength $g_\mathrm{OM}^{(2)}= \frac{1}{2} \left. \frac{d^2\omega_\mathrm{C}}{dx^2}\right|_{x=0} \propto \frac{ 4t_\E^2-4 \varepsilon^2}{(4t_\E^2+\varepsilon^2)^{7/2}}$ (Eq.~\ref{gom_2}). 

% ====================================================================================================
% ====================================================================================================
\clearpage
\newpage
\subsection{Calibration of mechanical displacement}\label{Details on calibration of mechanical displacement}
\vspace{-2mm}
In the main text, we show that the mechanical displacement squared relates to the cavity frequency shift through main text Eq.~3, that is $\Delta\omega_\mathrm{C} =  6\frac{g_\mathrm{EC}^2g_\mathrm{EM}^2}{(2t_\mathrm{E})^3} \left(\frac{x_\mathrm{0}}{x_\mathrm{zpf}}\right)^2$, valid in the regime $\frac{x_\mathrm{0}}{x_\mathrm{zpf}}\ll \frac{2t_\mathrm{E}}{2g_\mathrm{EM}}$. To accurately extend this relation to large displacement amplitudes, or when using the amplitude modulation detection scheme (Section~\ref{mechanical oscillator measurements}), the system dynamics are solved numerically. Here, we describe the numerical integration method used to quantify the displacement of the mechanical oscillator from the cavity frequency shift, which for small displacements reduces to main text Eq.~3, derived from the general Eq.~\ref{eq:omega_c_R}.

\begin{figure}[b!]
    \vspace{-10mm}
    \centering
    \includegraphics[width=0.95\linewidth]{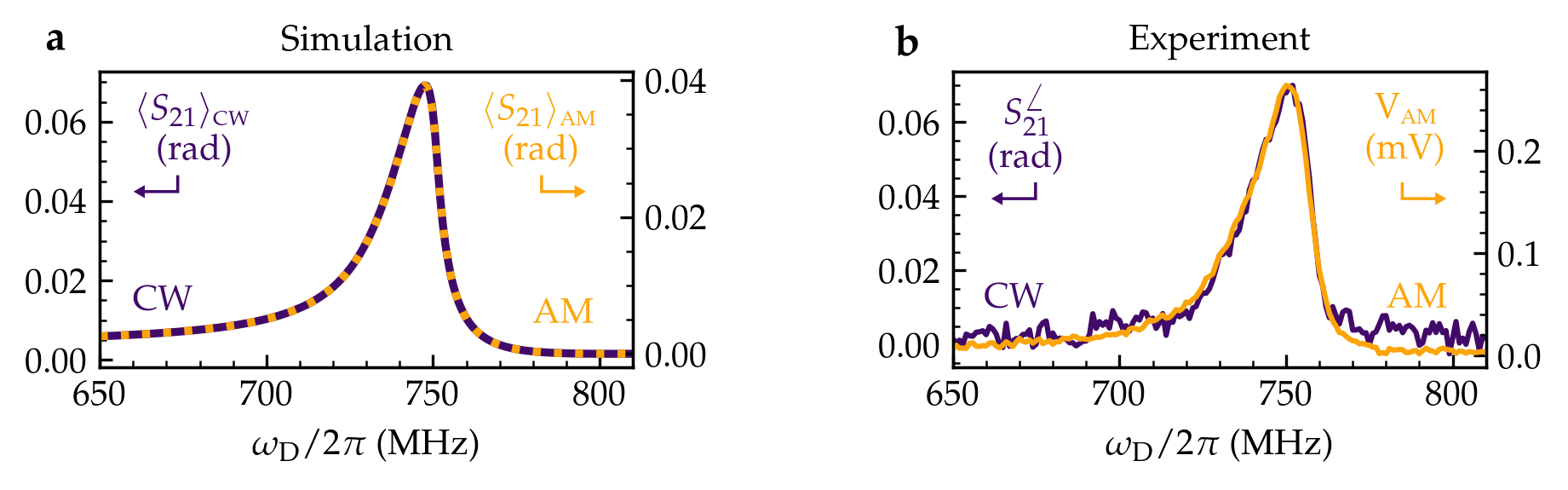}
    \vspace{-3mm}
     \customcaption{
        \textbf{Mechanical displacement spectrum.}
        \textbf{a.} Numerically calculated CW and AM mechanical spectra for detuning drive amplitudes $\varepsilon_\D/2\pi = \SI{550}{MHz}$ and \SI{300}{MHz}, whose ratio is $1.8$.
        \textbf{b.} Experimentally measured CW and AM mechanical spectra for (unmodulated) RF signal generator output powers of \SI{-20}{dBm} and \SI{-25}{dBm}, respectively, corresponding to gate drive amplitudes $\varepsilon_\D$ differing by a factor of $\simeq 1.8$. 
        All panels are plotted versus drive frequency $\omega_\D$. The numerical calculations use the experimental parameters of $\alpha_\mathrm{\varepsilon} = \SI{0.45}{\electronvolt / \volt}$, $\gamma/2\pi = \SI{15}{\mega\hertz}$, $\omega^0_\mathrm{M}/2\pi = \SI{0.801}{\giga\hertz}$, $g_\mathrm{EM}/2\pi = \SI{0.46}{\giga\hertz}$, $\omega_\mathrm{C}/2\pi = \omega_\mathrm{P}/2\pi = \SI{1.359}{\giga\hertz}$, $\kappa/2\pi = \SI{3.7}{\mega\hertz}$, $g_\mathrm{EC}/2\pi = \SI{49.8}{\mega\hertz}$, $2t_\mathrm{E}/2\pi = \SI{7.42}{\giga\hertz}$.}   \label{fig:SI_mech_spectrum_disp_calib}
\end{figure}

The numerical method calculates the driven displacement of the mechanical oscillator and the averaged cavity response. The oscillator stationary periodic trajectory $x(t)$ is obtained by numerically integrating its equation of motion under a CW detuning drive $\varepsilon_\D(t) = \varepsilon_\D\cos({\omega_\D t})$ as described in Section~\ref{sec:S21_thermal}. 
Since the cavity bandwidth is much slower than the mechanical dynamics ($\kappa\ll \omega_\M$), the cavity dynamics are averaged over one driven period using 
\vspace{1mm}
\begin{equation}
    \langle S_{21}\rangle_{\text{\tiny CW}}\left(\varepsilon_\D\right)
    = \frac{2\pi}{\omega_\D}
    \int_{0}^{2\pi/\omega_{\text{\tiny D}}}
    \!\!\!\!\!\!\!\!\!\!\!\!\!\!
    S_{21}(\varepsilon_\D,t)~dt .
    \nonumber
    \vspace{3mm}
\end{equation}
according to Eq.~\ref{eq:S21_mech_driven} to yield the CW cavity response of Fig.~\ref{fig:SI_mech_spectrum_disp_calib}a. In two-tone AM spectroscopy we modulate the detuning drive, $\varepsilon_\D(t) =\varepsilon_\D\left[1 + \cos({\omega_\mathrm{AM}t})\right]\cos({\omega_\D t})$. Since the AM modulation frequency $\omega_\mathrm{AM}$ is much slower than both the mechanical and cavity responses, we construct the modulated cavity response from steady-state CW spectra at different drive amplitudes. The AM cavity response is then found by numerically demodulating the now slowly time varying CW response at $\omega_\mathrm{AM}$ (Fig.~\ref{fig:SI_mech_spectrum_disp_calib}a). 
\vspace{-5mm}

\newpage
\noindent\mbox{}\vspace*{-1cm}   

We now describe a direct calibration of the AM cavity response, which is recorded as the lock-in amplifier output in volt units. For this, we tune the AM drive amplitude until the measured CW and AM nonlinear spectra acquire the same shape (Fig.~\ref{fig:SI_mech_spectrum_disp_calib}b), corresponding to a measured conversion of \SI{0.27}{mV/rad}. 
The spectral overlap occurs when the CW drive amplitude is larger than the AM drive by a factor $\beta\simeq 1.8$. This results from the combined influence of spectral averaging, and the fact that for the same $\varepsilon_\mathrm{D}$ amplitude the AM generates a peak drive amplitude that is twice that of the CW tone. The measured $\beta$ matches the numerical prediction in Fig.~\ref{fig:SI_mech_spectrum_disp_calib}a, where the CW and AM spectra overlap, and it remains close to $1.8$ across a wide range of drive amplitudes. Note that the numerically predicted vertical scales of the CW and AM cavity response in Fig.~\ref{fig:SI_mech_spectrum_disp_calib}a also differ by the factor $\beta$ as expected from a simple analysis, and that the numerical calculation in Fig.~\ref{fig:SI_mech_spectrum_disp_calib}a quantitatively predicts the measured CW spectrum in Fig.~\ref{fig:SI_mech_spectrum_disp_calib}b. Under these conditions, the AM demodulated peak displacement maps directly to the CW value through the factor $\beta$, that is 
$\left(x_\mathrm{0}/x_\mathrm{zpf}\right)_\mathrm{CW}^2 = \beta\left(x_\mathrm{0}/x_\mathrm{zpf}\right)_\mathrm{AM}^2$.
Together, this calibration and numerical method yields a unified quantitative link between $\Delta\omega_{\mathrm{C}}$, $\Delta\omega_{\mathrm{M}}$, and $\left(x_0/x_{\mathrm{zpf}}\right)^2$ for both AM and CW detection schemes.

\begin{figure}[h!]
    \vspace{-4mm}
    \centering\includegraphics[width=\linewidth]{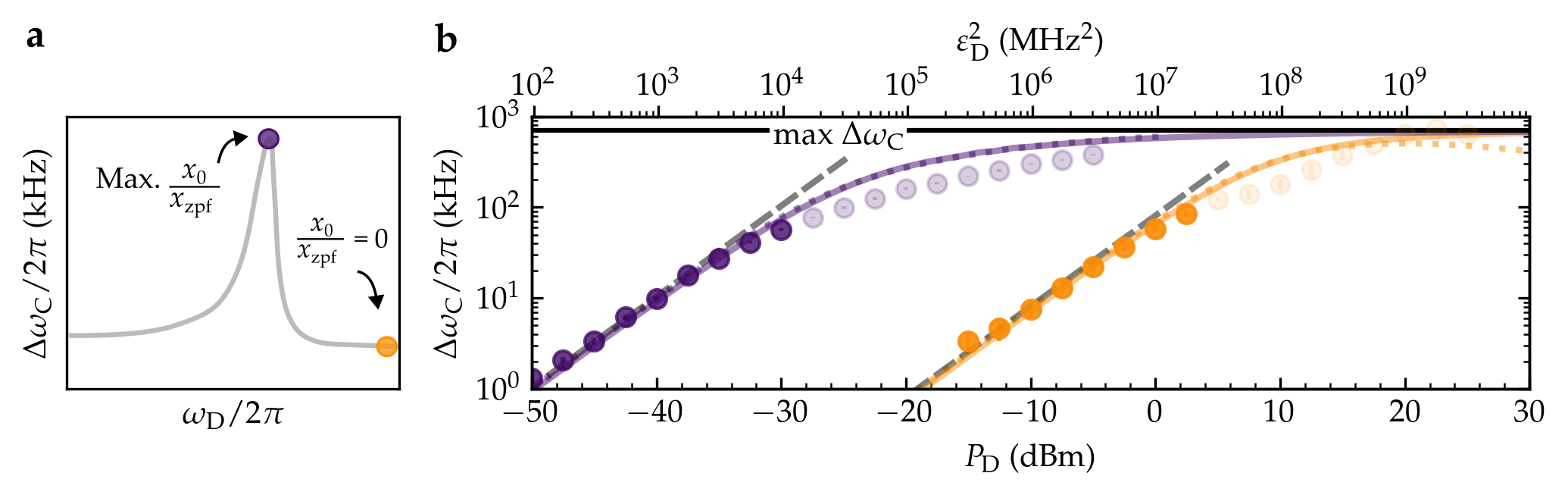}
    \vspace{-6mm}
    \customcaption{
            \textbf{Cavity frequency shift from mechanical displacement and ETLS modulation.}
            \textbf{a.} Schematic of a driven mechanical spectrum, with purple (orange) indicating configurations of maximal (negligible) mechanical displacement.  
            \textbf{b.} Cavity frequency shift $\Delta\omega_\mathrm{C}$ versus driving power at room temperature (bottom axis) and corresponding detuning drive (top axis), calculated using $\alpha_\mathrm{\varepsilon} = \SI{0.45}{\electronvolt/\volt}$ and measured RF line attenuation of \SI{87}{dB}.
            Solid (AM) and transparent (CW) points are offset by \SI{5}{dB} corresponding to the drive amplitude factor $\beta$.
            Solid and dotted curves (CW and AM drive) are calculated using the same parameters as Fig.~\ref{fig:SI_mech_spectrum_disp_calib}. Dashed grey lines: manual fit to the $\varepsilon_\mathrm{D}^2$ dependence from the semiclassical model. Black solid horizontal line is the maximum cavity shift, corresponding to the bare cavity frequency.
            }
             \label{fig:SI_cavity_freq_shift_vs_power__model_vs_data}
    \vspace{-12mm}
\end{figure}

At low drive powers, the cavity frequency shift exhibits a robust quadratic dependence $\Delta\omega_\mathrm{C} \propto \varepsilon_\mathrm{D}^2$ (Fig.~\ref{fig:SI_cavity_freq_shift_vs_power__model_vs_data}b), observed both when the mechanical mode is driven on resonance and when the drive is far detuned. On mechanical resonance, this $\varepsilon_\mathrm{D}^2$ scaling directly reflects the predicted $x^2$-readout sensitivity (main text Fig.~3b), arising from the linear relation between displacement and drive amplitude, $x_\mathrm{0} \propto \varepsilon_\mathrm{D}$. The same quadratic dependence also appears for a drive at \SI{1}{GHz} drive, well outside the mechanical bandwidth, where no mechanical motion is generated. In this off-resonant regime, the frequency shift originates purely from the modulation of the ETLS, and is fully quantitatively reproduced by the numerical model without any free parameters (Fig.~\ref{fig:SI_cavity_freq_shift_vs_power__model_vs_data}b).

%%%%%%%%%%%%%%%%%%%%%%%%%%%%%%%%%%%%%%%%%%%%%%%%%%%%%%%%%%%%%%%%%%%%%%%%%%%%%%%%%%%%%%%%%%%%%%%%%%%%%%%%%%%%%%%%%%%%%%%%%%%%

% ====================================================================================================

% ====================================================================================================
\clearpage
\newpage
\subsection{Similar results in a second device}\label{sec:second_device}

\begin{figure}[h!]
    \centering
    \includegraphics[width=0.95\linewidth]{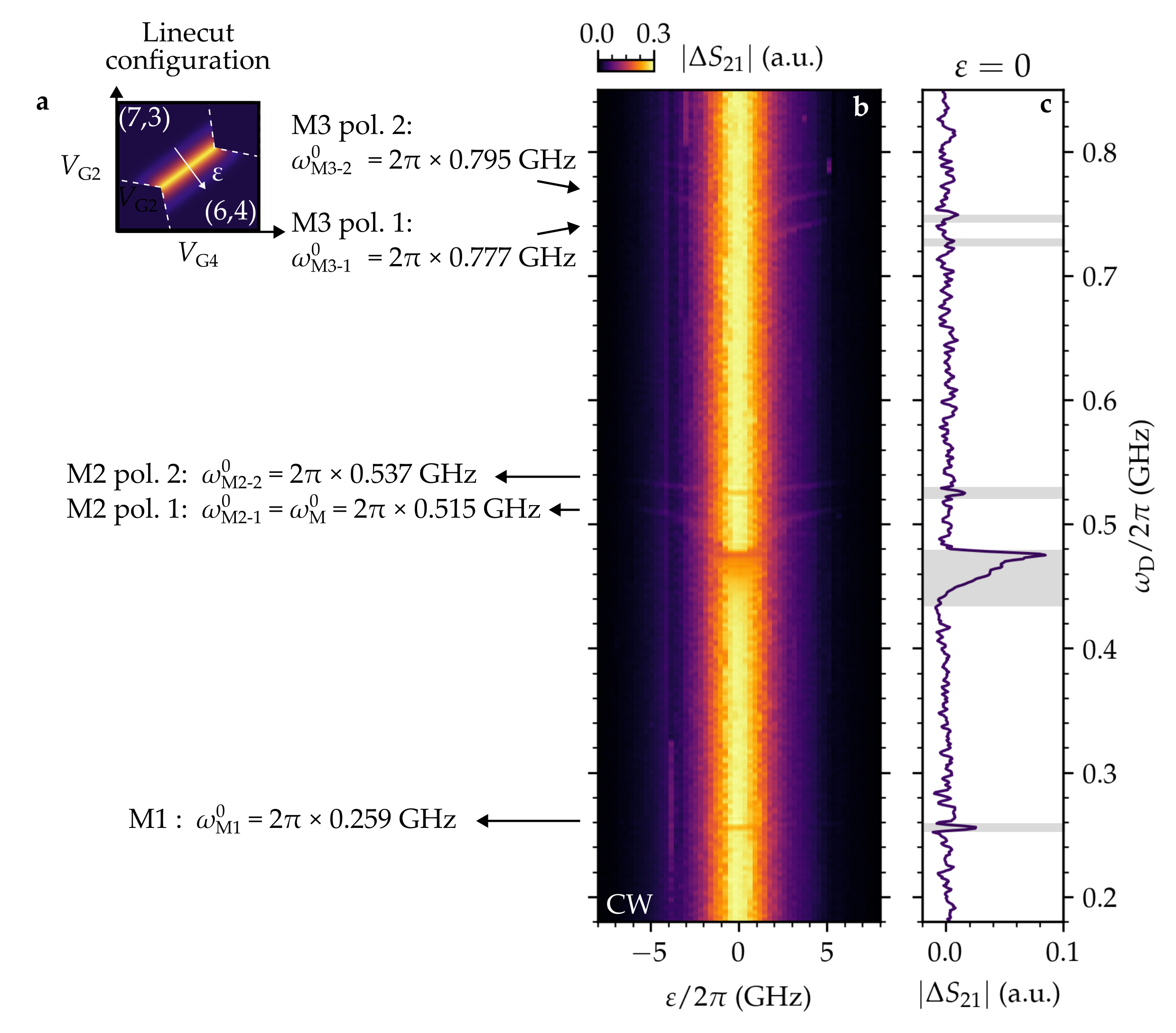}
    \customcaption{
        \textbf{Mechanical mode spectrum of a second device.}
        \textbf{a.}~Schematic of the measurement configuration.
        \textbf{b.}~Two-tone CW spectroscopy at the $(7,3) \leftrightarrow (6,4)$ ICT as a function of ETLS detuning $\varepsilon$.
        The barrier gate configuration $[V_\mathrm{G1},\, V_\mathrm{G3},\, V_\mathrm{G5}] = [-0.26, 0.54, -0.26]\,\text{V}$ results in a ETLS frequency at zero detuning of $2t_\mathrm{E}/2\pi = \SI{7.4}{\giga\hertz}$.
        Multiple resonances are observed in the cavity phase response $|\Delta S_{21}|$, corresponding to mechanical modes $\omega_{\mathrm{M}i-j}^0$, where $i = 1, 2, 3$ corresponds to the flexural mode number and $j = 1, 2$ to the mode polarization.
        \textbf{c.}~Linecut of the spectrum in \textbf{(b)}, recorded at zero detuning $\varepsilon = 0$. Distinct mechanical modes are labelled on the left-hand side of panel \textbf{(b)}.
        Data in all panels are recorded at \SI{30}{mK}.
    }
    \label{fig:SI_mechanical_mode_spectrum_thesis}
\end{figure}

The key results presented in the main text were reproduced on another device. This device was fabricated using a different method, see Section~\ref{Off-chip grown}. In this second device, the cavity frequency $\omega_\mathrm{C}^0/2\pi = \SI{1.476}{\giga\hertz}$ and the linewidth $\kappa/2\pi = \SI{2.6}{\mega\hertz}$ were comparable to those of the device presented in the main text. The ETLS was also tunable through gate voltages, allowing fine control in the $2t_\mathrm{E} \sim 1$–$\SI{100}{\giga\hertz}$ range. The mechanical modes were identified by two-tone CW spectroscopy (Fig.~\ref{fig:SI_mechanical_mode_spectrum_thesis}a–c). The second flexural mode was found at $\omega_\mathrm{M}^0/2\pi = \SI{0.515}{\giga\hertz}$ under a continuous drive on gate G3.

\begin{figure}[h!]
    \centering
    \includegraphics[width=1\linewidth]{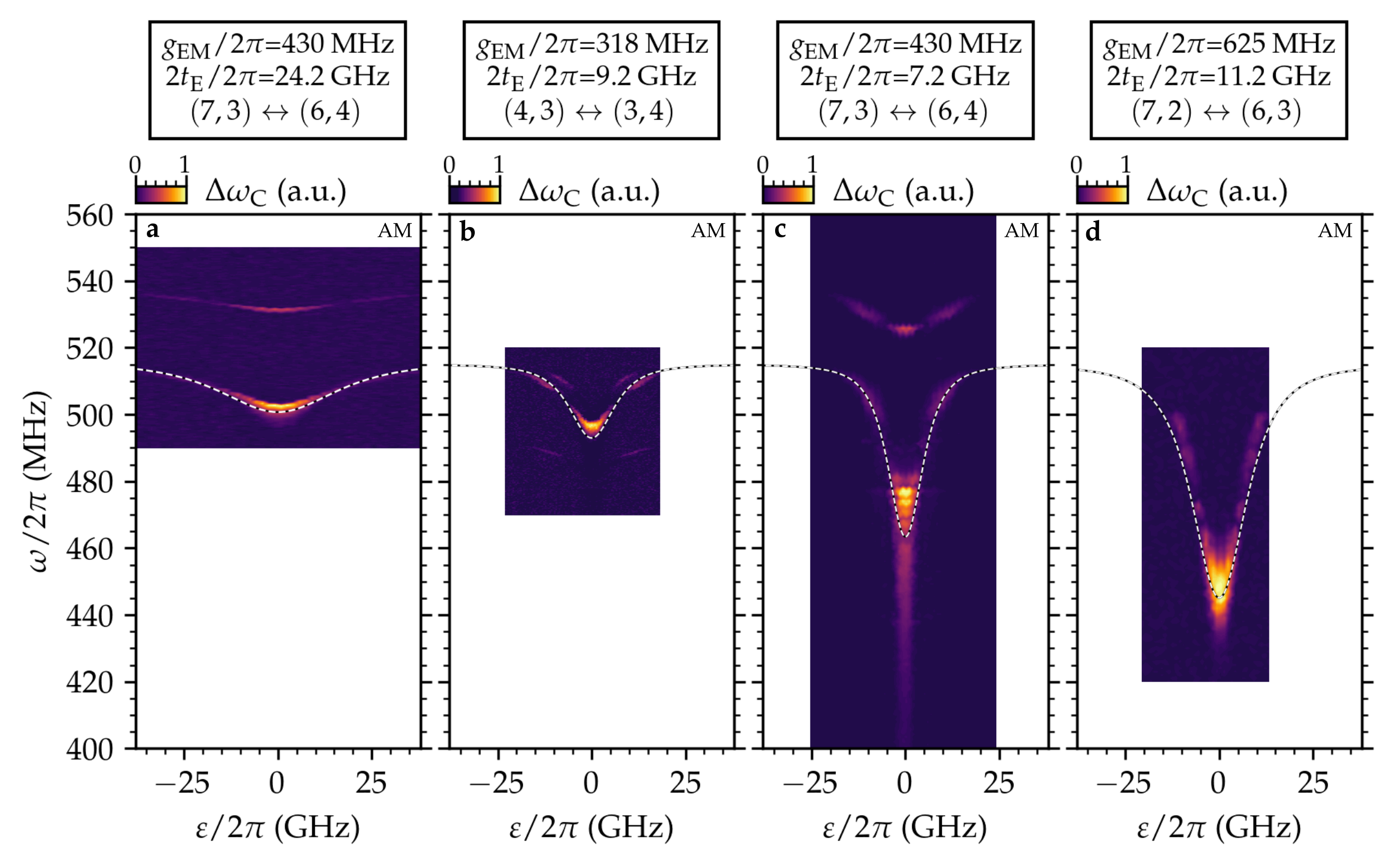}
    \customcaption{
            \textbf{Mechanical frequency softening for different double-quantum dot configurations.}
            \textbf{a--d}. Mechanical frequency softening measured with the AM technique as a function of detuning $\varepsilon$ and drive frequency $\omega$ for four different DQD ETLS frequencies. The bare mechanical frequency of the mode of interest is $\omega_\mathrm{M}^0/2\pi = \SI{0.515}{\giga\hertz}$. From left to right, the interdot charge transitions studied were $(7,3) \leftrightarrow (6,4)$, $(7,2) \leftrightarrow (6,3)$, $(4,3) \leftrightarrow (3,4)$, and $(7,3) \leftrightarrow (6,3)$. The respective barrier voltage configurations $[V_\mathrm{G1}, V_\mathrm{G3}, V_\mathrm{G5}]$ were $[-0.50,\  -0.46,\ -0.50]$V, $[-0.20,\  -0.35,\  -0.20]$V, $[-0.26,\ -0.54,\ -0.26]$V, and $[-0.50,\  -0.46,\  -1.00]$V. The corresponding ETLS frequencies at zero detuning are $2t_\mathrm{E}/2\pi = [24.2,\ 9.2,\ 7.2,\ 11.2]$GHz, and the electromechanical coupling rates are $g_\mathrm{EM}/2\pi = [430,\ 318,\ 430,\ 625]$MHz.
            White dashed lines are a manual fit to Eq.~4  of the main text.
            Data in all panels are recorded at \SI{30}{\milli\kelvin}.
            }
    \label{fig:SI_second_mode_different_configs_thesis}
\end{figure}

\newpage
Using the AM technique, we characterized the mechanical mode of interest for different ETLS configurations (Fig.~\ref{fig:SI_second_mode_different_configs_thesis}a–d). By adjusting the gate voltages we were able to modify the tunnel coupling $2t_\mathrm{E}$ as well as the DQD charge occupation, which in turn allowed us to modify the electromechanical coupling from $g_\mathrm{EM}/2\pi = \SI{0.318}{\giga\hertz}$ (Fig.~\ref{fig:SI_second_mode_different_configs_thesis}b) up to $\SI{0.625}{\giga\hertz}$ (Fig.~\ref{fig:SI_second_mode_different_configs_thesis}d).
In the ETLS configuration with $2t_\mathrm{E}/2\pi = \SI{24}{\giga\hertz}$ we observe a mechanical linewidth of $\delta \omega_M/2\pi = \SI{100}{\kilo\hertz}$ at zero detuning ($\varepsilon = 0$), narrowing to $\SI{3}{\kilo\hertz}$ at large detuning ($\varepsilon \sim 10\times 2t_\mathrm{E}$). This strong dependence of the linewidth on the ETLS frequency is consistent with the hypothesis that ETLS dephasing contributes largely to the mechanical linewidth.
 
\begin{figure}[h!]
    \centering
    \includegraphics[width=\linewidth]{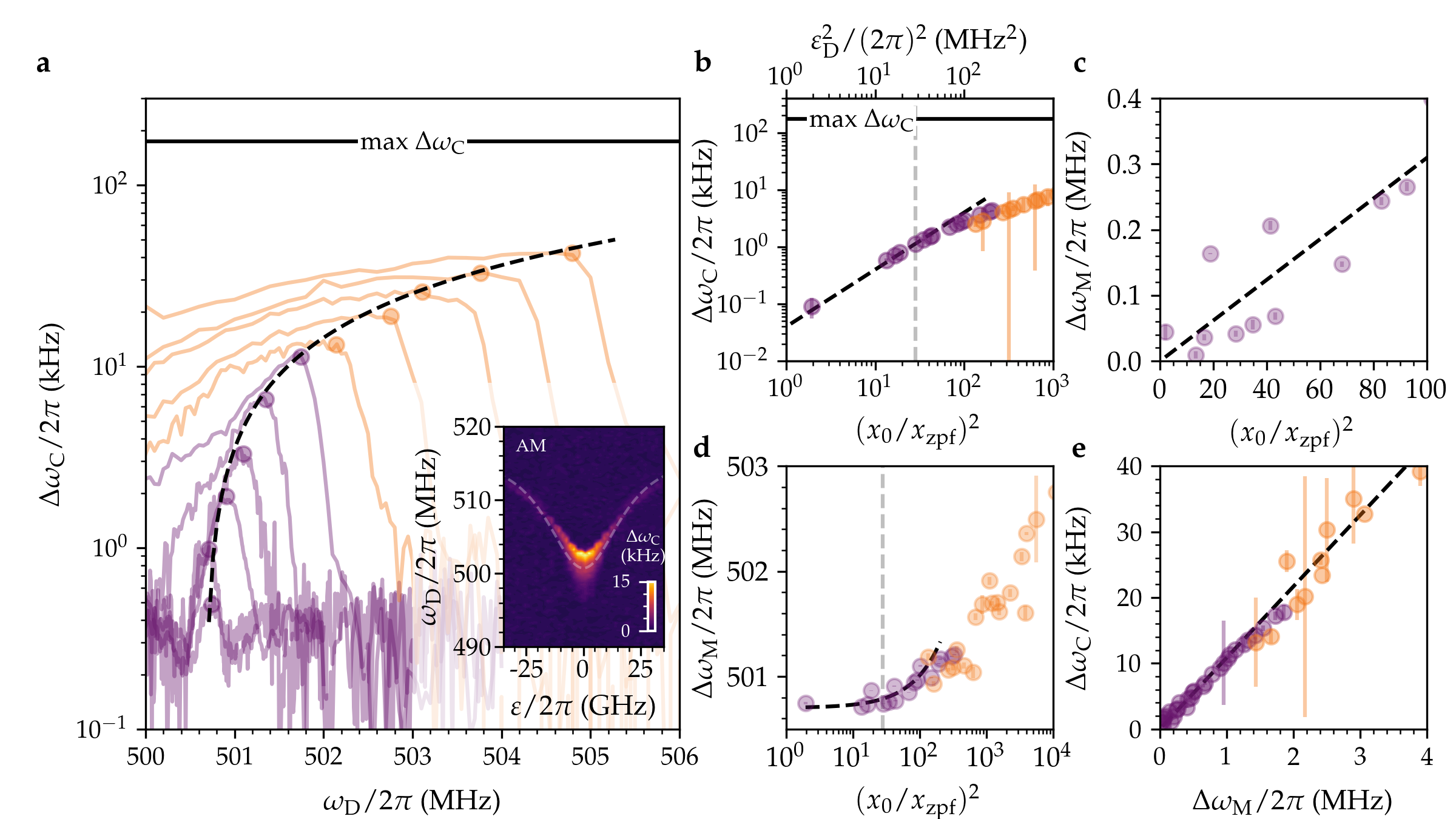}
		\customcaption{\textbf{Nonlinearity in both the mechanical motion and its readout.}        %
        \textbf{a.} Mechanical spectra for different drives at $\varepsilon=0$ measured by two-tone spectroscopy with an AM (orange) and CW (purple) homodyne detection scheme. 
        Solid horizontal line is given by the maximum cavity frequency shift $\Delta\omega_\mathrm{C}$.
        Inset shows the mechanical frequency $\omega_\mathrm{M}$ suppression as a function of detuning. Dashed white line shows theoretical prediction of the suppression without any Kerr contribution.      
        \textbf{b.} Dependence of $\Delta\omega_\mathrm{C}$ on calibrated displacement in units of the zero-point motion. The measured $\Delta\omega_\mathrm{C}$ depends quadratically on the detuning drive (dashed line) consistent with a $x^2$-readout. The scaling persists until $2g_\mathrm{EM}x/x_\mathrm{zpf}\simeq 2t_\mathrm{E}$ (gray dashed line) where the mechanical oscillation washes out the effect of the ETLS.       
        \textbf{c, d.} $x^2$-dependence of $\Delta\omega_\mathrm{M}$ at low displacements (dashed line, Eq.~2) indicating Kerr nonlinearity at the zero-point motion scale. $\Delta\omega_\mathrm{M}$ is no longer described by Eq.~2 of the main text at large drive when $2g_\mathrm{EM}x/x_\mathrm{zpf}> 2t_\mathrm{E}$.
        \textbf{e.} Dependence of $\Delta\omega_\mathrm{C}$ on $\Delta\omega_\mathrm{M}$ quantified at the peaks of the mechanical spectra in (\textbf{a}) at $\varepsilon=0$. 
        This linear dependence is also represented in (\textbf{a}) (dashed black line) with $\omega_\mathrm{M}$ reaching \SI{0.501}{GHz} at small drive where the Kerr contribution becomes zero.   
        Data in all panels are recorded at \SI{30}{mK}.
        }
    \label{fig:SI_calibrated_displacement_thesis}
\end{figure}

\newpage
Both Kerr nonlinearity and the QND-like readout of the mechanical motion were observed in this device for displacements close to $x_0/x_\mathrm{zpf} \approx 1$  (Figs.~\ref{fig:SI_calibrated_displacement_thesis}a--e). The large ETLS frequency chosen in this study compared to the device presented in the main text resulted in the observation of these two effects at slightly larger displacements. The displacement was calibrated using the method described in Section~\ref{Details on calibration of mechanical displacement}.

% ====================================================================================================

\clearpage

\SectionNoPage{Theoretical section}

\newcommand{\beq}{\begin{equation}}
\newcommand{\eeq}{\end{equation}}

\newcommand{\beqa}{\begin{eqnarray}}
\newcommand{\eeqa}{\end{eqnarray}}

\newcommand{\av}[1]{\langle #1 \rangle }

\newcommand{\refe}[1]{Eq.~(\ref{#1})}

\renewcommand{\P}{\mathrm{P}}

In this section we will present the model and the theoretical approach that is used to describe and interpret part of the experimental results. The equations used in the main text are derived in this theory section:
\begin{itemize}[noitemsep, topsep=0pt]
    \item Main text Eq. (1): SI \refe{H_EMC}
    \item Main text Eq. (2): SI \refe{eq:SI_version_of_maintext_eq2}
    \item Main text Eq. (3): SI \refe{eq:omega_c_R}
    \item Main text Eq. (4): SI \refe{eq:softening}
    \item Main text mechanical frequency renormalization with ETLS detuning: SI \refe{second_order_shift}
    \item Main text optomechanical coupling rates: SI \refe{eq:SI_definition_OM_coupling_rates}
    \item Main text Kerr term: SI \refe{eq:Kerr_term_simple}
    \item Main text anharmonicity: SI \refe{eq:anharmonicity_to_Kerr} 
    \item Main text mechanical frequency shift to Kerr-nonlinearity: SI~\refe{mech_freq_shift_to_Kerr}
    \item Main text mechanical thermal broadening: SI \refe{eq:SI_version_mechanical_thermal_broadening}
\end{itemize}

\subsection{System Hamiltonian}
We begin by introducing the Hamiltonian of the system.
The electronic part consists of a double quantum dot that forms an effective electronic two-level system (ETLS).
This ETLS can be described either in terms of the localized electronic states, corresponding to an electron occupying one of the two dots, or in terms of the bonding and antibonding states arising from the coherent delocalization of the electronic charge over both dots (see Fig. 1 in the main text).
We adopt the latter representation, so that the ETLS Hamiltonian is given by
\begin{align}
    H_\E  = t_\E  \sigma_z + \frac{\varepsilon}{2}\, \sigma_x \,,
\end{align}
where $t_\E $ denotes the tunneling amplitude between the two dots, $\varepsilon$ is the energy detuning between the left and right localized electronic states, and $\sigma_x$ and $\sigma_z$ are Pauli matrices.
For $\varepsilon = 0$, the two eigenstates of $\sigma_z$ represent the bonding and antibonding delocalized electronic states.
A periodic drive can be applied to the gate electrodes, resulting primarily in a time dependence of the detuning parameter $\varepsilon$:
\begin{align}
    \varepsilon(t) = \varepsilon_0 + \varepsilon_\D \cos(\omega_\D t) \,,
\end{align}
where $\varepsilon_0$ is the static detuning set by the gate voltage, $\varepsilon_\D$ is the driving amplitude, and $\omega_\D$ is the driving frequency.

The ETLS is coupled electrostatically to the flexural modes of the suspended carbon nanotube.
We assume that these modes can be described as harmonic under-damped oscillators.
By symmetry the double quantum dot couples strongly to the second flexural mode. 
We then single out this mode from the others as the main mode and we define its resonating frequency as $\omega_\M$. 
In Sect.~\ref{SectionBroadening} we will consider the effect of the other modes on the time evolution of the main one. 
In the remainder of the section we will neglect their contribution. 
The free Hamiltonian of the main mode is then given by 
\beq
    H_\M/\hbar = \omega_\M a^\dag a
\eeq
where $a$ is the annihilation operator for the main mode and it is coupled to the ETLS by 
\begin{align}
     V_\EM/\hbar = g_\EM  ( a^\dag + a) \sx \,.
     \label{VEM}
\end{align}
The central system of interest thus consists of the main mode $\omega_\M$ coupled to the ETLS. As a read-out device, a driven superconducting microwave cavity is additionally weakly coupled to the ETLS. 
We hence have the full coupled systems as 
\begin{align}
    H &= H_\E  + H_\M + V_\EM + H_\C + V_\EC ,
    \label{H_EMC}
\end{align}
where we defined
\begin{align}
    H_\C/\hbar &= \omega_\C b^\dag b + \Omega_\C ( b^\dag + b) \cos( \omega_\P t ),
    \\
    V_\EC/\hbar 
    &= g_\EC ( b^\dag + b ) \sx . 
\end{align}
Here, $b$ is the annihilation operator for photons in the cavity, $\omega_\C$ is the cavity frequency, $\Omega_\C$ denotes the cavity probe driving amplitude, $\omega_\P \simeq \omega_\C$ is the probe drive frequency, and $g_\EC \ll \omega_\C$ is the coupling constant between the cavity and the ETLS.  
Together with the previous terms, this defines the full Hamiltonian of the system, including the external drives. 
In the following, we analyze the different contributions to this Hamiltonian in order to derive the relevant physical quantities.

\subsection{Diagonalization of the ETLS–Mechanical mode Hamiltonian}\label{digonalization_hamiltonian}

The central system of interest consists of the mechanical mode with the largest coupling strength ($g_\EM$) to the double quantum dot forming the ETLS. 
In this section we analyze it in the quantum regime as a closed system, before including the drive and the coupling to both the environment and the microwave cavity.
Focusing on the undriven central system only, the Hamiltonian is simply given by $H=H_\E +H_\M+V_\EM$, or explicitly:
\beqa
    H/\hbar &=& {t_\E } \sigma_z+ 
         {\varepsilon_0\over 2}\sigma_x
         + \omega_\M a^\dag a 
        +   g_\EM (a+a^\dag)
          \sigma_x  \,.
        \label{eq:H_ETLS_mech}
\eeqa 

It can be diagonalized approximately up to fourth order in $g_\EM/t_\E  \ll 1$ via standard perturbation theory as 
\begin{align}
    H/\hbar 
    &= 
    \left( \frac{\omega_\E}{2} + \delta \omega_\E \right) \sigma_z 
    + ( \omega_\M + \delta \omega^{(1)} +   \delta \omega^{(2)} \sigma_z) a^\dag a 
    - {K \over 2} \sigma_z a^\dag a^\dag a a 
    \,,
    \label{H_4th_order}
\end{align}
where we omitted constant energy shifts, and the operators $\sigma_z$, $a^{(\dag)}$ are new operators that act on the approximated eigen-basis $\{ \ket{\Psi_{n,\sigma}} \}_{n \in \mathbb{N},\sigma = \pm 1}$, such that $\sz \ket{\Psi_{n,\sigma}} = \sigma \ket{\Psi_{n,\sigma}}$ and $a \ket{\Psi_{n,\sigma}} = \sqrt{n} \ket{\Psi_{n-1,\sigma}}$ and analogously for $a^\dag$. 
The approximated eigenstates $\ket{\Psi_{n\sigma}}$ refer to a set of entangled basis states that are linear combinations of the physical, uncoupled eigenbasis $\ket{n \sigma} = \ket{n}\ket{\sigma}$ such that $\ket{\Psi_{n\sigma}} \rightarrow \ket{n \sigma}$ for $g_\EM \rightarrow 0$. 
And the uncoupled eigenbasis is defined by $(H_\E + H_\M )\ket{n \sigma} = ( \omega_\E\sigma /2 + n \omega_\M) \ket{n \sigma}$. 

The energy terms in Eq.~\eqref{H_4th_order} are given explicitly by 
\begin{align}
    \omega_\E & = \sqrt{ (2 t_\E )^2 + \varepsilon_0^2}
    \\
    \delta \omega_\E 
    & = 
    \frac{(2t_\E )^2 g_\EM^2}{\omega_\E (\omega_\E^2-\omega _\M^2) } 
    -
    \frac{(2t_\E )^2 g_\EM^4 }{\omega_\E^5 \omega_\M^2 \left(\omega_\E^2 - \omega _\M^2\right){}^3}
    \big(-4 \varepsilon _0^2 \omega_\E^6+\omega_\E^2 \left((2t_\E )^2-12 \varepsilon _0^2\right) \omega _\M^4
    \nonumber
    \\ & \qquad \qquad\qquad\qquad \qquad\qquad\qquad \qquad\qquad
    +3 \omega_\E^4 \left(4 \varepsilon _0^2+(2t_\E )^2\right) \omega _\M^2+4 \varepsilon _0^2 \omega _\M^6\big)
    \,,
    \\
    \delta \omega^{(1)}
    &= 
    -
    \frac{2 g_\EM^4 (2t_\E )^4 \left(3 \omega _\M^2+\omega_\E ^2\right)}{\omega_\E ^2 \omega _\M \left(\omega_\E ^2-\omega _\M^2\right){}^3} 
    \,,
    \\
    \delta \omega^{(2)}
    &=
    \frac{2 g_\EM^2 (2t_\E )^2}{\omega_\E ^3-\omega_\E  \omega _\M^2}
    - 
    \frac{4 g_\EM^4 (2t_\E )^4   \left(\omega_\M^2+3 \omega_\E ^2\right)}{\left(\omega_\E ^3-\omega_\M^2 \omega_\E \right)^3} 
    \,,
    \\
    K
    &=  
    \frac{4 g_\EM^4 (2t_\E )^4   \left(\omega_\M^2+3 \omega_\E ^2\right)}{\left(\omega_\E ^3-\omega_\M^2 \omega_\E \right)^3}
    \,,
\end{align}
where $\omega_\E$ is the bare eigenenergy of the ETLS and $K$ is the ETLS-dependent Kerr-term of the mechanical oscillator. 
For $\omega_\M \ll \omega_\E$ and at zero detuning $\varepsilon_0 =0$, we can approximate it by the much more compact form 
\begin{align}
    K \approx \frac{12 g_\EM^4}{(2t_\E)^3}
    \,.
    \label{eq:Kerr_term_simple}
\end{align}

It is very important to investigate the anharmonicity of the mechanical mode induced by the coupling to the ETLS.
For the energy differences of the lower ETLS branch $ \omega_{n+1,n} = (E_{n+1 , -} - E_{n , -})/\hbar$, where $E_{n,\sigma}$ is the eigenenergy of eigenstate $\ket{\Psi_{n,\sigma}}$,
we find 
\begin{align}
     \omega_{n+1,n}
    &= \tilde  \omega_\M
    + n K 
    \,,
    \label{anharm_Delta_omega_M}
\end{align}
which consists of the renormalized frequency
$\tilde \omega_\M =  \omega_\M + \delta \omega^{(1)} - \delta \omega^{(2)}$ and the anharmonic frequency shift induced by the Kerr-term. 
The renormalized frequency $\tilde \omega_\M$ is well approximated by its leading order:
\begin{align}
    \tilde \omega_\M
    &=
    \omega_\M \left[ 1 
     -  {g_\EM^2 (2t_\E)^2 \over \omega_\M  \omega_\E^2 }  \left( 
    \frac{1}{\omega_\E -\omega_\M } + \frac{1}{\omega_\E + \omega_\M }
    \right) \right] \,,
    \label{second_order_shift}
\end{align}
where we rearranged the terms to explicitly display the included counter-rotating term. 
At higher orders of $g_\EM$, the Kerr-term introduces anharmonicity into the mechanical modes, such that $\omega_{n+1,n} \neq \omega_{m+1,m}$ for $n \neq m$. 
One way to quantify this effect is to define the anharmonicity as
\begin{align}\label{eq:anharmonicity_to_Kerr}
    \alpha &= \frac{ \omega_{2,1} - \omega_{1,0}}{\omega_{1,0}}  
    \\
    &= \frac{K}{\tilde \omega_\M}
    \,.
\end{align}  

\begin{figure}
    \centering
    \includegraphics[width=0.95\linewidth]{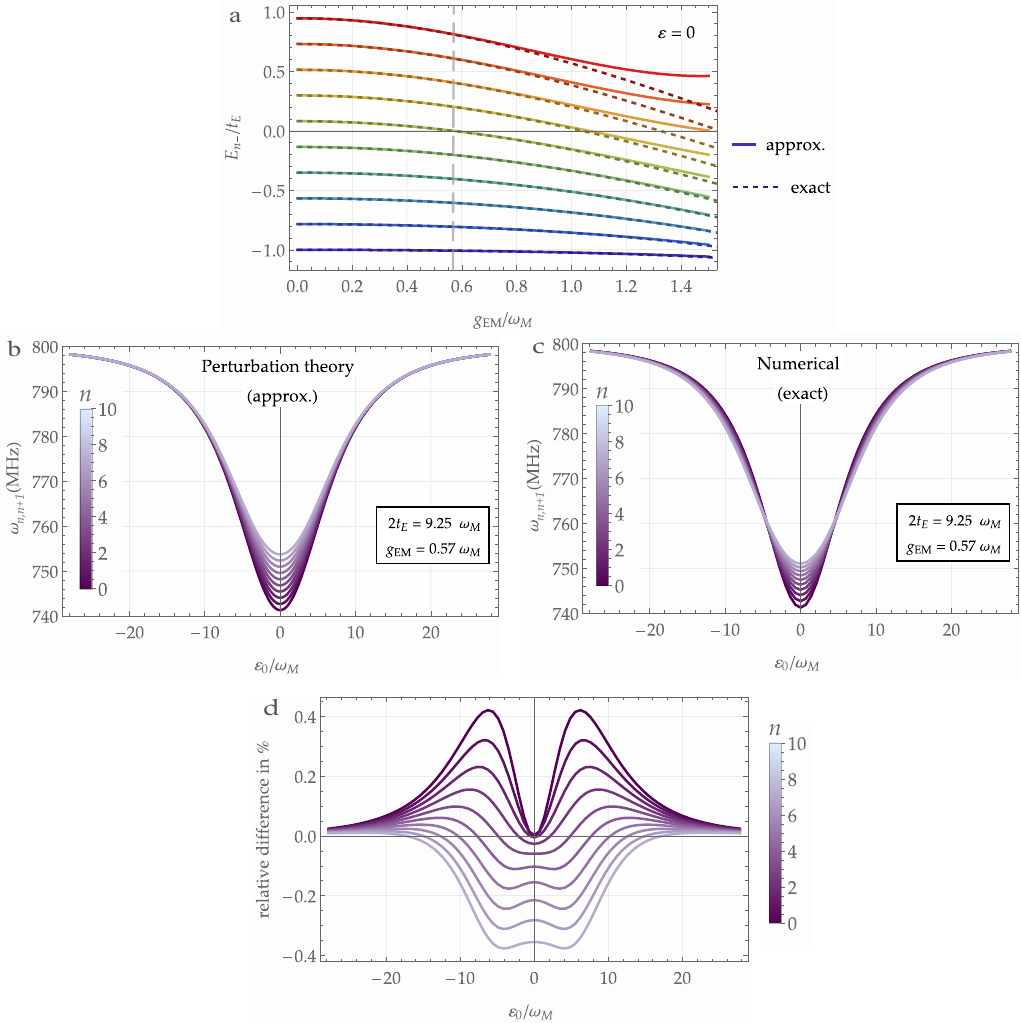}
    \customcaption{a) Analytically approximated and numerically exact eigenenergies of the Hamiltonian~\eqref{H_ETLS_flex} with $2t_\E/2\pi = 7.4$~GHz, $\omega_\M/2\pi = 800$~MHz and $\varepsilon = 0$ for different values of $g_\EM$ for the lowest 10 energylevels. The gray dashed line indicates the coupling constant of the experiment. 
    The differences between these energies give the transition frequency $\Delta \omega$~\eqref{anharm_Delta_omega_M}, which are shown below for $g_\EM/2\pi = 456$~MHz as a function of the detuning $\varepsilon$ in b) from fourth-order standard perturbation theory and in c) from the numerical diagonalisation. In both cases we find a frequency softening which is maximal for $\varepsilon =0$ and that the transition frequency between adjacent levels depend on the level itself.
    In d) we show the relative difference between the numerical and approximated frequency softening defined by $(\Delta \omega_\mathrm{num.} - \Delta \omega_\mathrm{approx.})/ \Delta \omega_\mathrm{num.}$. The relative difference increases 
    up to 0.4\% for $\varepsilon_0 \sim t_\E$ and increases as well (but with opposite sign) with increasing $n$.}
    \label{fig:frequ_soft_full_qu}
\end{figure}

As demonstrated in  Fig.~\ref{fig:frequ_soft_full_qu},  the applied perturbative expansion used to approximately diagonalize the Hamiltonian  holds best for $ 2t_\E  \gg \varepsilon_0$ and $\varepsilon_0 \gg  2t_\E $.
In the figure, the numerically
exact solution obtained from diagonalizing the full Hamiltonian numerically is compared to the analytical approximation for the first few phonon transitions. As one can see from the numerical diagonalization, for $\varepsilon_0 = t_\E$, the anharmonicity is lifted, such that $\omega_{n+1,n} = \omega_{m+1,m}$ and varying the detuning from $\varepsilon_0 < t_\E$ to $\varepsilon_0 > t_\E$, changes the sign of the anharmonicity. This result will be recovered analytically in the semi-classical limit.

At maximum frequency softening, i.e.~$\varepsilon_0 =0$ the energy differences of the lower branch yield,
\begin{align}
    \omega_{n+1,n} 
    = 
    \tilde \omega_\M  
   +
    n g_\EM^4
    \frac{4 (2t_\E) \left(3 (2t_\E)^2+\omega _\M^2\right)}{\left((2t_\E)^2-\omega _\M^2\right){}^3}
    \,,
\end{align}
where $\tilde \omega_\M$ is the renormalized frequency of the oscillator, given in leading order at maximum frequency softening as
\begin{align}
        \tilde \omega_\M
    &=
    \omega_\M \left[ 1 
     -  {g_\EM^2  \over \omega_\M   }  \left( 
    \frac{1}{2t_\E -\omega_\M } + \frac{1}{2t_\E + \omega_\M }
    \right) \right]
    \,.
     \label{eq:softening}
\end{align}

\begin{figure}[h!]
    \centering
    \includegraphics[width=0.8\linewidth]{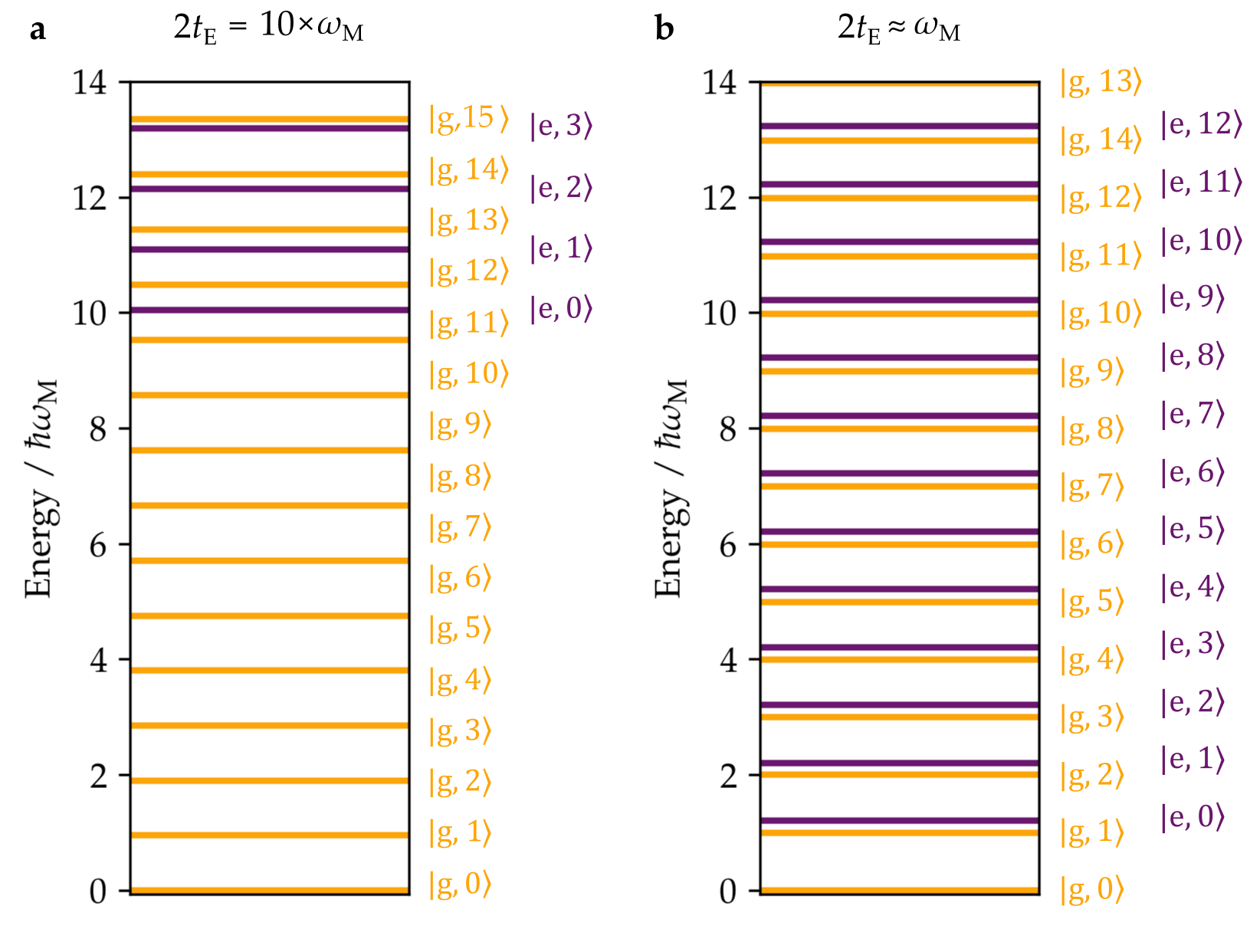}
    \customcaption{
            \textbf{Simulated energy states.}
            \textbf{a,b.} Energy spectra obtained by numerical diagonalization of the Hamiltonian in Eq.~\ref{eq:H_ETLS_mech} using the Python-based quantum toolbox QuTiP \cite{Qutip2013}. 
            We consider two limits: (\textbf{a}) the ultrastrong coupling regime closely matching our experimental parameters with $g_\mathrm{EM} = \omega_\mathrm{M}/2$ and $2t_\mathrm{E} = 10\,\omega_\mathrm{M}$, satisfying  
            the far-detuned dispersive condition $\lvert 2t_\mathrm{E}- \omega_\mathrm{M} \rvert \gg g_\mathrm{EM}$; and (\textbf{b}) the strong coupling regime with $g_\mathrm{EM} = \omega_\mathrm{M}/50$ and $2t_\mathrm{E} = 1.2\omega_\mathrm{M}$, satisfying  
            the dispersive condition $\lvert 2t_\mathrm{E}- \omega_\mathrm{M} \rvert > g_\mathrm{EM}$. 
            Energy levels are labeled according to the uncoupled dressed states $\ket{\sigma,n}$, where $\sigma$ denotes the ETLS ground ($\ket{g}$, orange) or excited ($\ket{e}$, purple) state, and 
            $n$ denotes the phonon-number state.
            }
\label{fig:SI_simulation_energy_levels}
\end{figure}

The lowest-energy states can be kept predominantly mechanical with only weak electrical admixture by setting the ETLS frequency about an order of magnitude larger the mechanical frequency. This ensures that the low-lying eigenstates remain in the ETLS ground-state branch and are well separated from the excited branch (Fig.~\ref{fig:SI_simulation_energy_levels}a). 
This behavior is enabled by the ultrastrong-coupling regime ($g_\mathrm{EM}/\omega_\mathrm{M} > 0.1$), which allows to detune the ETLS frequency from the mechanical oscillator frequency by a large amount ($2t_\E \gg \omega_\mathrm{M}$). Hybridization between the ETLS ground- and excited-state energy branches can only emerge for large mechanical displacements when $n_\M\times \omega_\M \approx 2t_\E$, which in our experiment occurs above $n_\M = 10$. 

By contrast, when operating the hybrid system in the strong coupling regime and in the dispersive limit, the energy spectrum consists of densely packed mechanical and electrical states. This is because the detuning $|2t_\mathrm{E} - \omega_\mathrm{M}|$ has to remain small to keep the effect of the coupling sizable, including the mechanical anharmonicity.   

\subsection{Coupling to the environment}\label{Coupling to the environment}

 The mechanical oscillator, the cavity, and the ETLS are coupled each to specific environments, which we assume to be independent.
 The mechanical oscillator features spectra with a quality factor $Q_\M>10^5$ when the effect of the coupling to the ETLS is suppressed, which indicates that the thermal bath of the mechanical oscillator leads to a weak decay rate $\gamma\ll\omega_\M$, and frequency noise is low. As for the cavity, the spectra feature narrow linewidth as well, pointing to weak cavity decay rate $\kappa \ll \omega_\mathrm{C}$. 
 
 For the ETLS, we expect two main effects of the environment, the decay rate $\Gamma$ from the excited state to the ground state of the ETLS and the pure dephasing rate $\Gamma_\phi$, which does not imply energy dissipation.
These measurable effects of the environment can be described theoretically in a standard way by deriving a master equation for the system coupled to the environmental degrees of freedom, and then obtaining an equation for the (reduced) density matrix $\rho$ of the system after tracing out the environment. 
In our case this gives the equation of motion for $\rho$
\beq
    \dot \rho = -i[H,\rho]+{\cal L_D} \rho ,
    \label{MasterEquation}
\eeq
with the dissipator reading:
\beq
    {\cal L}_{\cal D} 
    = 
    \gamma {\cal L}_o(a,\omega_\M)
    +
    \kappa {\cal L}_o(b,\omega_\mathrm{C})
    +
    \Gamma {\cal L}_o(\sigma_-,(2t_\E))
    + {\Gamma_\phi\over 2} {\cal D}(\sigma_z) ,
    \label{ElleD}
\eeq
where we defined $\sigma_\pm = (\sigma_x\pm i \sigma_y)/2$,
\beq
    {\cal L}_o(A,\omega) = (n_B(\omega)+1){\cal D}(A) + n_B(\omega){\cal D}(A^\dag)
    ,
\eeq
with ${\cal D}(A) \rho = A\rho A^\dag-(A^\dag A \rho+\rho A^\dag A)/2$ and
$n_B(\omega)=1/(e^{\hbar \omega/k_B T}-1)$ the Bose occupation, with $T$ the temperature and $k_B$ the Boltzmann constant.

Some care is needed to define the dissipator for the ETLS. 
The expression is valid when $\varepsilon_0\ll (2t_\E)$ so that the ground state of the ETLS is the fictitious spin down. 
When $\varepsilon_0 \approx (2t_\E)$ or the amplitude of the oscillation is large, the dissipator should be modified taking into account that the ground state is a combination of the two ETLS states. 
In the quantum case, this can be taken into account if the damping $\Gamma$ is small with respect to the energy level splitting, by finding the dissipator in the fully interacting basis. 
Since we are mainly interested in the semiclassical approach for the purpose of describing the present experiment, we will discuss an alternative method that is more adapted in the case where the damping $\Gamma$  is large or comparable to the coupling constants, or when the driving amplitude is large. 

\subsubsection{Semiclassical decoupling}\label{Semiclassical decouplingf}

When the dissipation or decoherence rate of the electronic two-level system (or of the oscillator) is larger than the interaction between the two systems,
the oscillator state becomes disentangled from the ETLS and closely resembles a coherent state for a time evolution longer than the inverse of the larger dissipation rate. 
In this case, the quantum averages of the product operators can be factorized to a good approximation, e.g.~$\av{(a+a^\dag) \sigma_x} \approx \av{(a+a^\dag)} \av{\sigma_x}$.
This approximation allows us to write equations of motions for the averages of the operators that coincide with the classical equations of motion for the oscillators, {\em i.e.}~one can treat
the oscillator quantities classically directly in the Hamiltonian. 
Then we can proceed to diagonalize the ETLS part of the Hamiltonian as a function
of $x(t)=x_z(a+a^\dag)$, regarded as a classical variable dependent on time (here $x_z=\sqrt{\hbar/2 m \omega_\M}$ and $m$ is the effective mass of the main mode. 
This gives 
\beqa
    H'/\hbar&=& [U^\dag(t) H U(t) -iU^\dag \dot U ]/\hbar
    \nonumber \\
    &=& 
    {\tilde{\omega}_\E(t) \over 2} \sigma_z 
    + \dot \theta(t) \sigma_y 
    + \left[g_{x}(t)  \sigma_x + g_{z}(t) \sigma_z\right]  (b^\dag +b)
    \nonumber \\
    &&
    + \omega_\M a^\dag a + \omega_\mathrm{C} b^\dag b
    + \Omega_\C \cos(\omega_\D t) (b+b^\dag)
    \label{Hprime}
\eeqa
where we defined $g_{x}(t)=g_\EC (2t_\E)/\tilde{\omega}_\E(t)$, $g_{z}(t)=-g_\EC \varepsilon(t)/\tilde{\omega}_\E(t)$,
\beq
    \tilde{\omega}_\E(t)=\sqrt{(2t_\E)^2+\varepsilon^2(t)},
    \label{OmegaEq}
\eeq
with $\varepsilon(t)=\varepsilon_0+\varepsilon_\D \cos(\omega_\D t)+
2 g_\EM x(t)/x_z$, and $\tan \theta(t)=-\varepsilon(t)/(2t_\E)$.
We note the term $\dot\theta \approx \omega_\D \approx \omega_\M$ that leads to a
transverse term in the Hamiltonian. 
When $\omega_\M \ll (2t_\E)$, as is the case in the experiment, this term can be neglected, at least at lowest order in $\omega_\M/(2t_\E)$. 
If taken into account it leads to additional dissipative terms for the mechanical oscillator.

\subsubsection{Dissipator for the ETLS}

We can now discuss the dissipator for the ETLS. 
The basis introduced for the definiton of $H'$, Eq.~\eqref{Hprime} is particularly adapted to take into account the interaction with the environment of the ETLS. 
Let us assume that the time dependence is sufficiently slow that we can regard $x(t)$ as a static variable for the interaction with the environment. 
This assumes that the correlation functions of the environment have a range in time shorter than the period of the oscillator, which is a reasonable assumption.
Neglecting the small terms $\dot \theta$ and the weak coupling to the cavity, the ETLS Hamiltonian has the diagonal form $ \hbar \tilde{\omega}_\E \sigma_z/2$. 
The ETLS dissipator reads then in {\em this basis}:
\beq
    {\cal L}^\mathrm{ETLS}_{\cal D} 
    =
        \Gamma {\cal L}_o(\sigma_-,\tilde{\omega}_\E(t))
    +{ \Gamma_\phi\over 2}  {\cal D}(\sigma_z) 
    \label{DETLS}
    .
\eeq
Even if the form is similar to that of \refe{ElleD}, it implies a different meaning, since it is written in the diagonal basis of the ETLS Hamiltonian, and it implies a decay into the true ground state of the ETLS.
This is thus the correct term to insert in the Master equation (\ref{MasterEquation}). 
In principle, the coefficients
$\Gamma$ and $\Gamma_\phi$ could also have a time dependence, but since we do not have direct experimental access to this behavior, for simplicity we assume that $\Gamma$ and $\Gamma_\phi$ are time independent.

\subsection{Equations of motions}\label{Equations of motions}
In this subsection we briefly introduce the equations of motions of the system fundamental observables.
Using the Master \refe{MasterEquation} with the Hamiltonian \refe{Hprime} and the dissipator given by \refe{ElleD} with \refe{DETLS} for the ETLS, 
one can derive equations of motions for the averages of any operator $A$:
\beq
{d\over dt} \langle A \rangle=
    {d\over dt} {\rm Tr}\{ \rho(t) A\} 
    = i {\rm Tr}\{ \rho(t) [H,A] \}  + {\rm Tr}\{ A { \cal L_D} \rho(t) \}
    .
\eeq
As explained above in the Section~\ref{Coupling to the environment}, the equations of motion for any mechanical mode is:
\beqa
    \dot x 
    &=&  p/m - \gamma x/2  ,
    \label{x_dot}
    \\
     \dot p 
    &=& -m \omega_\M^2  x 
    - \gamma  p /2   
    - {\hbar \over 2}
    { \partial \tilde{\omega}_\E  \over \partial x } \av{\sigma_z}\,,
    \label{p_dot}
\eeqa 
where $\tilde{\omega}_\E$ is given by Eq.~\eqref{OmegaEq}.
The equations of motion for the ETLS degrees of freedom read:
\beqa 
    \av{ \dot\sigma_x} 
    &=& -\tilde{\omega}_\E \av{\sigma_y}-\Gamma_t \av{\sigma_x} 
    - 2g_{z}\av{(b+b^\dag) \sigma_y}
    ,
    \\    
    \av{ \dot\sigma_y} 
    &=& 
    \tilde{\omega}_\E\av{\sigma_x} -\Gamma_t \av{\sigma_y}
    +2g_{z} \av{ (b+b^\dag) \sigma_x}
    -2g_{x}\av{(b+b^\dag)\sigma_z}   
    ,
    \\
    \av{ \dot\sigma_z} &=& 
    -2i g_{x}\av{ (b+b^\dag) \sigma_y}
   -\Gamma(2n_B(\tilde{\omega}_\E)+1) \av{\sigma_z}-\Gamma ,
 \eeqa
 with $\Gamma_t=\Gamma_\phi+\Gamma(2n_B(\tilde{\omega}_\E)+1)$,
and for the cavity fields operator $b$:
\beq
    \langle \dot b \rangle 
    = 
    (-i\omega_\mathrm{C} -\kappa/2 ) \langle b \rangle
    -i\left[g_{x}  \langle \sigma_x \rangle 
    + g_{z} \langle \sigma_z \rangle \right] 
    + i \Omega_\C \cos(\omega_\P t) 
    \label{b_dot}
\eeq
where $\kappa  = \kappa_\mathrm{R} + \kappa_\mathrm{D} + \kappa_\mathrm{I}$ is the total loss coefficient of the cavity, and with the Hermitian conjugate of the equation for $\av{\dot b^\dag}$.

\subsection{Cavity response to a drive}\label{Cavity response to the drive}

When the drive of the microwave cavity is sufficiently large the electromagnetic field is close to a coherent state, allowing the mean-field decoupling 
$\av{(b+b^\dag) \sigma_i} \approx \av{b+b^\dag} \av{\sigma_i}$ (see also Ref.~\cite{Ranjan2015} for instance).
For convenience, we write the equations of motion for $\av{\sigma_-}$ (with its conjugate for $\av{\sigma_+}$) instead of $\av{\sigma_x}$ and $\av{\sigma_y}$ 
in the following.
This gives:
\beqa
    \av{\dot \sigma_-}
    &=& 
    (-i\tilde{\omega}_\E -\Gamma_t)\av{\sigma_-}+ig_{x} \av{b+b^\dag} \av{\sigma_z}
    -2ig_{z} \av{b+b^\dag} \av{\sigma_-} ,
    \\
    \av{\dot \sigma_z}
    &=& 
    -2i g_{x} \av{b+b^\dag} \av{\sigma_+-\sigma_-}-\Gamma(2n_B(\tilde{\omega}_\E)+1)\av{\sigma_z}-\Gamma
    \,,
\eeqa
where $\tilde{\omega}_\E = \tilde{\omega}_\E(x,t)$ is given by \eqref{OmegaEq}.
Since the cavity probe frequency, $\omega_\P$, is much smaller than the ETLS resonance frequency ($\omega_\P \approx \omega_\C \ll \omega_\E)$) we account for the ETLS counter-rotating terms in the Hamiltonian.
For this reason instead of using the rotating frame we write the averages of the 
operators as  
\beq
    \av{\sigma_i}(t) = \sum_n e^{-in\omega_\P t} \Sigma_n^{i}(t),
\eeq
where we assume that a slow time dependence is left in the Fourier coefficients
$\Sigma_n^{i}(t) $. This allows to take into account the slow time dependence of $x(t)$ on the scale of $\omega_\M$ and the time evolution on the scale of the dissipative coefficients. 
Note also that 
$(\Sigma_{-n}^-)^*=\Sigma_n^+$.

On the contrary, since $\kappa \ll \omega_\mathrm{C}$ the cavity responds only in a very narrow range of frequencies, we can then keep only the first appropriate harmonic  for the $b$ fields,
\beq
\langle b\rangle (t) = e^{- i \omega_\P t} B_1(t)\,.
\eeq
With these assumptions we obtain:
\beqa    
    \dot \Sigma_n^--
    (-i \tilde{\omega}_\E-\Gamma_t+in\omega_\P) \Sigma_n^- 
    &=&
    + i g_{x} (\Sigma_{n-1}^z B_1 - \Sigma_{n+1}^z B_1^*)
    \nonumber \\
    &&-2ig_{z} (\Sigma_{n-1}^-B_1+\Sigma_{n+1}^- B_1^*)
    ,
    \\
    \dot \Sigma_n^z+
    [-in\omega_\P+(2n_B(\tilde{\omega}_\E)+1)\Gamma ]\Sigma_n^z
    &=&
    -2i g_{x} (\Sigma_{n-1}^+ B_1 - \Sigma_{n+1}^- B_1^*
    \nonumber \\
    &&
    +\Sigma_{n+1}^+B_1^*-\Sigma_{n-1}^- B_1)
    -\Gamma \delta_{n,0}
    .
\eeqa
Solving for the quasi-stationary state we have:
\beqa 
    \Sigma_n^- 
    &=&
    { i g_x (\Sigma_{n-1}^z B_1 - \Sigma_{n+1}^z B_1^*)
    -2ig_z(\Sigma_{n-1}^-B_1+\Sigma_{n+1}^- B_1^*)
    \over i \tilde{\omega}_\E+\Gamma_t-in\omega_\P}
    \label{SigmaPM}
\\
    \Sigma_n^z
    &=&
    {-\Gamma \delta_{n,0}
    -2i g_x (\Sigma_{n-1}^+ B_1 - \Sigma_{n+1}^- B_1^*
    +\Sigma_{n+1}^+B_1^*-\Sigma_{n-1}^- B_1)
    \over 
    -in\omega_\P+(2n_t+1)\Gamma }
    ,
    \label{SigmaZ}
\eeqa
with $B_1$ given by 
\beq
    B_1
    = {-ig_x(\Sigma_1^++\Sigma_1^-)
    -ig_z \Sigma_1^z 
    + i \Omega_\C/2
    \over 
    i\omega_\mathrm{C}-i\omega_\P+\kappa/2 }
    \label{B1eq}
    .
\eeq

From \refe{SigmaZ} one sees that if $|B_1| g_{x} \approx |B_1| g_{c} \ll \Gamma$ the first term in the numerator dominates the expression for $n=0$. 
In lowest order, one then finds:
\beq
    \Sigma_0^z
    =-{1
    \over 
    2n_B(\tilde{\omega}_\E)+1}
    \equiv \langle \sigma_z \rangle_0  
\label{thermalSigmaz}
    ,
\eeq
which is just the thermal average of the $\sigma_z$ operator. 

Physically the condition $|B_1| g_{c} \ll \Gamma$ sets a limit on the driving intensity [$|B_1|\approx \Omega_\C/\kappa$ from \refe{B1eq}] 
that guarantees that the ETLS remains in the thermal state. 
Calling $\epsilon=|B_1| g_{c}/D$, with $D=\sqrt{\tilde{\omega}_\E^2+\Gamma_t^2}$, one can 
solve the set of equations as an expansion in $\epsilon$ for $\epsilon \ll 1$.
By inspection of \refe{SigmaPM} and \refe{SigmaZ} one finds that a hierarchy sets in with $\Sigma^z_{n}\sim \epsilon^{|n|}$ for $|n|>1$, $\Sigma^z_{\pm 1}\sim \epsilon^3$, $\Sigma^z_{0}\approx 1$ as given by \refe{thermalSigmaz},
$\Sigma^\pm_{n}\sim \epsilon^{|n|}$ for $|n|>0$, and $\Sigma^\pm_{0}\sim \epsilon$.

Keeping only the leading orders we then obtain:
\beq    
    \Sigma_1^- 
    =
    {i g_x \langle \sigma_z \rangle_0 B_1
    \over i \tilde{\omega}_\E+\Gamma_t-i\omega_\P}
    \quad,\qquad 
     \Sigma_1^+ = (\Sigma_{-1}^-)^* 
    =
    {i g_x \langle \sigma_z \rangle_0 B_1
    \over -i \tilde{\omega}_\E+\Gamma_t-i\omega_\P}   
    \label{Sigma1St}
    .
\eeq
\beq
B_1
    =  \frac{ i \Omega_\C/2 }{   i\omega_\mathrm{C}-i\omega_\P+\kappa /2 + \rchi_\EC }
\eeq
where 
\beq
    \rchi_\EC = -g_x^2 \langle \sigma_z \rangle_0 \left[
    {1\over \Gamma_t+i \tilde{\omega}_\E-i\omega_\P}
    - 
    {1\over \Gamma_t-i \tilde{\omega}_\E-i\omega_\P}
    \right] .
    \label{Delta_cavity}
\eeq
From this expression it is then clear that the cavity acquires a renormalization of the resonance frequency $\Delta \omega_\mathrm{C}={\rm Im}(\rchi_\EC)$ and of the damping $\Delta \kappa={\rm Re}(\rchi_\EC)$.
We thus find
\beq
    \Delta \omega_\mathrm{C} =
    g_\EC^2 \langle \sigma_z \rangle_0  \left(\frac{(2t_\E)}{\tilde{\omega}_\E}\right)^2\left[
    {\tilde{\omega}_\E-\omega_\P\over \Gamma_t^2+(\tilde{\omega}_\E-\omega_\P)^2}
    +{\tilde{\omega}_\E +\omega_\P \over \Gamma_t^2+(\tilde{\omega}_\E+\omega_\P)^2} 
    \right]
    ,
    \label{Eq47}
\eeq
where we've written out $g_x^2$ explicitedly. We recognize the first term as the standard correction in the rotating wave approximation (see Ref.~\cite{Ranjan2015}). 
It dominates the second if $|\tilde{\omega}_\E-\omega_\P|\ll \Gamma_t$, which is not necessarily the case in the experiment.
This expression depends on $x(t)$ and $\varepsilon_\D$ implicitly through $\tilde{\omega}_\E(t)$.
It will be used to measure the amplitude of oscillation of the mechanical oscillator as discussed in the next section.

From input-output theory we can derive the transmission scattering parameter for a cavity with two different ports for probe and read-out. 
The output field at drive frequency is then given by $a_\mathrm{out}(\omega_\D) = - \sqrt{\kappa_\mathrm{R}} b(\omega_\D) = -\sqrt{\kappa_\mathrm{R}} B_1 $. And the input field can be related to the semi-classical cosine-drive as $\sqrt{\kappa_\D} a_\mathrm{in}(\omega_\D) = i \Omega_\C /2 $, we hence find for the tansmission signal 
\begin{align}
S_{21} 
&= \frac{ a_\mathrm{out}(\omega_\D) }{a_\mathrm{in} (\omega_\D) } 
= 
-
\frac{ 
\sqrt{\kappa_\mathrm{R} \kappa_\mathrm{D}} }{\kappa/2+ i\Delta_\mathrm{C} + \rchi_\EC} 
\label{eq:S21_bare}, 
\end{align}
with $\rchi_\EC = \rchi_\EC(t)$ the mechanically compliant ETLS induced cavity shift given by Eq.~\eqref{Delta_cavity} and $\Delta_\mathrm{C} = \omega_\mathrm{C} - \omega_\P$ the cavity detuning. 

Experimentally we also account for the overall transmission (attentuation, probe power, amplifier gain) $T_0$, phase delay caused by the propagation time $\tau$ through the circuit, as well as an overall global phase shift for the full transmission, leading to  
\begin{equation}
    S_{21}(\omega, t) = T_0 \cdot \frac{ \sqrt{  \kappa_\mathrm{R} \kappa_\mathrm{D}}}{\kappa/2 + i \Delta_{\text{C}} + \rchi_\EC(t)} \cdot \exp\left[-i \{ (\omega_{\text{P}} - \omega_{\text{C}}) \tau + \phi \} \right],
    \label{eq:S21_VNA}
\end{equation}
where the time dependence indicates that the mechanics cause the system parameters to change dynamically.

\subsection{Cavity response averaged over mechanical motion}\label{sec:S21_thermal}

The variation of the cavity frequency can be detected by interference with a reference signal. 
Since typically $\kappa \ll \omega_\M$, the cavity does not have the bandwidth to react to a variation of $x(t)$ on the time scale of the oscillation period.
A simple way to take this slow reaction of the cavity into account is to average $S_{21}$ as given by \refe{eq:S21_bare} over one period of the ETLS drive. 
This allows to find the variation of the cavity frequency due to the time dependence of $x(t)$, which is what is sought to be measured (but also the variations due to the dependence of $\tilde{\omega}_\E$ on the drive itself).
The second important effect that has to be kept in mind is that the oscillator is anharmonic, due to the strong coupling to the ETLS. 
In order to fit the experimentally observed dependence of $S_{21}$ on the ETLS  driving frequency $\omega_\D$ and intensity $\varepsilon_\D$, we proceed by simulating the expected stationary periodic solution $x^{st}_1(t)$ of the nonlinear equation of motion for the oscillator:
\beq
    \ddot x = -\omega_\M^2 x -\gamma \dot x 
        -{\hbar \av{\sigma_z} \over 2 m } {\partial \tilde{\omega}_\E \over \partial x} 
        .
        \label{eom_class_mech}
\eeq
This expression takes the effect of the drive on the oscillator into full account: at lowest order in the drive intensity it is just a linear drive of the oscillator, but for larger drive higher order terms can contribute.
We use the thermal value for $\av{\sigma_z}$ given by \refe{thermalSigmaz}.
The result for the stationary periodic solution $x^{st}_1(t)$ can then be entered into the expression of $\tilde{\omega}_\E$ given by \refe{OmegaEq} and used to calculate numerically the average of $S_{21}$ over one period
\beq
    \av{S_{21}}
    =
    \frac{\omega_\D}{2 \pi} \int_0^{2\pi/\omega_\D}\!\!\!\!\!\!\!\!\!\!\!\!\!\!\!{dt}\,
    S_{21}(\tilde{\omega}_\E(t))
    .
    \label{eq:S21_mech_driven}
\eeq

Even in the absence of a mechanical drive, the signal $S_{21}(\tilde{\omega}_\E(x))$ is still affected by the thermal fluctuations of the oscillator mode. To estimate this temperature dependence of the signal, we assume the mechanical oscillator to be in a thermal state described by the Boltzmann distribution 
\begin{align}
    P_T(x) = \sqrt{\frac{m \tilde 
    \omega_\M^2}{2 \pi k_B T}}
    e^{ - m \tilde \omega_\M ^2 x^2 / 2k_BT } 
\end{align}
where $\tilde \omega_\M$ is the eigenfrequency of the mode, renormalized by the ETLS-interaction, see Eq.~\eqref{wm_renorm}, while we neglected here the induced anharmonicity. 
The averaged temperature-dependent cavity response is then given by 
\begin{align}
    \braket{S_{21}}_\mathrm{thermal}
    &= 
    \int_{- \infty}^\infty \dif  x P_T(x) S_{21}(\tilde{\omega}_\E(x)) 
    \,.
    \label{eq:S21_thermal}
\end{align}

\subsection{Effective optomechanical readout}
\label{sec:NEW__Effective_cavity-oscillator_coupling}
Here, we use the previously obtained result of Sec.~\ref{Cavity response to the drive} to analyze the effective coupling between the cavity and the mechanical oscillator.
We can analyze the effective optomechanical coupling by returning to Eq.~\eqref{b_dot}
\beq
    \langle \dot b \rangle 
    = 
    (-i\omega_\mathrm{C} -\kappa /2 ) \langle b \rangle
    -i\left[g_{x}  \langle \sigma_x \rangle 
    + g_{z} \langle \sigma_z \rangle \right] 
    + i \Omega_\C \cos(\omega_\P t) 
\eeq
which with $\braket{\sigma_{x(z)}}$
eliminated in second order of $g_\EC$, as derived in Sec.~\ref{Cavity response to the drive},
simplifies to 
\beq
    \langle \dot b \rangle 
    = 
    (-i\omega_\mathrm{C} -\kappa/2 ) \langle b \rangle
    - \rchi_\EC \langle b \rangle 
    + i \Omega_\C \cos(\omega_\P t) 
\eeq
with $\rchi_\EC = \rchi_\EC(x,t)$ given by Eq.~\eqref{Delta_cavity}. It yields the frequency  
shift $\Delta \omega_c = \im \rchi_\EC$, see Eq.~\eqref{Eq47}, and the additional damping constant $\re \rchi_\EC$ to the cavity. We get 
\beq
    \langle \dot b \rangle 
    = 
    i (\omega_\mathrm{C} + \Delta\omega_\C ) \langle b \rangle 
    - \tilde \kappa  \langle b \rangle /2
    + i \Omega_\C \cos(\omega_\P t) 
\eeq
with $\tilde \kappa = \kappa +2 \re \rchi_\EC$.
This corresponds to an effective Hamiltonian for the classical cavity field:
\begin{align}\label{eq:effective_classical_cavity_hamiltonian}
    H_\mathrm{eff}
    &= 
    ( \omega_\C + \Delta \omega_\C ) \braket{b^\dag b}  + \Omega_\C \braket{b+b^\dag}  \cos( \omega_\mathrm{Dc} t )
    .
\end{align} 

By expanding $\Delta \omega_\mathrm{C}$ given by Eq.~\eqref{Eq47} in the mechanical coupling constant $g_\EM$, we can find the explicit dependency of $\Delta \omega_\mathrm{C}$ on the square of the mechanical displacement $x(t)$. The full expression is rather cumbersome, however, applying the approximation $\Gamma_t \ll  2t_\E $ as well as $\omega_\P \ll 2t_\E$, we find in lowest order, 
\begin{align}
\Delta\omega_\mathrm{C}
\approx 
    \frac{2 g_\EC^2 (2t_\E)^2 \av{\sigma_z}}{\omega_\E^3}
    -
    g_\mathrm{OM}^{(1)}\left(\frac{x(t)}{x_z}\right)
    -
    g_\mathrm{OM}^{(2)}
    \left(\frac{x(t) }{x_z}\right)^2\,.
\end{align}
The first term is the ETLS induced static frequency shift, such as seen in Fig.~\ref{fig:SI_collage_fitting_results}, and $g_\mathrm{OM}^{(1)}$ and $g_\mathrm{OM}^{(2)}$ are the effective linear and quadratic optomechanical coupling rates, respectively, given by
\begin{subequations}
\label{eq:SI_definition_OM_coupling_rates}
\begin{align}
    g_\mathrm{OM}^{(1)} &= \left. \frac{d\Delta \omega_\mathrm{C}}{dx}\right|_{x=0} ~=  \frac{12 g_\EC^2 g_\EM (2t_\E)^2 \av{\sigma_z} \varepsilon_0}{\omega_\E^5}, \label{gom_1} \\
    g_\mathrm{OM}^{(2)} &= \frac{1}{2} \left. \frac{d^2\Delta \omega_\mathrm{C}}{dx^2}\right|_{x=0} = \frac{12 g_\EC^2 g_\EM^2 (2t_\E)^2 \av{\sigma_z} \left((2t_\E)^2-4 \varepsilon_0^2\right)}{\omega_\E^7} \label{gom_2} .
\end{align}
\end{subequations}
These optomechanical coupling rates are used to calculate the theoretical curves shown in main text Fig.~4e. Note that the linear coupling $g_\mathrm{OM}^{(1)}$ is suppressed to zero as $\varepsilon_0 \rightarrow0$, whereas the quadratic coupling $g_\mathrm{OM}^{(2)}$ is maximum at $\varepsilon_0=0$.

We first focus on the cavity frequency shift when driving the mechanical oscillator on resonance, such that $\varepsilon_\D \ll g_\EM x(t)/x_z$,
and we may hence omit the explicit driving term in $\tilde{\omega}_\E(t) \approx \sqrt{(2t_\E)^2 + ( \varepsilon_0 + 2 g_\EM x/x_z)^2}$.

We may find the effective amplitude-dependent frequency shift of the cavity by averaging over the fast oscillating $x(t)$. Approximating $x(t)$ to this end by a simple harmonic oscillation with amplitude $A$ yields
\begin{align}
\av{\Delta\omega_\mathrm{C}^\mathrm{R} }
\approx 
    \frac{2 g_\EC^2 (2t_\E)^2 \av{\sigma_z}}{\omega_\E^3} 
    -
    \frac{6 g_\EC^2 g_\EM^2 (2t_\E)^2 \av{\sigma_z} \left((2t_\E)^2-4 \varepsilon_0^2\right)}{\omega_\E^7}
    \left(\frac{A}{x_z}\right)^2 \,,
\end{align}
which simplifies further for $\varepsilon_0 = 0$ 
\begin{align}
\av{\Delta\omega_\mathrm{C}^\mathrm{R} }\big|_{\varepsilon_0 = 0}
    \approx
    -
    \frac{2 g_\EC^2 }{(2t_\E)}
    +
    \frac{6 g_\EC^2 g_\EM^2  }{(2t_\E)^3}
    \left(\frac{A }{x_z}\right)^2 \,,
    \label{eq:omega_c_R}
\end{align}
where we again set $\av{\sigma_z} = -1$ explicitly. 

In the contrary non-resonant driving regime, where the drive frequency $\omega_\D$ is set very far away from $\omega_\mathrm{M}$, we may approximate $g_\EM x(t)/x_z \ll \varepsilon_\D $ and hence $\tilde{\omega}_\E(t) \approx \sqrt{(2t_\E)^2 + ( \varepsilon_0 + \varepsilon_\D \cos(  \omega_\D t ))^2 }$. Applying again the approximations $\omega_\D \ll 2t_\E$ and $\Gamma \ll 2t_\E$, we find 
\begin{align}
\Delta\omega_\mathrm{C}^\mathrm{NR}
&=
    \frac{2 g_\EC^2 (2t_\E)^2 \braket{\sigma_z}
    }{\left((2t_\E)^2+(\varepsilon_\D \cos ( \omega_\D t )+\varepsilon_0)^2\right)^{3/2}}
\end{align}
which for small drives $\varepsilon_\D$ can be approximated in a Taylor series and averaged over a period of oscillations as before. We find
\begin{align}
\braket{\Delta\omega_\mathrm{C}^\mathrm{NR}}
&=
    \frac{2 g_\EC^2 (2t_\E)^2 \braket{\sigma_z}}{\omega_\E^3}-\frac{3 g_\EC^2 (2t_\E)^2 \braket{\sigma_z} \left((2t_\E)^2-4 \varepsilon_0^2\right)}{ 2 \omega_\E^7}\varepsilon_\D^2 
    \,,
\end{align}
and at $\varepsilon_0 = 0$ 
\begin{align}
\braket{\Delta\omega_\mathrm{C}^\mathrm{NR}}\big|_{\varepsilon_0 = 0}
&=
    \frac{2 g_\EC^2   \braket{\sigma_z}}{(2t_\E)}-\frac{3 g_\EC^2   \braket{\sigma_z}    }{ 2 (2t_\E)^3}\varepsilon_\D^2 
    \,.
    \label{eq:omega_c_NR}
\end{align} 

\subsection{Analysis of classical Kerr oscillator}
Here, we study the amplitude and resonance frequency of the driven classical Kerr (Duffing) oscillator resulting from the coupling of the mechanical oscillator to the ETLS.
The equation of motion for the mechanical oscillator~\refe{eom_class_mech} can be explicitly written as
\begin{align}\label{Eq:EOM_driven_mech}
    \ddot x = -\omega_\M^2 x -\gamma \dot x 
        -
        2 \hbar  g_\EM  \omega_\M x_z {\varepsilon(t) \over \tilde{\omega}_\E(t)}   \av{\sigma_z}  
\end{align}
where we used that $2 \omega_\M x_z = 1/  m x_z $ and we note again that $\tilde{\omega}_\E(t) = \sqrt{(2t_\E)^2 +\varepsilon(t)^2}
$
and
$
\varepsilon(t) = \varepsilon_0 + \varepsilon_\D \cos(\omega_\D t) + 2 g_\EM x /x_z$. 
For now we will neglect the driving of the DQD, i.e.~$\varepsilon_\D = 0$. 
Assuming that $g_\EM x/ x_z$ is sufficiently small compared to $2t_\E$, we may expand the right-hand side and find    
\begin{align}
        \ddot x \approx 
        - 
        ( \omega_\M^2 + g \alpha ) x 
       + \alpha_0 x_z 
        +g^2  \beta 
        \frac{x ^2}{x_z} 
        + g^3 \delta 
         \frac{x^3 }{x_z^2} 
        -\gamma \dot x     
\end{align}
where we introduced the small parameter $g = g_\EM / \omega_\E$.
This is an anharmonic Duffing oscillator with 
\begin{align}
\alpha_0 
&= -2 g_\EM \omega_\M  \av{\sigma_z}\frac{\varepsilon_0}{\omega_\E}  \,, 
\label{alpha0}
\\
    \alpha &= 4  g_\EM \omega_\M  \av{\sigma_z}
        \frac{ (2t_\E)^2}{\omega_\E^2}
        \,,
    \\ 
    \beta & = 
        12   g_\EM \omega_\M \av{\sigma_z}
        \frac{ (2t_\E)^2\varepsilon_0  }{ \omega_\E^3}
        \,,
    \\ 
    \delta &=
        8  g_\EM \omega_\M \av{\sigma_z} 
        \frac{ (2t_\E)^2 \left((2t_\E)^2 -4 \varepsilon_0^2\right)}{   \omega_\E^4}
        \,.
        \label{delta}
\end{align}
 At maximum frequency softening, i.e.~at $\varepsilon_0 = 0$, this simplifies to 
\begin{align}
          \ddot x \approx 
        -
        \left( 
        \omega_\M^2
        + 
        \frac{4 g_\EM^2 \omega_\M}{(2t_\E)}  \av{\sigma_z} 
        \right)
        x 
        +
        \frac{8 g_\EM^4 \omega_\M }{(2t_\E)^3  } \av{\sigma_z}   
        \frac{x^3 }{x_z^2} 
        -\gamma \dot x  
        \,.
\end{align}

Let us next include the drive of the ETLS. 
Assuming that the drive amplitude $\varepsilon_\D$ is small compared to the ETLS energy splitting $\omega_\E = \sqrt{(2t_\E)^2 + \varepsilon_0^2}$, we can approximate in lowest order of $\varepsilon_\D$, 
\begin{align}
          \ddot x \approx 
         - 
        ( \omega_\M^2 + g \alpha ) x 
       + \alpha_0 x_z 
        +g^2  \beta 
        \frac{x ^2}{x_z} 
        + g^3 \delta 
         \frac{x^3 }{x_z^2} 
        -\gamma \dot x    
        +   f_\D(x) \cos(\omega_\D t ) 
        \,.
        \label{ddotx}
\end{align}
with
\begin{align}
f_\D (x) 
&= \varepsilon_\D  \omega _\M \frac{(2t_\E)^2}{\omega_\E^2}\left( 
-\frac{2 g_\EM x_z}{\omega_\E}
+\frac{12\varepsilon_0 g_\EM^2}{\omega_\E^3}  x 
+\frac{12  g_\EM^3 \left((2t_\E)^2-4 \varepsilon_0^2\right)}{\omega_\E^5  } \frac{x^2}{x_z} \right) 
\end{align}
where the dominant term is $x$-independent and yields a standard driving term $\propto \cos(\omega_\D t)$ and we neglected terms with order $\varepsilon_\D g_\EM^4/\omega_\E^5$ and higher. 

Introducing a constant shift $\tilde{x}(t) = x_s + x(t)$, we can rewrite the equation of motion as 
\begin{align}
          \ddot x \approx 
         - 
        \left( \tilde \omega^\mathrm{sc}_\M
        \right)^2   x 
        + g^2 \tilde \beta 
        \frac{x ^2}{x_z} 
        + g^3 \delta 
         \frac{x^3 }{x_z^2} 
        -\gamma \dot x    
        +   f_\D(\tilde x) \cos(\omega_\D t ) 
        \,.
        \label{ddotx_shifted}
\end{align}
with $\tilde \beta = \beta+ 3g   \delta x_s /x_z$ and
the shift $x_s$ is determined by
\begin{align}
0
= 
 \alpha_0  x_z 
 + 
x_s \left(-
     g \alpha   - \omega _\M^2\right)
+
    g^2 \beta
\frac{ x_s^2}{x_z}
+
     g^3 \delta
\frac{ x_s^3}{x_z^2}
\,.
\end{align}
In leading order of $g = g_\EM /\omega_\E$, this is solved by 
\begin{align}
x_s \approx \left( \frac{\alpha _0  }{\omega _\M^2}-\frac{g_\EM  \alpha  \alpha _0 }{\omega_\E \omega _\M^4} \right) x_z 
\,,
\end{align}
and the renormalized semi-classical frequency $\tilde \omega^\mathrm{sc}_\M$ in Eq.~\eqref{ddotx_shifted} is then given by
\begin{align}
    \left(\tilde \omega^\mathrm{sc}_\M\right)^2 
    =
    \omega _\M^2 + g \alpha    
    - 2 g^2 \beta \frac{x_s}{x_z} 
    - 3 g^3 \delta \frac{x_s^2}{x_z^2}
\end{align}
which in leading order of $g$ and for the ETLS in its groundstate, yields
\begin{align}
    \tilde \omega_\M^\mathrm{sc} 
    =
   \omega _\M-\frac{2 g_\EM^2 (2 t_\E)^2}{\omega_\E^3} \,.
   \label{wm_renorm}
\end{align}
which is the same leading order renormalisation we found from the diagonalisation of the Hamiltonian, see Eq.~\eqref{second_order_shift}, when approximating additonally $\omega_\M \ll \omega_\E$.
\subsubsection{Effective resonance frequency}
For simplicity, here we only keep the linear driving term $f_\D = f_\D(x=0)$.
% % = - 2 \varepsilon_\D g_\EM x_z \omega_\M (2t_\E)^2 /\omega_\E^3}$, where $x_z = 1/\sqrt{2 m \omega_\M}$ 
% and consider the case of $\varepsilon_0 = 0$. 
%
We use as an ansatz for $x(t)$ a fast harmonic oscillation with $\omega_\D$ and slow varying amplitude $A(t)$ and phase $\phi(t)$, 
\begin{align}
    x(t) = A(t) \cos( \omega_\D t + \phi(t))  \,.
\end{align}
Plugging the ansatz $x(t)$ into the differential equation and applying $\dot A = \dot \phi = 0$ to solve for the stationary solution, we find 
\begin{align}
  \left[
  \omega_\D^2 
  -
  \left(\tilde \omega ^\mathrm{sc}_\M\right)^2  \right]
x  
  -  \gamma \dot x  
+
    g^2 \tilde{\beta}  
    \frac{ x^2}{x_z}
+
g^3 
    \frac{ x^3}{x_z^2} 
= 
- 
f_\D \cos \left(t \omega _\D\right)
\,.
\end{align}
 
Using this ansatz and keeping only first harmonics, i.e.\ effectively decouple from fast oscillating terms, such that inside of the differential equation we may approximate 
\begin{align}
        x^2 &=u^2  e^{2 i \omega_\D t}+ u^{*2} e^{-2 i \omega_\D t} + 2 |u |^2  \approx 0  \,,
    \\
    x^3 &= u^3 e^{3 i \omega_\D t}+ u^{*3} e^{-3 i \omega_\D t} +3 |u |^2 x \approx \frac{3}{4}A^3  \cos( \omega_\D t + \phi)\,,\end{align}
with $u = A  e^{i \phi }/2$,
we find from the equation of motion, 
\begin{multline}
     A \left[
  \omega_\D^2 
  -
  \left(\tilde \omega^\mathrm{sc} _\M\right)^2  
+
\frac{3 g^3 \delta  }{4  } \left(\frac{A}{x_z}\right)^2
    \right]
\cos( \omega_\D t + \phi)  
\\
+ \gamma \omega_\D A \sin( \omega_\D t + \phi ) 
= 
f_\D [\cos( \omega_\D t + \phi ) \cos \phi 
     + \sin( \omega_\D t + \phi) \sin \phi] \,,
\end{multline}
where we used $\cos(\omega_\D t) = \cos( \omega_\D t + \phi ) \cos \phi 
     + \sin( \omega_\D t + \phi) \sin \phi$ for the drive. 

Comparing the terms that oscillate with the same phase, we get two equations, 
\begin{align} 
f_\D   \cos \phi  
&=
         A \left[
  \omega_\D^2 
    -
  \left(\tilde \omega^\mathrm{sc} _\M\right)^2  
+
\frac{3 g^3 \delta  }{4  } \left(\frac{A}{x_z}\right)^2
    \right] 
     \\
f_\D  \sin \phi
&= 
 \gamma \omega_\D A   \,.
\end{align}
We can square these two equations and add them up to find the phase-independent relation between amplitude and drive as, 
\begin{align}
    A^2 \left[
  \omega_\D^2 
    -
  \left(\tilde \omega^\mathrm{sc} _\M\right)^2  
+
\frac{3 g^3 \delta  }{4  } \left(\frac{A}{x_z}\right)^2
    \right] ^2 
    +A^2 \gamma ^2 \omega _\D^2 =  f_\D^2 ,
\end{align}
which yields then the peak frequency ( at $A'(\omega_\D) = 0$ ) and by applying that $\gamma \ll \omega_\M$,
\begin{align}
    \omega_\D 
     & = 
    \tilde \omega^\mathrm{sc} _\M-\frac{3  g^3 \delta  }{8 \tilde \omega^\mathrm{sc} _\M \ } 
    \left(
    \frac{A}{x_z}
    \right)^2
    \\
    &= \tilde \omega^\mathrm{sc} _\M
    + 
    \frac{3   (2t_\E)^2 g_\EM^4 \left((2t_\E)^2-4 \varepsilon _0^2\right)}{  \omega _\E^7} 
    \left(
    \frac{A}{x_z}
    \right)^2
    \,,
\end{align}
where we used the explicit expression for $g = g_\EM / \omega_\E$ and $\delta$ (see Eq.~\eqref{delta}), as well as $\braket{\sz}_0 = -1$ in the last line.
From this we can see that the amplitude dependency vanishes for $\varepsilon_0 = t_\E$. 
At maximum frequency softening, i.e.~for $\varepsilon_0 =0$ the resonance frequency simplifies to 
\begin{align}\label{eq:SI_version_of_maintext_eq2}
    \omega_\D 
    &= 
    \tilde \omega^\mathrm{sc} _\M
    + \frac{3 g_\EM^4}{(2t_\E)^3}  \left( \frac{A}{x_z} \right)^2
\end{align}

We can compare this result to the one obtained for the transition frequencies in the quantized case, see Eq.~\eqref{anharm_Delta_omega_M}, where we found that the transition frequency and hence the resonance frequency for driving the transition, scales as 
\begin{align}
    \omega_\D = \tilde{\omega}_\M +  \braket{n} K
\end{align}
where $\braket{n}$ is the expectation value of the number operator. 
In the semiclassical limit ${\omega_\M \ll \omega_\E}$, we find that $K$ can be approximated by 
\begin{align}
    K \approx \frac{12 g_\EM ^4 (2 t_\E)^4}{\omega_\E^7}
\end{align}
and we may evaluate $\braket{n}$ in a coherent state $\ket{\alpha}$, with $a \ket{\alpha} = \alpha \ket{\alpha}$ as $\braket{n} = | \alpha|^2$, while in a coherent state the displacement oscillates with 
\begin{align}
    \braket{x(t)} 
    = 2 |\alpha|  x_z \cos( \omega t + \phi  ) 
\end{align}
i.e.~with an amplitude $A/x_z = 2 |\alpha|  = 2 \sqrt{\braket{n}}  $, such that we find from the quantum result in the semi-classical limit,
\begin{align}\label{mech_freq_shift_to_Kerr}
    \omega_\D = \tilde{\omega}_\M +   \frac{K}{4} \left(  \frac{A}{x_z}\right)^2
    \approx   \tilde{\omega}^\mathrm{sc}_\M + \frac{3 g_\EM^4 (2 t_\E)^4}{\omega_\E^7}
     \left(  \frac{A}{x_z}\right)^2
\end{align}
which coincides with the result obtained in this section from a classical Duffing oscillator approach.

\subsection{Thermal broadening of the main mechanical mode}
\label{SectionBroadening}

In this Section we investigate the effect of thermal fluctuations on the linewidth of the mode of interest induced by the interplay between its intrinsic Kerr nonlinearity, intermode nonlinear coupling (cross-Kerr), and thermal fluctuations.
In principle for a perfectly symmetric nanotube, one can expect that the flexural modes are double degenerate due to the polarization degree of freedom.
We will consider only polarization orthogonal to the gates for each mode,
assuming that the other one is negligibly coupled to the ETLS.

The free Hamiltonian for the flexural modes is then given by 
\beq
    H_\mathrm{flex.} 
    = H_\M + \sum_n \hbar \omega_n a_n^\dag a_n \,,
\eeq
where $H_\M = \omega_\M a^\dag a$ is the free Hamiltonian of the main mode introduced above, 
and we introduced the resonating frequencies $\omega_n$, and the annihilation operators for the secondary modes $a_n$.
We assume that the numbering of the mode is such that they are ordered 
by increasing frequency so that  
$\omega_1 < \omega_\M < \omega_n$ for $n>1$. 
The flexural modes are coupled to the ETLS via the interaction Hamiltonian 
\beq
    V_{EMs} = \sum_n g_n (a_n^\dag + a_n) \sx,
\eeq
where $g_n$ is the respective coupling constant and $a_n$ the respective ladder operator. Similarly the main mode is coupled by $V_\EM$, see Eq.~\eqref{VEM}, with a coupling constant $g_\EM \gg g_n$.
The Hamiltonian of the coupled ETLS-flexural-modes-system then reads (see \refe{eq:H_ETLS_mech})
\begin{align}
  H &= H_\E  + H_{\rm flex.} + V_\EM + V_{EMs} .
  \label{H_ETLS_flex}
\end{align} 
In order to define in a simple way the problem, we will consider the system in thermal equilibrium and calculate the spectrum of the $x$ operator defined as usual:
\beq
   S_{xx}(\omega)=\int dt e^{i\omega t} 
    {\rm Tr} \left\{ 
    e^{iH t} x e^{-i H t} x e^{-H/k_B T} 
    \right\} /{\rm Tr} \left\{ e^{-H/k_B T} 
    \right\} ,
    \label{S1definition}
\eeq
here $H$ is the ETLS-flexural-modes Hamiltonian defined in \refe{H_ETLS_flex}.
We assume here that the coupling of the main mode to the environment has a negligible effect on the mechanical linewidth of the main mode compared to the frequency noise broadening, given by the interplay of the thermal fluctuations with the anharmonicities.
This method allows us to write an explicit expression of the spectrum of the main mode and obtain an analytical estimation.
Performing the trace in \refe{S1definition} on the eigenstates of $H$ and the integration in time one obtains the standard Lehmann representation of the spectrum of the main mode:
\beq
   S_{xx}(\omega)= 2\pi x_z^2 
    \sum_{qp} P_{p} |\langle q| a+a^\dag | p\rangle|^2 \delta(\omega-(E_q-E_p)/\hbar)
    \label{Lehmann}
\eeq
where $|p\rangle$ and $E_p$ are the eigenvectors and eigenvalues of the Hamiltonian, respectively, and 
$P_p$ is the probability of occupation of state $p$.
By numerical diagonalization one can then calculate the spectrum. 

In order to proceed analytically we 
diagonalize the Hamiltonian in the ETLS sector, regarding formally the operators $\hat x_n=(a_n+a_n^\dag)$ as classical fields:
\beq
    H/\hbar = \omega_\M a^\dagger a 
    + 
    \sum_n \omega_n a_n^\dag a_n
    +
    {\sqrt{(2t_\E)^2+(\varepsilon_0+2 g_\EM \hat x + 2\sum_n g_n \hat x_n )^2 }\over 2} \sigma_z .
\eeq
We expect that this should be a good approximation as far as $2 t_\E \gg \omega_\M, \omega_n$.
We can now expand to fourth order in $\hat x_{n}$ and $x_\M$.
Keeping only the first two modes: the lowest mode $\omega_1$ and the main mode of interest $\omega_\M$, with $\omega_1 < \omega_\M$, we obtain for the lowest ETLS branch (for $ 2 t_\E\gg k_B T$ we may safely assume that only states corresponding to the ETLS ground state, $\braket{\sigma_z} =-1$,  contribute):
\beq
    H/\hbar=-t_\E+
    \omega_\M a^\dag a
    +
    \omega_1 a_1^\dag a_1
    -
    {\left[g_\EM \hat x +g_1 \hat x_1 \right]^2 \over  (2t_\E)}
    +
    {\left[g_\EM \hat x +g_1\hat x_1 \right]^4 \over (2t_\E)^3}\,,
\eeq
where we again set $\varepsilon_0 = 0$ for simplicity. 
We expand now the last two terms neglecting the fast rotating terms with an odd number of the field $a$ or $a_1$. 
 We obtain then for the effective Hamiltonian for the two modes:
\beq
    H/\hbar = \sum_{i \in \{1, \M \}}\left[ \omega_i \hat n_i
    -{g_i^2 \over (2t_\E)} (2 \hat n_i+1)
    +{g_i^4 \over (2t_\E)^3}(6 {\hat n_i}^2+ 6\hat n_i +3)
    \right]
    +{6g_\EM^2 g_1^2 \over (2t_\E)^3}(2 \hat n_\M+1)(2 \hat n_1+1)
    ,
\eeq
where $g_\M$ is a short notation for $g_\EM$ inside the sum,
$\hat n_\M=a^\dag a$, and  $\hat n_1=a_1^\dag a_1$\,.
Collecting the powers of $\hat n_i$ and discarding constant terms, we have:
\beq
    H/\hbar = \sum_{i \in \{1, \M \}}
    \left[ \tilde \omega_{i} \hat n_i
    + \frac{K_{ii}}{2} {\hat n}_i^2
    \right]
    + \frac{K_{\M1}}{2} \hat n_\M\hat n_1,
\eeq
with eigenenergies 
\beq
    E_{n_\M,n_1}/\hbar = 
    \sum_{i=1}^2 [\tilde{\omega}_{i}
    n_i+\frac{K_{ii}}{2} n_i^2]+ K_{\M1} n_\M n_1 ,
\eeq
where $\tilde{\omega}_{i}$ are the renormalized resonance frequencies and $K_{ij}$ the Kerr nonlinearities defined by 
\beq
    \tilde{\omega}_{i} 
    = 
    \omega_i-2 {g_i^2\over (2t_\E)} + {6 g_i^4 \over (2t_\E)^3}
    \,,
    \quad
    K_{ii} = 
    {12 g_i^4 \over (2t_\E)^3}
\,,       
\quad
    K_{\M1} = 
    {24 g_\EM^2 g_1^2 \over (2t_\E)^3}
    .
    \label{coefficientsChi}
\eeq 
We apply now the expression of the Lehmann representation \refe{Lehmann} to this 
Hamiltonian considering only positive frequencies (absorption spectrum) of the main mode.
\beq
   S_{xx}(\omega>0) = 2 \pi x_z^2 \sum_{n_\M,n_1} P_{n_\M,n_1} (n_\M+1)  
    \delta(\omega-\omega_{n_\M,n_1} ) 
\eeq
where $\omega_{n_\M,n_1}=(E_{n_\M+1,n_1}-E_{n_\M,n_1})/\hbar$ 
yielding 
\beq
    \omega_{n_\M,n_1}
    =
    \tilde{\omega}_\M + K_{\M\M}/2 + K_{\M\M}  n_\M+K_{\M1} n_1 ,
\eeq
the factor
$(n_\M+1)$ results from evaluating the square of $\hat x$,
and 
\beq
P_{n_\M,n_1}=e^{-\beta E_{n_\M,n_1}}/\sum_{n_\M,n_1} e^{-\beta E_{n_\M,n_1}}
\eeq
are the Boltzmann weights and $\beta=\hbar/k_B T$.
Note that at this level of approximation the eigenstates are the same of the non-interacting system, which allows to evaluate the matrix element of $\hat x$ on the states $|n_\M\rangle$. This also selects only the states differing of one phonon of the main mode.

We can now estimate the width of the resonance by evaluating the average of $\omega$
and its variance from the spectral density \refe{Lehmann} as
\beq
    \av{\omega_\M^k} = \int_0^\infty d \omega \omega^kS_{xx}(\omega) / \int_0^\infty d\omega S_{xx}(\omega)
    .
\eeq
This gives 
\beq
    \av{\omega_\M^k}
    = \sum_{n_\M,n_1} P_{n_\M,n_1}(n_\M+1) \omega^k_{n_\M,n_1} /\sum_{n_\M,n_1} P_{n_\M,n_1}(n_\M+1) .
\eeq
The variance of the resonance frequency reads:
\begin{align}
\av{\delta \omega_\M^2 }
&=
    \av{\omega_\M^2}-\av{\omega_\M}^2
    \nonumber
    \\
   &=
   K_{\M\M}^2 (\av{n_\M^2}-\av{n_\M}^2)
    +K_{\M1}^2(\av{n_1^2}-\av{n_1}^2)
    \nonumber
    \\& \qquad\qquad\qquad\qquad
    +2 K_{\M\M} K_{\M1} (\av{n_\M n_1}-\av{n_\M}\av{n_1}) .
    \label{variances}    
\end{align}
This expression can be easily evaluated numerically, but when neglecting the $K$ terms in the exponentials of $P_{n_\M,n_1}$ we can obtain a closed expression for the variance of the resonance frequency. 
In particular, neglecting the term proportional to $\rchi_{M1}$, the spectral distributions for $n_\M$ and $n_1$ become disjoint $P_{n_\M, n_1 } = P_{n_\M} P_{n_1}$, and we can perform their averages independently. 
In this case the last term of \refe{variances} vanishes, and the average of the $n_1$ term is performed only with the Boltzmann distribution. 
Performing the sums one finds: 
\beq
    \av{n_1^2}-\av{n_1}^2 = 2 n^2_B(\tilde{\omega}_{1})
    ,
\eeq
where $n_B(\omega)=1/(e^{\beta \omega}-1)$ is the Bose distribution, while for $n_\M$ we are left with the  distribution additionally weighted with $(n_\M+1)$, yielding
\beq
    \av{n_\M^2}-\av{n_\M}^2 = 2 n^2_B(\tilde{\omega}_{\M}) e^{\beta \tilde{\omega}_{\M}} 
    .
\eeq

We thus finally have
\beq
    \av{\delta \omega_\M^2}
   =
   2K_{\M\M}^2  n^2_B(\tilde{\omega}_{\M}) e^{\beta \tilde{\omega}_{\M}} 
    +2 K_{\M1}^2 n^2_B(\tilde{\omega}_{1}) ,
    \label{varianceOm}
\eeq
where the expressions for the coefficients are given in \refe{coefficientsChi}.
This is the main result of this Section. The square root of \refe{varianceOm} gives a good estimate of the thermal linewidth of the main mode.

\begin{figure}
    \centering
    \includegraphics[width=0.4\linewidth]{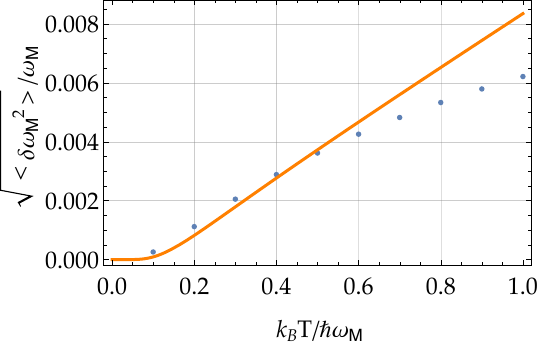}
    \includegraphics[width=0.4\linewidth]{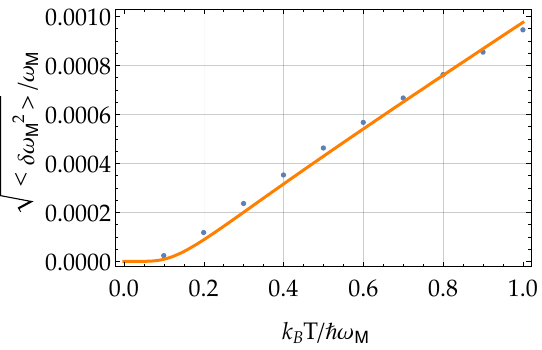}
    \vspace{-.4cm}
    \customcaption{Comparing of the numerical calculation with the Lehmann representation from the exact diagonalization of the Hamiltonian with the ETLS degrees of freedom and the analytical expression. These two plots are obtained for $t_\E=5 \omega_\M$ on the left and $t_\E=10 \omega_\M$ on the right. For both figures $\omega_1/\omega_\M=0.5$, $g_\EM=0.8 \omega_\M$ and $g_1=0.3 \omega_\M$.
    In the numerical calculation of the width one has to use a cut off in the frequencies ($\omega<\omega_\M$) to avoid including the contribution of the upper branch that lead to a small peak symmetric to the lower branch peak. These calculations are performed with $20 \times 20 \times 2$ states.}
    \label{fig:TempDep}
    \vspace{-0.3cm}
\end{figure}
\begin{figure}
    \centering
    \includegraphics[width=0.4\linewidth]{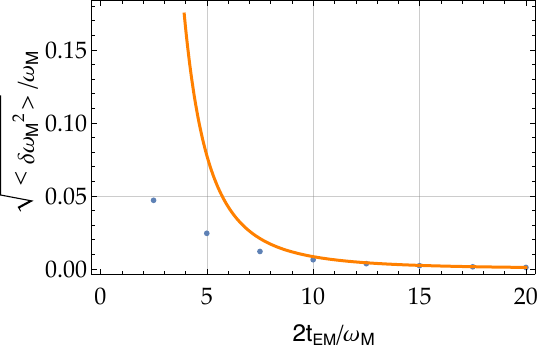}
    \vspace{-.4cm}
    \customcaption{Same as Fig. \ref{fig:TempDep} as a function of $t_\E$. 
    The other parameters are $\omega_1/\omega_\M=0.5$, $g_\EM=0.8 \omega_\M$ and $g_1=0.3 \omega_\M$, $k_B T/\hbar \omega_\M=1$.
    The difference at low value of $t_\E$ is probably due to the breaking down of the simple diagonalization approach. 
    }
    \label{fig:Jdep}
    \vspace{-0.3cm}
\end{figure}

In order to verify the validity of the approximations used, we performed the exact numerical calculation of the spectrum using \refe{Lehmann} with the full Hamiltonian \refe{H_ETLS_flex} with two modes. 
The results are shown in Figs. \ref{fig:TempDep}-\ref{fig:Jdep} and compared 
with the  analytical expression \refe{varianceOm} as a function of the temperature and $t_\E$. 
We have chosen the other parameters (indicated in the caption) in the typical range of the experiment. 
One clearly sees that the analytical expression works well for $\omega_\M\ll (2t_\E)$, but shows important variations when $t_\E<5 \omega_\M$. 
Nevertheless the analytical expression gives the right order of magnitude for the full range of interest, and can be used as a first estimate of the broadening induced by the thermal fluctuations. 
Note that the broadening vanishes at low temperature, as expected from the Lehmann expression: only one state contributes in the vanishing $T$ limit, the ground state, so there cannot be a finite width. 
Of course if the system is driven the situation can be more complex.

In conclusion we have found a simple expression for the broadening induced on the main mode by the thermal fluctuations in the quantum and classical regime of the mode itself (term in $K_{\M\M})$ and from the coupling to another mode (the $K_{\M1}$ term).
This broadening can be an important limitation if the lower mode is strongly coupled (like in this case) and the temperature is not sufficiently low. 

\begin{figure}
    \centering
    \includegraphics[width=0.4\linewidth]{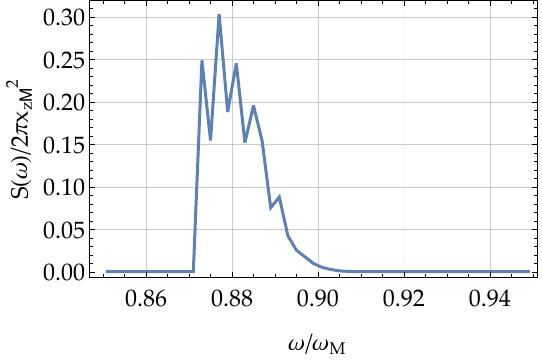}
    \customcaption{Shape of the resonance for $t_\E=5 \omega_\M$, $\omega_1/\omega_\M=0.5$, $g_\EM=0.8 \omega_\M$ and $g_1=0.3 \omega_\M$, $k_B T/\hbar \omega_\M=1$.
    The oscillations are due to the finite number of states considered ($30\times40\times2$ states) and the absence of dissipation in the model. 
    A simple binning with $\Delta \omega=0.002 \omega_\M$ is used to reproduce the figure from  \refe{Lehmann}.
}
    \label{fig:Som}
\end{figure}

As a byproduct of this investigation we find also the shape of the line that can be asymmetric. 
We show a typical form of the spectrum in Fig.~\ref{fig:Som}.

\subsubsection{Scaling of the mechanical linewidth induced by thermal fluctuations}

In this subsection, we will derive the scaling $ {\delta \omega}_\M 
     \propto {g_\EM^4/ (2t_\E)^3}$ of the mechanical linewidth induced by thermal fluctuations used in the main text. For this, we will first analyze the expected scaling of the electromechanical coupling constant at different ICTs with $(N+1,N)\leftrightarrow(N,N+1)$ of the charge stability diagram  so that the charge on the carbon nanotube is
$Q=-(2N+1)e$, where $N$ is an integer and $e$ the electron charge. 
Taking a discrete picture along the suspended carbon nanotube one can associate a gate capacitance to each gate $i$ defined by $C_{gi}$. 
The gate capacitance depends on the displacement of the nanotube that for simplicity we assume in the $z$ direction and we define the nanotube displacement in gate $i$ section as:
\beq
    z_{i}= \sum_n A_n u_n(s_i) ,
\eeq
where $u_n(s)$ gives the shape of the mode $n$ and $s_i$ ($0<s_i<1$) indicates the position of the $i$ capacitance along the nanotube as a fraction $s$ of the total length. 
For instance, we have in the simplest case sinusoidal modes $u_n(s)=\sin(n s \pi)$, and in devices with five gate electrodes as in the main text 
$s_i\approx i/6$ with $i=1,\dots , 5$.
The mode displacement amplitude is $A_n$. 

We define now the charge distribution on the nanotube double quantum dot where the left ($Q_i^+$) or right dot 
($Q_i^-$) is occupied by $Q_i^\pm=Q_i^{(0)}+\delta Q_i^\pm$, with the condition $\sum_i \delta Q^\pm_i=-e c_g $ and $\sum_i Q_i^{(0)}=-N c_g e$,
where $0<c_g<1$ is the fraction of the charge localized on the dot-gate capacitances.
In this way one takes into account that even when the electron is localized on one of the two quantum dot, the charge can be distributed on the different segments of the nanotube.
A fraction of the charge is distributed on the left and right lead capacitance, the rest is on the gate capacitances.
In the following we will assume for simplicity that the fraction localized on the leads and gates does not depend on $N$
since we expect that the shape of the electrostatic configuration does not change for different ICTs.

The force acting on the mode $n$ of the nanotube in one of the two states reads:
\beq
    F^\pm_n = 
    -\sum_i 
    {(Q_i^\pm)^2\over 2}
    {\partial \over \partial A_n} (C_{gi}(z_i(A_n)))^{-1}
    .
    \label{Force}
\eeq
The coupling constant for the mechanical mode $n$ coupled to the charge imbalance in the double quantum dot reads:
\beq
    \hbar g_n = (F^+_n-F^-_n)z^{\rm zpm}_{n} ,
\eeq
where $z^{\rm zpm}_{n}$ is the zero point motion of the mode $n$.
Expanding the square in \refe{Force} 
one obtains
\beq
    \hbar g_n = -z^{\rm zpm}_{n}
     \sum_i 
    [
    2 Q_i^{(0)} (\delta Q_i^+ -\delta Q_i^-)
    +
    (\delta Q_i^+)^2 -(\delta Q_i^-)^2
    ]   
    {\partial \over \partial A_n} (C_{gi}(z_i))^{-1} /2.
\eeq
Note that for a symmetric charge distribution, only asymmetric modes will contribute to $g_n$.
We can now assume that $Q_i^{(0)}=-2Ne c_g f_i$, with
$\sum_i f_i=1$ and $f_i$ indicate the distribution of the charge on the different gate capacitances that we assume independent of $N$.
With this simplifying hypothesis one obtains:
\beq
    \hbar g_n = z^{\rm zpm}_{n}
     \sum_i 
    [
    4N e c_g f_i (\delta Q_i^+ -\delta Q_i^-)
    -
    (\delta Q_i^+)^2 +(\delta Q_i^-)^2
    ]   
    {\partial \over \partial A_n} (C_{gi}(z_i))^{-1} /2.
\eeq
In principle, assuming that the charge distribution is not strongly modified from one ICT to another one, the coupling $g_n$ depends primarly on the first term that contains $N$.
For large $N$ the second term is negligible and one finds a linear scaling. 
This does not apply to the first values of $N$, in particular $N=0$ considered in the main text. 

Nevertheless a scaling of the different coupling constants for the same ICT point can be explained by assuming that the charge is localized on top of the two main gates below the two quantum dot, the gates 2 and  4 (see main text).
In this case we can assume that 
$\delta Q_2^+=\delta Q_4^-=-e c_g $, and $\delta Q^\pm_i=0$ for the other gates. 
Then the expression for $g_n$ becomes:
\beq
    \hbar g_n = -z^{\rm zpm}_{n} e^2 c_g^2
    (
    4Nf 
    +
    1
     )   
    [
    {\partial \over \partial A_n} (C_{g2}(z_2))^{-1}
    -
    {\partial \over \partial A_n} (C_{g4}(z_4))^{-1}
    ]
    /2,
\eeq
where for simplicity we assumed also $f_2=f_4=f$.
This gives 
\beq
    g_n(N)/g_n(N=0)
    =
    (
    4Nf
    +
    1
     ),
\eeq
which does no longer depend on $n$. 
Another way of writing the same relation is that 
$g_n(N)=\alpha_n g_\EM(N)$, with $\alpha_n$ independent of $N$.
Thus within these (rather strong) approximations one finds a simple scaling of the coupling constant. 
For $f \approx 0.1$, 
$N=4$ one could explain the variation of the coupling constant observed in the main text $g_\EM^{45}/g_\EM^{01}=0.46/0.18\approx 2.5$.

We now look at the linewidth of the main mode induced by the thermal mechanical  fluctuations. When we consider the expression of the variation of $\delta \omega_i^2$ (\refe{varianceOm}) and 
the expression of the Kerr terms (\refe{coefficientsChi}), we see that all the terms entering $\sqrt{\delta \omega_i^2}$
are combinations of coupling constant at the fourth power, for instance 
$g_\EM^4$ or $g_1^2 g_\EM^2$. 
Since $g_n=\alpha_n g_\EM$, the mechanical linewidth induced by thermal fluctuations is expected to scale from one ICT to another as the ratio of the fourth power of the two main coupling constants   $\sqrt{\delta \omega_i^2} \sim g_\EM^4(N)$, as discussed in the main text.
More explicitly from \refe{varianceOm} one can write:
\beq
\delta {\omega}_\M 
= 
K_{\M\M} {\tilde n}_B
\eeq
where 
\beq
{\tilde n}_B
=
\left(
2 n^2_B(\tilde{\omega}_\M)
e^{\hbar \omega_\M^\mathrm{R}/k_B T} + 
2 \sum_n  (K_{\M n}/K_{\M\M})^2 n^2_B(\tilde{\omega}_n)
\right)^{1/2},
\eeq
and where we reintroduced the sum over the other mechanical modes $n$.
Using the scaling of the Kerr factors 
the ratios $K_{\M n}/K_{\M\M}$ 
does not depend on the ICT leading to the simple scaling
\beq\label{eq:SI_version_mechanical_thermal_broadening}
    {\delta \omega}_\M 
     = {g_\EM^4\over (2t_\E)^3} {\tilde n}_B
\eeq
as a function of $g_\EM$ and $t_\E$.
The observed $1/t_\E^3$ scaling is shown in Fig.~4c of the main text.  

It is worth noting that the same $1/t_\E^3$ scaling is expected for linewidth broadening arising from detuning noise, such as charge fluctuators in the surrounding dielectrics or gate-array voltage noise. In this case, the induced mechanical linewidth scales as
\beq
\delta {\omega}_\M \propto
\frac{g_\EM^2}{(2t_\E)^3}\,\delta\varepsilon^2,
\eeq
where $\delta\varepsilon^2$ is the variance of the detuning noise. The mechanisms can however be distinguished by their strong dependence on the electromechanical coupling strength, with $\delta\omega_\M \propto g_\EM^4$ for thermomechanical broadening, whereas $\delta\omega_\M \propto g_\EM^2$ for electrical detuning noise.

\putbibNoTOC{bibliography}

\end{bibunit}

\end{document}